\documentclass[a4paper]{paper_cw}

\usepackage{helvet}
\usepackage{amsmath}
\usepackage{amssymb}
\usepackage{mathrsfs}
\usepackage{graphics}
\usepackage[numbers, sort&compress]{natbib}
\usepackage{color}
\usepackage{fancyhdr} % must be loaded for header/footer
\usepackage{colortbl} % must be loaded for \columncolor
\usepackage[dvips]{graphicx} % must be loaded to scale graphics
\usepackage{marvosym}
\usepackage{float}
\usepackage{placeins}
\usepackage{hyperref}
\usepackage[labelfont=bf, figurename=Fig.]{caption}
\usepackage{subcaption}
\usepackage{multirow}

\usepackage{array} 
\usepackage{framed}

\usepackage{pgfplots}
\pgfplotsset{compat=1.18}
\usepackage{tcolorbox}
\usepackage{multirow}
\usepackage{enumitem}
\usepackage{algorithm}
\usepackage{algpseudocode}
\usepackage{xcolor}
\usepackage{textcase}

\newcolumntype{C}[1]{>{\centering\arraybackslash\hspace{0pt}}p{#1}} %Cenetred column type

\begin{document}

%==================================================================================
% LATEX MACRO COMMANDS
%==================================================================================

% Comments
\newcommand{\HW}{\textcolor{blue}}
\newcommand{\DA}{\textcolor{orange}}
\newcommand{\JPF}{\textcolor{green}}

% Syntax
\newcommand{\lrap}[1]{``#1''}

% A different caption: single spaced and italic
\newcommand{\Caption}[1]{ \begin{singlespace} \em{\caption{#1}} \end{singlespace} }

% A different footnote (with a star or another symbol you choose): call with \symbolfootnote
\long\def\symbolfootnote[#1]#2{\begingroup \def\thefootnote{\fnsymbol{footnote}}\footnote[#1]{#2} \endgroup} 

% Fundamental quantities
\renewcommand{\vec}[1]{ \ensuremath{ \mathbf{ #1 } } }
\newcommand{\ten}[1]{ \ensuremath {\mathbf{#1} } }

% Fundamental quantities
\newcommand{\gvec}[1]{ \ensuremath{ \boldsymbol{ #1 } } }
\newcommand{\gten}[1]{ \ensuremath {\boldsymbol{#1} } }

%Text words

\newcommand{\mn}{_{\mbox{\tiny{N}}}}
\newcommand{\mt}{_{\mbox{\tiny{T}}}}

\newcommand{\dint}{\mbox{d}}
\newcommand{\de}{\,\mbox{det}\,}
\newcommand{\gra}{\,\mbox{grad}\,}
\newcommand{\Gra}{\,\mbox{Grad}\,}
\newcommand{\tra}{\,\mbox{tr}\,}
\newcommand{\Div}{\mbox{Div}\,}
\renewcommand{\div}{\mbox{div}\,}
\newcommand{\sign}{\mbox{sign}}

\newcommand{\ncdot}{\hspace{-0.14cm}\cdot}

% Hat quantities
\newcommand{\hata}{\ensuremath{ \widehat{a} }}
\newcommand{\hatb}{\ensuremath{ \widehat{b} }}
\newcommand{\hatc}{\ensuremath{ \widehat{c} }}
\newcommand{\hatd}{\ensuremath{ \widehat{d} }}
\newcommand{\hate}{\ensuremath{ \widehat{e} }}
\newcommand{\hatf}{\ensuremath{ \widehat{f} }}
\newcommand{\hatg}{\ensuremath{ \widehat{g} }}
\newcommand{\hath}{\ensuremath{ \widehat{h} }}
\newcommand{\hati}{\ensuremath{ \widehat{i} }}
\newcommand{\hatj}{\ensuremath{ \widehat{j} }}
\newcommand{\hatk}{\ensuremath{ \widehat{k} }}
\newcommand{\hatl}{\ensuremath{ \widehat{l} }}
\newcommand{\hatm}{\ensuremath{ \widehat{m} }}
\newcommand{\hatn}{\ensuremath{ \widehat{n} }}
\newcommand{\hato}{\ensuremath{ \widehat{o} }}
\newcommand{\hatp}{\ensuremath{ \widehat{p} }}
\newcommand{\hatq}{\ensuremath{ \widehat{q} }}
\newcommand{\hatr}{\ensuremath{ \widehat{r} }}
\newcommand{\hats}{\ensuremath{ \widehat{s} }}
\newcommand{\hatt}{\ensuremath{ \widehat{t} }}
\newcommand{\hatu}{\ensuremath{ \widehat{u} }}
\newcommand{\hatv}{\ensuremath{ \widehat{v} }}
\newcommand{\hatw}{\ensuremath{ \widehat{w} }}
\newcommand{\hatx}{\ensuremath{ \widehat{x} }}
\newcommand{\haty}{\ensuremath{ \widehat{y} }}
\newcommand{\hatz}{\ensuremath{ \widehat{z} }}

\newcommand{\hatA}{\ensuremath{ \widehat{A} }}
\newcommand{\hatB}{\ensuremath{ \widehat{B} }}
\newcommand{\hatC}{\ensuremath{ \widehat{C} }}
\newcommand{\hatD}{\ensuremath{ \widehat{D} }}
\newcommand{\hatE}{\ensuremath{ \widehat{E} }}
\newcommand{\hatF}{\ensuremath{ \widehat{F} }}
\newcommand{\hatG}{\ensuremath{ \widehat{G} }}
\newcommand{\hatH}{\ensuremath{ \widehat{H} }}
\newcommand{\hatI}{\ensuremath{ \widehat{I} }}
\newcommand{\hatJ}{\ensuremath{ \widehat{J} }}
\newcommand{\hatK}{\ensuremath{ \widehat{K} }}
\newcommand{\hatL}{\ensuremath{ \widehat{L} }}
\newcommand{\hatM}{\ensuremath{ \widehat{M} }}
\newcommand{\hatN}{\ensuremath{ \widehat{N} }}
\newcommand{\hatO}{\ensuremath{ \widehat{O} }}
\newcommand{\hatP}{\ensuremath{ \widehat{P} }}
\newcommand{\hatQ}{\ensuremath{ \widehat{Q} }}
\newcommand{\hatR}{\ensuremath{ \widehat{R} }}
\newcommand{\hatS}{\ensuremath{ \widehat{S} }}
\newcommand{\hatT}{\ensuremath{ \widehat{T} }}
\newcommand{\hatU}{\ensuremath{ \widehat{U} }}
\newcommand{\hatV}{\ensuremath{ \widehat{V} }}
\newcommand{\hatW}{\ensuremath{ \widehat{W} }}
\newcommand{\hatX}{\ensuremath{ \widehat{X} }}
\newcommand{\hatY}{\ensuremath{ \widehat{Y} }}
\newcommand{\hatZ}{\ensuremath{ \widehat{Z }}}

% Tilde quantities
\newcommand{\tila}{\ensuremath{ \widetilde{a} }}
\newcommand{\tilb}{\ensuremath{ \widetilde{b} }}
\newcommand{\tilc}{\ensuremath{ \widetilde{c} }}
\newcommand{\tild}{\ensuremath{ \widetilde{d} }}
\newcommand{\tile}{\ensuremath{ \widetilde{e} }}
\newcommand{\tilf}{\ensuremath{ \widetilde{f} }}
\newcommand{\tilg}{\ensuremath{ \widetilde{g} }}
\newcommand{\tilh}{\ensuremath{ \widetilde{h} }}
\newcommand{\tili}{\ensuremath{ \widetilde{i} }}
\newcommand{\tilj}{\ensuremath{ \widetilde{j} }}
\newcommand{\tilk}{\ensuremath{ \widetilde{k} }}
\newcommand{\till}{\ensuremath{ \widetilde{l} }}
\newcommand{\tilm}{\ensuremath{ \widetilde{m} }}
\newcommand{\tiln}{\ensuremath{ \widetilde{n} }}
\newcommand{\tilo}{\ensuremath{ \widetilde{o} }}
\newcommand{\tilp}{\ensuremath{ \widetilde{p} }}
\newcommand{\tilq}{\ensuremath{ \widetilde{q} }}
\newcommand{\tilr}{\ensuremath{ \widetilde{r} }}
\newcommand{\tils}{\ensuremath{ \widetilde{s} }}
\newcommand{\tilt}{\ensuremath{ \widetilde{t} }}
\newcommand{\tilu}{\ensuremath{ \widetilde{u} }}
\newcommand{\tilv}{\ensuremath{ \widetilde{v} }}
\newcommand{\tilw}{\ensuremath{ \widetilde{w} }}
\newcommand{\tilx}{\ensuremath{ \widetilde{x} }}
\newcommand{\tily}{\ensuremath{ \widetilde{y} }}
\newcommand{\tilz}{\ensuremath{ \widetilde{z} }}

\newcommand{\tilA}{\ensuremath{ \widetilde{A} }}
\newcommand{\tilB}{\ensuremath{ \widetilde{B} }}
\newcommand{\tilC}{\ensuremath{ \widetilde{C} }}
\newcommand{\tilD}{\ensuremath{ \widetilde{D} }}
\newcommand{\tilE}{\ensuremath{ \widetilde{E} }}
\newcommand{\tilF}{\ensuremath{ \widetilde{F} }}
\newcommand{\tilG}{\ensuremath{ \widetilde{G} }}
\newcommand{\tilH}{\ensuremath{ \widetilde{H} }}
\newcommand{\tilI}{\ensuremath{ \widetilde{I} }}
\newcommand{\tilJ}{\ensuremath{ \widetilde{J} }}
\newcommand{\tilK}{\ensuremath{ \widetilde{K} }}
\newcommand{\tilL}{\ensuremath{ \widetilde{L} }}
\newcommand{\tilM}{\ensuremath{ \widetilde{M} }}
\newcommand{\tilN}{\ensuremath{ \widetilde{N} }}
\newcommand{\tilO}{\ensuremath{ \widetilde{O} }}
\newcommand{\tilP}{\ensuremath{ \widetilde{P} }}
\newcommand{\tilQ}{\ensuremath{ \widetilde{Q} }}
\newcommand{\tilR}{\ensuremath{ \widetilde{R} }}
\newcommand{\tilS}{\ensuremath{ \widetilde{S} }}
\newcommand{\tilT}{\ensuremath{ \widetilde{T} }}
\newcommand{\tilU}{\ensuremath{ \widetilde{U} }}
\newcommand{\tilV}{\ensuremath{ \widetilde{V} }}
\newcommand{\tilW}{\ensuremath{ \widetilde{W} }}
\newcommand{\tilX}{\ensuremath{ \widetilde{X} }}
\newcommand{\tilY}{\ensuremath{ \widetilde{Y} }}
\newcommand{\tilZ}{\ensuremath{ \widetilde{Z }}}

% Calligraphic letters
\newcommand{\calA}{\ensuremath{ \mathcal{A} }}
\newcommand{\calB}{\ensuremath{ \mathcal{B} }}
\newcommand{\calC}{\ensuremath{ \mathcal{C} }}
\newcommand{\calD}{\ensuremath{ \mathcal{D} }}
\newcommand{\calE}{\ensuremath{ \mathcal{E} }}
\newcommand{\calF}{\ensuremath{ \mathcal{F} }}
\newcommand{\calG}{\ensuremath{ \mathcal{G} }}
\newcommand{\calH}{\ensuremath{ \mathcal{H} }}
\newcommand{\calI}{\ensuremath{ \mathcal{I} }}
\newcommand{\calJ}{\ensuremath{ \mathcal{J} }}
\newcommand{\calK}{\ensuremath{ \mathcal{K} }}
\newcommand{\calL}{\ensuremath{ \mathcal{L} }}
\newcommand{\calM}{\ensuremath{ \mathcal{M} }}
\newcommand{\calN}{\ensuremath{ \mathcal{N} }}
\newcommand{\calO}{\ensuremath{ \mathcal{O} }}
\newcommand{\calP}{\ensuremath{ \mathcal{P} }}
\newcommand{\calQ}{\ensuremath{ \mathcal{Q} }}
\newcommand{\calR}{\ensuremath{ \mathcal{R} }}
\newcommand{\calS}{\ensuremath{ \mathcal{S} }}
\newcommand{\calT}{\ensuremath{ \mathcal{T} }}
\newcommand{\calU}{\ensuremath{ \mathcal{U} }}
\newcommand{\calV}{\ensuremath{ \mathcal{V} }}
\newcommand{\calW}{\ensuremath{ \mathcal{W} }}
\newcommand{\calX}{\ensuremath{ \mathcal{X} }}
\newcommand{\calY}{\ensuremath{ \mathcal{Y} }}
\newcommand{\calZ}{\ensuremath{ \mathcal{Z} }}

% Blackboard letters
\newcommand{\bbA}{\ensuremath{ \mathbb{A} }}
\newcommand{\bbB}{\ensuremath{ \mathbb{B} }}
\newcommand{\bbC}{\ensuremath{ \mathbb{C} }}
\newcommand{\bbD}{\ensuremath{ \mathbb{D} }}
\newcommand{\bbE}{\ensuremath{ \mathbb{E} }}
\newcommand{\bbF}{\ensuremath{ \mathbb{F} }}
\newcommand{\bbG}{\ensuremath{ \mathbb{G} }}
\newcommand{\bbH}{\ensuremath{ \mathbb{H} }}
\newcommand{\bbI}{\ensuremath{ \mathbb{I} }}
\newcommand{\bbJ}{\ensuremath{ \mathbb{J} }}
\newcommand{\bbK}{\ensuremath{ \mathbb{K} }}
\newcommand{\bbL}{\ensuremath{ \mathbb{L} }}
\newcommand{\bbM}{\ensuremath{ \mathbb{M} }}
\newcommand{\bbN}{\ensuremath{ \mathbb{N} }}
\newcommand{\bbO}{\ensuremath{ \mathbb{O} }}
\newcommand{\bbP}{\ensuremath{ \mathbb{P} }}
\newcommand{\bbQ}{\ensuremath{ \mathbb{Q} }}
\newcommand{\bbR}{\ensuremath{ \mathbb{R} }}
\newcommand{\bbS}{\ensuremath{ \mathbb{S} }}
\newcommand{\bbT}{\ensuremath{ \mathbb{T} }}
\newcommand{\bbU}{\ensuremath{ \mathbb{U} }}
\newcommand{\bbV}{\ensuremath{ \mathbb{V} }}
\newcommand{\bbW}{\ensuremath{ \mathbb{W} }}
\newcommand{\bbX}{\ensuremath{ \mathbb{X} }}
\newcommand{\bbY}{\ensuremath{ \mathbb{Y} }}
\newcommand{\bbZ}{\ensuremath{ \mathbb{Z} }}

% Tensor quantities
\newcommand{\tenA}{\ensuremath{ \ten{A} }}
\newcommand{\tenB}{\ensuremath{ \ten{B} }}
\newcommand{\tenC}{\ensuremath{ \ten{C} }}
\newcommand{\tenD}{\ensuremath{ \ten{D} }}
\newcommand{\tenE}{\ensuremath{ \ten{E} }}
\newcommand{\tenF}{\ensuremath{ \ten{F} }}
\newcommand{\tenG}{\ensuremath{ \ten{G} }}
\newcommand{\tenH}{\ensuremath{ \ten{H} }}
\newcommand{\tenI}{\ensuremath{ \ten{I} }}
\newcommand{\tenJ}{\ensuremath{ \ten{J} }}
\newcommand{\tenK}{\ensuremath{ \ten{K} }}
\newcommand{\tenL}{\ensuremath{ \ten{L} }}
\newcommand{\tenM}{\ensuremath{ \ten{M} }}
\newcommand{\tenN}{\ensuremath{ \ten{N} }}
\newcommand{\tenO}{\ensuremath{ \ten{O} }}
\newcommand{\tenP}{\ensuremath{ \ten{P} }}
\newcommand{\tenQ}{\ensuremath{ \ten{Q} }}
\newcommand{\tenR}{\ensuremath{ \ten{R} }}
\newcommand{\tenS}{\ensuremath{ \ten{S} }}
\newcommand{\tenT}{\ensuremath{ \ten{T} }}
\newcommand{\tenU}{\ensuremath{ \ten{U} }}
\newcommand{\tenV}{\ensuremath{ \ten{V} }}
\newcommand{\tenW}{\ensuremath{ \ten{W} }}
\newcommand{\tenX}{\ensuremath{ \ten{X} }}
\newcommand{\tenY}{\ensuremath{ \ten{Y} }}
\newcommand{\tenZ}{\ensuremath{ \ten{Z} }}

\newcommand{\tena}{\ensuremath{ \ten{a} }}
\newcommand{\tenb}{\ensuremath{ \ten{b} }}
\newcommand{\tenc}{\ensuremath{ \ten{c} }}
\newcommand{\tend}{\ensuremath{ \ten{d} }}
\newcommand{\tene}{\ensuremath{ \ten{e} }}
\newcommand{\tenf}{\ensuremath{ \ten{f} }}
\newcommand{\teng}{\ensuremath{ \ten{g} }}
\newcommand{\tenh}{\ensuremath{ \ten{h} }}
\newcommand{\teni}{\ensuremath{ \ten{i} }}
\newcommand{\tenj}{\ensuremath{ \ten{j} }}
\newcommand{\tenk}{\ensuremath{ \ten{k} }}
\newcommand{\tenl}{\ensuremath{ \ten{l} }}
\newcommand{\tenm}{\ensuremath{ \ten{m} }}
\newcommand{\tenn}{\ensuremath{ \ten{n} }}
\newcommand{\teno}{\ensuremath{ \ten{o} }}
\newcommand{\tenp}{\ensuremath{ \ten{p} }}
\newcommand{\tenq}{\ensuremath{ \ten{q} }}
\newcommand{\tenr}{\ensuremath{ \ten{r} }}
\newcommand{\tens}{\ensuremath{ \ten{s} }}
\newcommand{\tent}{\ensuremath{ \ten{t} }}
\newcommand{\tenu}{\ensuremath{ \ten{u} }}
\newcommand{\tenv}{\ensuremath{ \ten{v} }}
\newcommand{\tenw}{\ensuremath{ \ten{w} }}
\newcommand{\tenx}{\ensuremath{ \ten{x} }}
\newcommand{\teny}{\ensuremath{ \ten{y} }}
\newcommand{\tenz}{\ensuremath{ \ten{z} }}

% Time derivative + Tensor quantities
\newcommand{\ttena}{\dot{\tena}}
\newcommand{\ttenb}{\dot{\tenb}}
\newcommand{\ttenc}{\dot{\tenc}}
\newcommand{\ttend}{\dot{\tend}}
\newcommand{\ttene}{\dot{\tene}}
\newcommand{\ttenf}{\dot{\tenf}}

\newcommand{\ttenE}{\dot{\tenE}}
\newcommand{\ttenF}{\dot{\tenF}}
\newcommand{\ttenS}{\dot{\tenS}}
\newcommand{\ttenT}{\dot{\tenT}}
\newcommand{\ttenU}{\dot{\tenU}}
\newcommand{\ttenV}{\dot{\tenV}}
\newcommand{\ttenW}{\dot{\tenW}}
\newcommand{\ttenX}{\dot{\tenX}}
\newcommand{\ttenY}{\dot{\tenY}}
\newcommand{\ttenZ}{\dot{\tenZ}}

% Linearization + Tensor quantities
\newcommand{\ltenA}{\stackrel{\triangle}{\tenA}}
\newcommand{\ltenD}{\stackrel{\triangle}{\tenD}}
\newcommand{\ltenM}{\stackrel{\triangle}{\tenM}}

% Bar + Tensor quantities
\newcommand{\btena}{\bar{\tena}}
\newcommand{\btenb}{\bar{\tenb}}
\newcommand{\btenc}{\bar{\tenc}}
\newcommand{\btend}{\bar{\tend}}
\newcommand{\btene}{\bar{\tene}}
\newcommand{\btenf}{\bar{\tenf}}
\newcommand{\bteng}{\bar{\teng}}

\newcommand{\btenA}{\bar{\tenA}}
\newcommand{\btenB}{\bar{\tenB}}
\newcommand{\btenC}{\bar{\tenC}}
\newcommand{\btenD}{\bar{\tenD}}
\newcommand{\btenE}{\bar{\tenE}}
\newcommand{\btenF}{\bar{\tenF}}
\newcommand{\btenK}{\bar{\tenK}}
\newcommand{\btenL}{\bar{\tenL}}
\newcommand{\btenM}{\bar{\tenM}}
\newcommand{\btenN}{\bar{\tenN}}
\newcommand{\btenO}{\bar{\tenO}}
\newcommand{\btenP}{\bar{\tenP}}
\newcommand{\btenQ}{\bar{\tenQ}}
\newcommand{\btenR}{\bar{\tenR}}
\newcommand{\btenS}{\bar{\tenS}}
\newcommand{\btenT}{\bar{\tenT}}

% Linearization + Bar + Tensor quantities

\newcommand{\lbtenE}{\Delta\bar{\tenE}}

% Hat + Tensor quantities
\newcommand{\hattena}{\ensuremath{ \widehat{\ten{a}} }}
\newcommand{\hattenb}{\ensuremath{ \widehat{\ten{b}} }}
\newcommand{\hattenc}{\ensuremath{ \widehat{\ten{c}} }}
\newcommand{\hattend}{\ensuremath{ \widehat{\ten{d}} }}
\newcommand{\hattene}{\ensuremath{ \widehat{\ten{e}} }}
\newcommand{\hattenf}{\ensuremath{ \widehat{\ten{f}} }}
\newcommand{\hatteng}{\ensuremath{ \widehat{\ten{g}} }}
\newcommand{\hattenh}{\ensuremath{ \widehat{\ten{h}} }}
\newcommand{\hatteni}{\ensuremath{ \widehat{\ten{i}} }}
\newcommand{\hattenj}{\ensuremath{ \widehat{\ten{j}} }}
\newcommand{\hattenk}{\ensuremath{ \widehat{\ten{k}} }}
\newcommand{\hattenl}{\ensuremath{ \widehat{\ten{l}} }}
\newcommand{\hattenm}{\ensuremath{ \widehat{\ten{m}} }}
\newcommand{\hattenn}{\ensuremath{ \widehat{\ten{n}} }}
\newcommand{\hatteno}{\ensuremath{ \widehat{\ten{o}} }}
\newcommand{\hattenp}{\ensuremath{ \widehat{\ten{p}} }}
\newcommand{\hattenq}{\ensuremath{ \widehat{\ten{q}} }}
\newcommand{\hattenr}{\ensuremath{ \widehat{\ten{r}} }}
\newcommand{\hattens}{\ensuremath{ \widehat{\ten{s}} }}
\newcommand{\hattent}{\ensuremath{ \widehat{\ten{t}} }}
\newcommand{\hattenu}{\ensuremath{ \widehat{\ten{u}} }}
\newcommand{\hattenv}{\ensuremath{ \widehat{\ten{v}} }}
\newcommand{\hattenw}{\ensuremath{ \widehat{\ten{w}} }}
\newcommand{\hattenx}{\ensuremath{ \widehat{\ten{x}} }}
\newcommand{\hatteny}{\ensuremath{ \widehat{\ten{y}} }}
\newcommand{\hattenz}{\ensuremath{ \widehat{\ten{z}} }}

\newcommand{\hattenA}{\ensuremath{ \widehat{\ten{A}} }}
\newcommand{\hattenB}{\ensuremath{ \widehat{\ten{B}} }}
\newcommand{\hattenC}{\ensuremath{ \widehat{\ten{C}} }}
\newcommand{\hattenD}{\ensuremath{ \widehat{\ten{D}} }}
\newcommand{\hattenE}{\ensuremath{ \widehat{\ten{E}} }}
\newcommand{\hattenF}{\ensuremath{ \widehat{\ten{F}} }}
\newcommand{\hattenG}{\ensuremath{ \widehat{\ten{G}} }}
\newcommand{\hattenH}{\ensuremath{ \widehat{\ten{H}} }}
\newcommand{\hattenI}{\ensuremath{ \widehat{\ten{I}} }}
\newcommand{\hattenJ}{\ensuremath{ \widehat{\ten{J}} }}
\newcommand{\hattenK}{\ensuremath{ \widehat{\ten{K}} }}
\newcommand{\hattenL}{\ensuremath{ \widehat{\ten{L}} }}
\newcommand{\hattenM}{\ensuremath{ \widehat{\ten{M}} }}
\newcommand{\hattenN}{\ensuremath{ \widehat{\ten{N}} }}
\newcommand{\hattenO}{\ensuremath{ \widehat{\ten{O}} }}
\newcommand{\hattenP}{\ensuremath{ \widehat{\ten{P}} }}
\newcommand{\hattenQ}{\ensuremath{ \widehat{\ten{Q}} }}
\newcommand{\hattenR}{\ensuremath{ \widehat{\ten{R}} }}
\newcommand{\hattenS}{\ensuremath{ \widehat{\ten{S}} }}
\newcommand{\hattenT}{\ensuremath{ \widehat{\ten{T}} }}
\newcommand{\hattenU}{\ensuremath{ \widehat{\ten{U}} }}
\newcommand{\hattenV}{\ensuremath{ \widehat{\ten{V}} }}
\newcommand{\hattenW}{\ensuremath{ \widehat{\ten{W}} }}
\newcommand{\hattenX}{\ensuremath{ \widehat{\ten{X}} }}
\newcommand{\hattenY}{\ensuremath{ \widehat{\ten{Y}} }}
\newcommand{\hattenZ}{\ensuremath{ \widehat{\ten{Z}} }}

% Tilde + Tensor quantities
\newcommand{\tiltena}{\ensuremath{ \widetilde{\ten{a}} }}
\newcommand{\tiltenb}{\ensuremath{ \widetilde{\ten{b}} }}
\newcommand{\tiltenc}{\ensuremath{ \widetilde{\ten{c}} }}
\newcommand{\tiltend}{\ensuremath{ \widetilde{\ten{d}} }}
\newcommand{\tiltene}{\ensuremath{ \widetilde{\ten{e}} }}
\newcommand{\tiltenf}{\ensuremath{ \widetilde{\ten{f}} }}
\newcommand{\tilteng}{\ensuremath{ \widetilde{\ten{g}} }}
\newcommand{\tiltenh}{\ensuremath{ \widetilde{\ten{h}} }}
\newcommand{\tilteni}{\ensuremath{ \widetilde{\ten{i}} }}
\newcommand{\tiltenj}{\ensuremath{ \widetilde{\ten{j}} }}
\newcommand{\tiltenk}{\ensuremath{ \widetilde{\ten{k}} }}
\newcommand{\tiltenl}{\ensuremath{ \widetilde{\ten{l}} }}
\newcommand{\tiltenm}{\ensuremath{ \widetilde{\ten{m}} }}
\newcommand{\tiltenn}{\ensuremath{ \widetilde{\ten{n}} }}
\newcommand{\tilteno}{\ensuremath{ \widetilde{\ten{o}} }}
\newcommand{\tiltenp}{\ensuremath{ \widetilde{\ten{p}} }}
\newcommand{\tiltenq}{\ensuremath{ \widetilde{\ten{q}} }}
\newcommand{\tiltenr}{\ensuremath{ \widetilde{\ten{r}} }}
\newcommand{\tiltens}{\ensuremath{ \widetilde{\ten{s}} }}
\newcommand{\tiltent}{\ensuremath{ \widetilde{\ten{t}} }}
\newcommand{\tiltenu}{\ensuremath{ \widetilde{\ten{u}} }}
\newcommand{\tiltenv}{\ensuremath{ \widetilde{\ten{v}} }}
\newcommand{\tiltenw}{\ensuremath{ \widetilde{\ten{w}} }}
\newcommand{\tiltenx}{\ensuremath{ \widetilde{\ten{x}} }}
\newcommand{\tilteny}{\ensuremath{ \widetilde{\ten{y}} }}
\newcommand{\tiltenz}{\ensuremath{ \widetilde{\ten{z}} }}

\newcommand{\tiltenA}{\ensuremath{ \widetilde{\ten{A}} }}
\newcommand{\tiltenB}{\ensuremath{ \widetilde{\ten{B}} }}
\newcommand{\tiltenC}{\ensuremath{ \widetilde{\ten{C}} }}
\newcommand{\tiltenD}{\ensuremath{ \widetilde{\ten{D}} }}
\newcommand{\tiltenE}{\ensuremath{ \widetilde{\ten{E}} }}
\newcommand{\tiltenF}{\ensuremath{ \widetilde{\ten{F}} }}
\newcommand{\tiltenG}{\ensuremath{ \widetilde{\ten{G}} }}
\newcommand{\tiltenH}{\ensuremath{ \widetilde{\ten{H}} }}
\newcommand{\tiltenI}{\ensuremath{ \widetilde{\ten{I}} }}
\newcommand{\tiltenJ}{\ensuremath{ \widetilde{\ten{J}} }}
\newcommand{\tiltenK}{\ensuremath{ \widetilde{\ten{K}} }}
\newcommand{\tiltenL}{\ensuremath{ \widetilde{\ten{L}} }}
\newcommand{\tiltenM}{\ensuremath{ \widetilde{\ten{M}} }}
\newcommand{\tiltenN}{\ensuremath{ \widetilde{\ten{N}} }}
\newcommand{\tiltenO}{\ensuremath{ \widetilde{\ten{O}} }}
\newcommand{\tiltenP}{\ensuremath{ \widetilde{\ten{P}} }}
\newcommand{\tiltenQ}{\ensuremath{ \widetilde{\ten{Q}} }}
\newcommand{\tiltenR}{\ensuremath{ \widetilde{\ten{R}} }}
\newcommand{\tiltenS}{\ensuremath{ \widetilde{\ten{S}} }}
\newcommand{\tiltenT}{\ensuremath{ \widetilde{\ten{T}} }}
\newcommand{\tiltenU}{\ensuremath{ \widetilde{\ten{U}} }}
\newcommand{\tiltenV}{\ensuremath{ \widetilde{\ten{V}} }}
\newcommand{\tiltenW}{\ensuremath{ \widetilde{\ten{W}} }}
\newcommand{\tiltenX}{\ensuremath{ \widetilde{\ten{X}} }}
\newcommand{\tiltenY}{\ensuremath{ \widetilde{\ten{Y}} }}
\newcommand{\tiltenZ}{\ensuremath{ \widetilde{\ten{Z}} }}

% Vector quantities
\newcommand{\veca}{\ensuremath{ \vec{a} }}
\newcommand{\vecb}{\ensuremath{ \vec{b} }}
\newcommand{\vecc}{\ensuremath{ \vec{c} }}
\newcommand{\vecd}{\ensuremath{ \vec{d} }}
\newcommand{\vece}{\ensuremath{ \vec{e} }}
\newcommand{\vecf}{\ensuremath{ \vec{f} }}
\newcommand{\vecg}{\ensuremath{ \vec{g} }}
\newcommand{\vech}{\ensuremath{ \vec{h} }}
\newcommand{\veci}{\ensuremath{ \vec{i} }}
\newcommand{\vecj}{\ensuremath{ \vec{j} }}
\newcommand{\veck}{\ensuremath{ \vec{k} }}
\newcommand{\vecl}{\ensuremath{ \vec{l} }}
\newcommand{\vecm}{\ensuremath{ \vec{m} }}
\newcommand{\vecn}{\ensuremath{ \vec{n} }}
\newcommand{\veco}{\ensuremath{ \vec{o} }}
\newcommand{\vecp}{\ensuremath{ \vec{p} }}
\newcommand{\vecq}{\ensuremath{ \vec{q} }}
\newcommand{\vecr}{\ensuremath{ \vec{r} }}
\newcommand{\vecs}{\ensuremath{ \vec{s} }}
\newcommand{\vect}{\ensuremath{ \vec{t} }}
\newcommand{\vecu}{\ensuremath{ \vec{u} }}
\newcommand{\vecv}{\ensuremath{ \vec{v} }}
\newcommand{\vecw}{\ensuremath{ \vec{w} }}
\newcommand{\vecx}{\ensuremath{ \vec{x} }}
\newcommand{\vecy}{\ensuremath{ \vec{y} }}
\newcommand{\vecz}{\ensuremath{ \vec{z} }}

\newcommand{\vecA}{\ensuremath{ \vec{A} }}
\newcommand{\vecB}{\ensuremath{ \vec{B} }}
\newcommand{\vecC}{\ensuremath{ \vec{C} }}
\newcommand{\vecD}{\ensuremath{ \vec{D} }}
\newcommand{\vecE}{\ensuremath{ \vec{E} }}
\newcommand{\vecF}{\ensuremath{ \vec{F} }}
\newcommand{\vecG}{\ensuremath{ \vec{G} }}
\newcommand{\vecH}{\ensuremath{ \vec{H} }}
\newcommand{\vecI}{\ensuremath{ \vec{I} }}
\newcommand{\vecJ}{\ensuremath{ \vec{J} }}
\newcommand{\vecK}{\ensuremath{ \vec{K} }}
\newcommand{\vecL}{\ensuremath{ \vec{L} }}
\newcommand{\vecM}{\ensuremath{ \vec{M} }}
\newcommand{\vecN}{\ensuremath{ \vec{N} }}
\newcommand{\vecO}{\ensuremath{ \vec{O} }}
\newcommand{\vecP}{\ensuremath{ \vec{P} }}
\newcommand{\vecQ}{\ensuremath{ \vec{Q} }}
\newcommand{\vecR}{\ensuremath{ \vec{R} }}
\newcommand{\vecS}{\ensuremath{ \vec{S} }}
\newcommand{\vecT}{\ensuremath{ \vec{T} }}
\newcommand{\vecU}{\ensuremath{ \vec{U} }}
\newcommand{\vecV}{\ensuremath{ \vec{V} }}
\newcommand{\vecW}{\ensuremath{ \vec{W} }}
\newcommand{\vecX}{\ensuremath{ \vec{X} }}
\newcommand{\vecY}{\ensuremath{ \vec{Y} }}
\newcommand{\vecZ}{\ensuremath{ \vec{Z} }}

% Time derivative vector quantities
\newcommand{\tveca}{\ensuremath{ \dot{\vec{a}} }}
\newcommand{\tvecb}{\dot{\vecb}}
\newcommand{\tvecc}{\ensuremath{ \vec{c} }}
\newcommand{\tvecd}{\ensuremath{ \vec{d} }}
\newcommand{\tvece}{\ensuremath{ \vec{e} }}
\newcommand{\tvecf}{\ensuremath{ \vec{f} }}
\newcommand{\tvecg}{\ensuremath{ \dot{\vec{g}} }}
\newcommand{\tvech}{\ensuremath{ \vec{h} }}
\newcommand{\tveci}{\ensuremath{ \vec{i} }}
\newcommand{\tvecj}{\ensuremath{ \vec{j} }}
\newcommand{\tveck}{\ensuremath{ \vec{k} }}
\newcommand{\tvecl}{\ensuremath{ \vec{l} }}
\newcommand{\tvecm}{\ensuremath{ \vec{m} }}
\newcommand{\tvecn}{\dot{\ensuremath{ \vec{n} }}}
\newcommand{\tveco}{\dot{\ensuremath{ \vec{o} }}}
\newcommand{\tvecp}{\dot{\ensuremath{ \vec{p} }}}
\newcommand{\tvecq}{\dot{\ensuremath{ \vec{q} }}}
\newcommand{\tvecr}{\ensuremath{ \vec{r} }}
\newcommand{\tvecs}{\ensuremath{ \vec{s} }}
\newcommand{\tvect}{\ensuremath{ \vec{t} }}
\newcommand{\tvecu}{\dot{\ensuremath{ \vec{u} }}}
\newcommand{\tvecv}{\dot{\ensuremath{ \vec{v} }}}
\newcommand{\tvecw}{\dot{\ensuremath{ \vec{w} }}}
\newcommand{\tvecx}{\dot{\ensuremath{ \vec{x} }}}
\newcommand{\tvecy}{\dot{\ensuremath{ \vec{y} }}}
\newcommand{\tvecz}{\dot{\ensuremath{ \vec{z} }}}

\newcommand{\tvecA}{\ensuremath{ \vec{A} }}
\newcommand{\tvecB}{\ensuremath{ \vec{B} }}
\newcommand{\tvecC}{\ensuremath{ \vec{C} }}
\newcommand{\tvecD}{\ensuremath{ \vec{D} }}
\newcommand{\tvecE}{\ensuremath{ \vec{E} }}
\newcommand{\tvecF}{\ensuremath{ \vec{F} }}
\newcommand{\tvecG}{\ensuremath{ \vec{G} }}
\newcommand{\tvecH}{\ensuremath{ \vec{H} }}
\newcommand{\tvecI}{\ensuremath{ \vec{I} }}
\newcommand{\tvecJ}{\ensuremath{ \vec{J} }}
\newcommand{\tvecK}{\ensuremath{ \vec{K} }}
\newcommand{\tvecL}{\ensuremath{ \vec{L} }}
\newcommand{\tvecM}{\ensuremath{ \vec{M} }}
\newcommand{\tvecN}{\ensuremath{ \vec{N} }}
\newcommand{\tvecO}{\ensuremath{ \vec{O} }}
\newcommand{\tvecP}{\ensuremath{ \vec{P} }}
\newcommand{\tvecQ}{\ensuremath{ \vec{Q} }}
\newcommand{\tvecR}{\ensuremath{ \vec{R} }}
\newcommand{\tvecS}{\ensuremath{ \vec{S} }}
\newcommand{\tvecT}{\ensuremath{ \vec{T} }}
\newcommand{\tvecU}{\ensuremath{ \vec{U} }}
\newcommand{\tvecV}{\ensuremath{ \vec{V} }}
\newcommand{\tvecW}{\ensuremath{ \vec{W} }}
\newcommand{\tvecX}{\ensuremath{ \vec{X} }}
\newcommand{\tvecY}{\ensuremath{ \vec{Y} }}
\newcommand{\tvecZ}{\ensuremath{ \vec{Z} }}

%Special commands
\newcommand{\no}{\hspace{-0.14cm}\,}

% Linearization + vector quantities
\newcommand{\lveca}{\Delta\veca}
\newcommand{\lvecb}{\Delta\vecb}
\newcommand{\lvecc}{\Delta\vecc}
\newcommand{\lvecd}{\Delta\vecd}
\newcommand{\lvece}{\Delta\vece}
\newcommand{\lvecf}{\Delta\vecf}
\newcommand{\lvecg}{\Delta\vecg}
\newcommand{\lvech}{\Delta\vech}
\newcommand{\lveci}{\Delta\veci}
\newcommand{\lvecj}{\Delta\vecj}
\newcommand{\lveck}{\Delta\veck}
\newcommand{\lvecl}{\Delta\vecl}
\newcommand{\lvecm}{\Delta\vecm}
\newcommand{\lvecn}{\Delta\vecn}
\newcommand{\lveco}{\Delta\veco}
\newcommand{\lvecp}{\Delta\vecp}
\newcommand{\lvecq}{\Delta\vecq}
\newcommand{\lvecr}{\Delta\vecr}
\newcommand{\lvecs}{\Delta\vecs}
\newcommand{\lvect}{\Delta\vect}
\newcommand{\lvecu}{\Delta\vecu}
\newcommand{\lvecv}{\Delta\vecv}
\newcommand{\lvecw}{\Delta\vecw}
\newcommand{\lvecx}{\Delta\vecx}
\newcommand{\lvecy}{\Delta\vecy}
\newcommand{\lvecz}{\Delta\vecz}

\newcommand{\lvecN}{\Delta\vecN}

% Variation + vector quantities

\newcommand{\dveca}{\delta\veca}
\newcommand{\dvecb}{\delta\vecb}
\newcommand{\dvecc}{\delta\vecc}
\newcommand{\dvecd}{\delta\vecd}
\newcommand{\dvece}{\delta\vece}
\newcommand{\dvecf}{\delta\vecf}
\newcommand{\dvecg}{\delta\vecg}
\newcommand{\dvech}{\delta\vech}
\newcommand{\dveci}{\delta\veci}
\newcommand{\dvecj}{\delta\vecj}
\newcommand{\dveck}{\delta\veck}
\newcommand{\dvecl}{\delta\vecl}
\newcommand{\dvecm}{\delta\vecm}
\newcommand{\dvecu}{\delta\vecu}
\newcommand{\dvecv}{\delta\vecv}
\newcommand{\dvecw}{\delta\vecw}
\newcommand{\dvecx}{\delta\vecx}
\newcommand{\dvecy}{\delta\vecy}
\newcommand{\dvecz}{\delta\vecz}

% Bar + Vector quantities
\newcommand{\bveca}{\bar{\veca}}
\newcommand{\bvecb}{\bar{\vecb}}
\newcommand{\bvecc}{\bar{\vecc}}
\newcommand{\bvecd}{\bar{\vecd}}
\newcommand{\bvece}{\bar{\vece}}
\newcommand{\bvecf}{\bar{\vecf}}
\newcommand{\bvecg}{\bar{\vecg}}
\newcommand{\bvecm}{\bar{\vecm}}
\newcommand{\bvecn}{\bar{\vecn}}
\newcommand{\bveco}{\bar{\veco}}
\newcommand{\bvecp}{\bar{\vecp}}
\newcommand{\bvecq}{\bar{\vecq}}
\newcommand{\bvecr}{\bar{\vecr}}
\newcommand{\bvecs}{\bar{\vecs}}
\newcommand{\bvect}{\bar{\vect}}
\newcommand{\bvecu}{\bar{\vecu}}
\newcommand{\bvecv}{\bar{\vecv}}
\newcommand{\bvecw}{\bar{\vecw}}
\newcommand{\bvecx}{\bar{\vecx}}
\newcommand{\bvecy}{\bar{\vecy}}
\newcommand{\bvecz}{\bar{\vecz}}

\newcommand{\bvecA}{\bar{\vecA}}
\newcommand{\bvecB}{\bar{\vecB}}
\newcommand{\bvecC}{\bar{\vecC}}
\newcommand{\bvecD}{\bar{\vecD}}
\newcommand{\bvecE}{\bar{\vecE}}
\newcommand{\bvecF}{\bar{\vecF}}
\newcommand{\bvecG}{\bar{\vecG}}
\newcommand{\bvecH}{\bar{\vecH}}
\newcommand{\bvecI}{\bar{\vecI}}
\newcommand{\bvecJ}{\bar{\vecJ}}
\newcommand{\bvecK}{\bar{\vecK}}
\newcommand{\bvecL}{\bar{\vecL}}
\newcommand{\bvecM}{\bar{\vecM}}
\newcommand{\bvecN}{\bar{\vecN}}
\newcommand{\bvecO}{\bar{\vecO}}
\newcommand{\bvecP}{\bar{\vecP}}
\newcommand{\bvecR}{\bar{\vecR}}
\newcommand{\bvecS}{\bar{\vecS}}
\newcommand{\bvecT}{\bar{\vecT}}
\newcommand{\bvecU}{\bar{\vecU}}

% Linearization + Bar + Vector quantities

\newcommand{\lbveca}{\Delta\bar{\veca}}
\newcommand{\lbvecb}{\Delta\bar{\vecb}}
\newcommand{\lbvecc}{\Delta\bar{\vecc}}
\newcommand{\lbvecd}{\Delta\bar{\vecd}}
\newcommand{\lbvece}{\Delta\bar{\vece}}
\newcommand{\lbvecf}{\Delta\bar{\vecf}}
\newcommand{\lbvecg}{\Delta\bar{\vecg}}
\newcommand{\lbvech}{\Delta\bar{\vech}}
\newcommand{\lbveci}{\Delta\bar{\veci}}
\newcommand{\lbvecj}{\Delta\bar{\vecj}}
\newcommand{\lbvect}{\Delta\bar{\vect}}
\newcommand{\lbvecu}{\Delta\bar{\vecu}}
\newcommand{\lbvecx}{\Delta\bar{\vecx}}

% Variation + Bar + Vector quantities

\newcommand{\dbveca}{\delta\bar{\veca}}
\newcommand{\dbvecb}{\delta\bar{\vecb}}
\newcommand{\dbvecc}{\delta\bar{\vecc}}
\newcommand{\dbvecd}{\delta\bar{\vecd}}
\newcommand{\dbvece}{\delta\bar{\vece}}
\newcommand{\dbvecf}{\delta\bar{\vecf}}
\newcommand{\dbvecg}{\delta\bar{\vecg}}
\newcommand{\dbvech}{\delta\bar{\vech}}
\newcommand{\dbveci}{\delta\bar{\veci}}
\newcommand{\dbvecj}{\delta\bar{\vecj}}
\newcommand{\dbvect}{\delta\bar{\vect}}
\newcommand{\dbvecu}{\delta\bar{\vecu}}
\newcommand{\dbvecx}{\delta\bar{\vecx}}

% Hat+Vector quantities
\newcommand{\hatveca}{\ensuremath{ \widehat{\vec{a}} }}
\newcommand{\hatvecb}{\ensuremath{ \widehat{\vec{b}} }}
\newcommand{\hatvecc}{\ensuremath{ \widehat{\vec{c}} }}
\newcommand{\hatvecd}{\ensuremath{ \widehat{\vec{d}} }}
\newcommand{\hatvece}{\ensuremath{ \widehat{\vec{e}} }}
\newcommand{\hatvecf}{\ensuremath{ \widehat{\vec{f}} }}
\newcommand{\hatvecg}{\ensuremath{ \widehat{\vec{g}} }}
\newcommand{\hatvech}{\ensuremath{ \widehat{\vec{h}} }}
\newcommand{\hatveci}{\ensuremath{ \widehat{\vec{i}} }}
\newcommand{\hatvecj}{\ensuremath{ \widehat{\vec{j}} }}
\newcommand{\hatveck}{\ensuremath{ \widehat{\vec{k}} }}
\newcommand{\hatvecl}{\ensuremath{ \widehat{\vec{l}} }}
\newcommand{\hatvecm}{\ensuremath{ \widehat{\vec{m}} }}
\newcommand{\hatvecn}{\ensuremath{ \widehat{\vec{n}} }}
\newcommand{\hatveco}{\ensuremath{ \widehat{\vec{o}} }}
\newcommand{\hatvecp}{\ensuremath{ \widehat{\vec{p}} }}
\newcommand{\hatvecq}{\ensuremath{ \widehat{\vec{q}} }}
\newcommand{\hatvecr}{\ensuremath{ \widehat{\vec{r}} }}
\newcommand{\hatvecs}{\ensuremath{ \widehat{\vec{s}} }}
\newcommand{\hatvect}{\ensuremath{ \widehat{\vec{t}} }}
\newcommand{\hatvecu}{\ensuremath{ \widehat{\vec{u}} }}
\newcommand{\hatvecv}{\ensuremath{ \widehat{\vec{v}} }}
\newcommand{\hatvecw}{\ensuremath{ \widehat{\vec{w}} }}
\newcommand{\hatvecx}{\ensuremath{ \widehat{\vec{x}} }}
\newcommand{\hatvecy}{\ensuremath{ \widehat{\vec{y}} }}
\newcommand{\hatvecz}{\ensuremath{ \widehat{\vec{z}} }}

\newcommand{\hatvecA}{\ensuremath{ \widehat{\vec{A}} }}
\newcommand{\hatvecB}{\ensuremath{ \widehat{\vec{B}} }}
\newcommand{\hatvecC}{\ensuremath{ \widehat{\vec{C}} }}
\newcommand{\hatvecD}{\ensuremath{ \widehat{\vec{D}} }}
\newcommand{\hatvecE}{\ensuremath{ \widehat{\vec{E}} }}
\newcommand{\hatvecF}{\ensuremath{ \widehat{\vec{F}} }}
\newcommand{\hatvecG}{\ensuremath{ \widehat{\vec{G}} }}
\newcommand{\hatvecH}{\ensuremath{ \widehat{\vec{H}} }}
\newcommand{\hatvecI}{\ensuremath{ \widehat{\vec{I}} }}
\newcommand{\hatvecJ}{\ensuremath{ \widehat{\vec{J}} }}
\newcommand{\hatvecK}{\ensuremath{ \widehat{\vec{K}} }}
\newcommand{\hatvecL}{\ensuremath{ \widehat{\vec{L}} }}
\newcommand{\hatvecM}{\ensuremath{ \widehat{\vec{M}} }}
\newcommand{\hatvecN}{\ensuremath{ \widehat{\vec{N}} }}
\newcommand{\hatvecO}{\ensuremath{ \widehat{\vec{O}} }}
\newcommand{\hatvecP}{\ensuremath{ \widehat{\vec{P}} }}
\newcommand{\hatvecQ}{\ensuremath{ \widehat{\vec{Q}} }}
\newcommand{\hatvecR}{\ensuremath{ \widehat{\vec{R}} }}
\newcommand{\hatvecS}{\ensuremath{ \widehat{\vec{S}} }}
\newcommand{\hatvecT}{\ensuremath{ \widehat{\vec{T}} }}
\newcommand{\hatvecU}{\ensuremath{ \widehat{\vec{U}} }}
\newcommand{\hatvecV}{\ensuremath{ \widehat{\vec{V}} }}
\newcommand{\hatvecW}{\ensuremath{ \widehat{\vec{W}} }}
\newcommand{\hatvecX}{\ensuremath{ \widehat{\vec{X}} }}
\newcommand{\hatvecY}{\ensuremath{ \widehat{\vec{Y}} }}
\newcommand{\hatvecZ}{\ensuremath{ \widehat{\vec{Z}} }}

% Tilde+Vector quantities
\newcommand{\tilveca}{\ensuremath{ \widetilde{\vec{a}} }}
\newcommand{\tilvecb}{\ensuremath{ \widetilde{\vec{b}} }}
\newcommand{\tilvecc}{\ensuremath{ \widetilde{\vec{c}} }}
\newcommand{\tilvecd}{\ensuremath{ \widetilde{\vec{d}} }}
\newcommand{\tilvece}{\ensuremath{ \widetilde{\vec{e}} }}
\newcommand{\tilvecf}{\ensuremath{ \widetilde{\vec{f}} }}
\newcommand{\tilvecg}{\ensuremath{ \widetilde{\vec{g}} }}
\newcommand{\tilvech}{\ensuremath{ \widetilde{\vec{h}} }}
\newcommand{\tilveci}{\ensuremath{ \widetilde{\vec{i}} }}
\newcommand{\tilvecj}{\ensuremath{ \widetilde{\vec{j}} }}
\newcommand{\tilveck}{\ensuremath{ \widetilde{\vec{k}} }}
\newcommand{\tilvecl}{\ensuremath{ \widetilde{\vec{l}} }}
\newcommand{\tilvecm}{\ensuremath{ \widetilde{\vec{m}} }}
\newcommand{\tilvecn}{\ensuremath{ \widetilde{\vec{n}} }}
\newcommand{\tilveco}{\ensuremath{ \widetilde{\vec{o}} }}
\newcommand{\tilvecp}{\ensuremath{ \widetilde{\vec{p}} }}
\newcommand{\tilvecq}{\ensuremath{ \widetilde{\vec{q}} }}
\newcommand{\tilvecr}{\ensuremath{ \widetilde{\vec{r}} }}
\newcommand{\tilvecs}{\ensuremath{ \widetilde{\vec{s}} }}
\newcommand{\tilvect}{\ensuremath{ \widetilde{\vec{t}} }}
\newcommand{\tilvecu}{\ensuremath{ \widetilde{\vec{u}} }}
\newcommand{\tilvecv}{\ensuremath{ \widetilde{\vec{v}} }}
\newcommand{\tilvecw}{\ensuremath{ \widetilde{\vec{w}} }}
\newcommand{\tilvecx}{\ensuremath{ \widetilde{\vec{x}} }}
\newcommand{\tilvecy}{\ensuremath{ \widetilde{\vec{y}} }}
\newcommand{\tilvecz}{\ensuremath{ \widetilde{\vec{z}} }}

\newcommand{\tilvecA}{\ensuremath{ \widetilde{\vec{A}} }}
\newcommand{\tilvecB}{\ensuremath{ \widetilde{\vec{B}} }}
\newcommand{\tilvecC}{\ensuremath{ \widetilde{\vec{C}} }}
\newcommand{\tilvecD}{\ensuremath{ \widetilde{\vec{D}} }}
\newcommand{\tilvecE}{\ensuremath{ \widetilde{\vec{E}} }}
\newcommand{\tilvecF}{\ensuremath{ \widetilde{\vec{F}} }}
\newcommand{\tilvecG}{\ensuremath{ \widetilde{\vec{G}} }}
\newcommand{\tilvecH}{\ensuremath{ \widetilde{\vec{H}} }}
\newcommand{\tilvecI}{\ensuremath{ \widetilde{\vec{I}} }}
\newcommand{\tilvecJ}{\ensuremath{ \widetilde{\vec{J}} }}
\newcommand{\tilvecK}{\ensuremath{ \widetilde{\vec{K}} }}
\newcommand{\tilvecL}{\ensuremath{ \widetilde{\vec{L}} }}
\newcommand{\tilvecM}{\ensuremath{ \widetilde{\vec{M}} }}
\newcommand{\tilvecN}{\ensuremath{ \widetilde{\vec{N}} }}
\newcommand{\tilvecO}{\ensuremath{ \widetilde{\vec{O}} }}
\newcommand{\tilvecP}{\ensuremath{ \widetilde{\vec{P}} }}
\newcommand{\tilvecQ}{\ensuremath{ \widetilde{\vec{Q}} }}
\newcommand{\tilvecR}{\ensuremath{ \widetilde{\vec{R}} }}
\newcommand{\tilvecS}{\ensuremath{ \widetilde{\vec{S}} }}
\newcommand{\tilvecT}{\ensuremath{ \widetilde{\vec{T}} }}
\newcommand{\tilvecU}{\ensuremath{ \widetilde{\vec{U}} }}
\newcommand{\tilvecV}{\ensuremath{ \widetilde{\vec{V}} }}
\newcommand{\tilvecW}{\ensuremath{ \widetilde{\vec{W}} }}
\newcommand{\tilvecX}{\ensuremath{ \widetilde{\vec{X}} }}
\newcommand{\tilvecY}{\ensuremath{ \widetilde{\vec{Y}} }}
\newcommand{\tilvecZ}{\ensuremath{ \widetilde{\vec{Z}} }}

\newcommand{\tilveclam}{\ensuremath{ \widetilde{\gvec{\lambda}} }}

% Double Dot+Vector quantities
\newcommand{\ttvecd}{\ensuremath{ \ddot{\vec{d}} }}
\newcommand{\ttvect}{\ensuremath{ \ddot{\vec{t}} }}
\newcommand{\ttvecu}{\ensuremath{ \ddot{\vec{u}} }}
\newcommand{\ttvecv}{\ensuremath{ \ddot{\vec{v}} }}
\newcommand{\ttvecw}{\ensuremath{ \ddot{\vec{w}} }}
\newcommand{\ttvecx}{\ensuremath{ \ddot{\vec{x}} }}
\newcommand{\ttvecy}{\ensuremath{ \ddot{\vec{y}} }}
\newcommand{\ttvecz}{\ensuremath{ \ddot{\vec{z}} }}

% Scalar quantities
\newcommand{\scaa}{\ensuremath{\mathrm{a}}}
\newcommand{\scab}{\ensuremath{\mathrm{b}}}
\newcommand{\scac}{\ensuremath{\mathrm{c}}}
\newcommand{\scad}{\ensuremath{\mathrm{d}}}
\newcommand{\scae}{\ensuremath{\mathrm{e}}}
\newcommand{\scaf}{\ensuremath{\mathrm{f}}}
\newcommand{\scag}{\ensuremath{\mathrm{g}}}
\newcommand{\scah}{\ensuremath{\mathrm{h}}}
\newcommand{\scai}{\ensuremath{\mathrm{i}}}
\newcommand{\scaj}{\ensuremath{\mathrm{j}}}
\newcommand{\scak}{\ensuremath{\mathrm{k}}}
\newcommand{\scal}{\ensuremath{\mathrm{l}}}
\newcommand{\scam}{\ensuremath{\mathrm{m}}}
\newcommand{\scan}{\ensuremath{\mathrm{n}}}
\newcommand{\scao}{\ensuremath{\mathrm{o}}}
\newcommand{\scap}{\ensuremath{\mathrm{p}}}
\newcommand{\scaq}{\ensuremath{\mathrm{q}}}
\newcommand{\scar}{\ensuremath{\mathrm{r}}}
\newcommand{\scas}{\ensuremath{\mathrm{s}}}
\newcommand{\scat}{\ensuremath{\mathrm{t}}}
\newcommand{\scau}{\ensuremath{\mathrm{u}}}
\newcommand{\scav}{\ensuremath{\mathrm{v}}}
\newcommand{\scaw}{\ensuremath{\mathrm{w}}}
\newcommand{\scax}{\ensuremath{\mathrm{x}}}
\newcommand{\scay}{\ensuremath{\mathrm{y}}}
\newcommand{\scaz}{\ensuremath{\mathrm{z}}}

\newcommand{\scaA}{\ensuremath{\mathrm{A}}}
\newcommand{\scaB}{\ensuremath{\mathrm{B}}}
\newcommand{\scaC}{\ensuremath{\mathrm{C}}}
\newcommand{\scaD}{\ensuremath{\mathrm{D}}}
\newcommand{\scaE}{\ensuremath{\mathrm{E}}}
\newcommand{\scaF}{\ensuremath{\mathrm{F}}}
\newcommand{\scaG}{\ensuremath{\mathrm{G}}}
\newcommand{\scaH}{\ensuremath{\mathrm{H}}}
\newcommand{\scaI}{\ensuremath{\mathrm{I}}}
\newcommand{\scaJ}{\ensuremath{\mathrm{J}}}
\newcommand{\scaK}{\ensuremath{\mathrm{K}}}
\newcommand{\scaL}{\ensuremath{\mathrm{L}}}
\newcommand{\scaM}{\ensuremath{\mathrm{M}}}
\newcommand{\scaN}{\ensuremath{\mathrm{N}}}
\newcommand{\scaO}{\ensuremath{\mathrm{O}}}
\newcommand{\scaP}{\ensuremath{\mathrm{P}}}
\newcommand{\scaQ}{\ensuremath{\mathrm{Q}}}
\newcommand{\scaR}{\ensuremath{\mathrm{R}}}
\newcommand{\scaS}{\ensuremath{\mathrm{S}}}
\newcommand{\scaT}{\ensuremath{\mathrm{T}}}
\newcommand{\scaU}{\ensuremath{\mathrm{U}}}
\newcommand{\scaV}{\ensuremath{\mathrm{V}}}
\newcommand{\scaW}{\ensuremath{\mathrm{W}}}
\newcommand{\scaX}{\ensuremath{\mathrm{X}}}
\newcommand{\scaY}{\ensuremath{\mathrm{Y}}}
\newcommand{\scaZ}{\ensuremath{\mathrm{Z}}}

% Bar+Scalar  quantities
\newcommand{\bscaa}{\bar{\scaa}}
\newcommand{\bscab}{\bar{\scab}}
\newcommand{\bscac}{\bar{\scac}}
\newcommand{\bscad}{\bar{\scad}}
\newcommand{\bscae}{\bar{\scae}}
\newcommand{\bscaf}{\bar{\scaf}}
\newcommand{\bscag}{\bar{\scag}}
\newcommand{\bscah}{\bar{\scah}}
\newcommand{\bscai}{\bar{\scai}}
\newcommand{\bscaj}{\bar{\scaj}}
\newcommand{\bscak}{\bar{\scak}}
\newcommand{\bscal}{\bar{\scal}}
\newcommand{\bscam}{\bar{\scam}}
\newcommand{\bscan}{\bar{\scan}}
\newcommand{\bscao}{\bar{\scao}}
\newcommand{\bscap}{\bar{\scap}}
\newcommand{\bscaq}{\bar{\scaq}}
\newcommand{\bscar}{\bar{\scar}}
\newcommand{\bscas}{\bar{\scas}}
\newcommand{\bscat}{\bar{\scat}}

\newcommand{\bscaA}{\bar{\scaA}}
\newcommand{\bscaB}{\bar{\scaB}}
\newcommand{\bscaC}{\bar{\scaC}}
\newcommand{\bscaD}{\bar{\scaD}}
\newcommand{\bscaE}{\bar{\scaE}}
\newcommand{\bscaF}{\bar{\scaF}}
\newcommand{\bscaG}{\bar{\scaG}}
\newcommand{\bscaH}{\bar{\scaH}}
\newcommand{\bscaI}{\bar{\scaI}}
\newcommand{\bscaJ}{\bar{\scaJ}}
\newcommand{\bscaK}{\bar{\scaK}}
\newcommand{\bscaL}{\bar{\scaL}}
\newcommand{\bscaM}{\bar{\scaM}}
\newcommand{\bscaN}{\bar{\scaN}}
\newcommand{\bscaO}{\bar{\scaO}}
\newcommand{\bscaP}{\bar{\scaP}}
\newcommand{\bscaQ}{\bar{\scaQ}}
\newcommand{\bscaR}{\bar{\scaR}}
\newcommand{\bscaS}{\bar{\scaS}}
\newcommand{\bscaT}{\bar{\scaT}}

% Hat+Scalar  quantities
\newcommand{\hatscan}{\ensuremath{\widehat{\scan}}}

% Hat+Scalar  quantities
\newcommand{\tilscag}{\ensuremath{\widetilde{\scag}}}
\newcommand{\tilscat}{\ensuremath{\widetilde{\scat}}}

\newcommand{\tillam}{\ensuremath{\widetilde{\lambda}}}

% Dot+Scalar  quantities
\newcommand{\tscag}{\ensuremath{\dot{\scag}}}

% Linearization + Scalar quantities
\newcommand{\lscaa}{\Delta\scaa}
\newcommand{\lscab}{\Delta\scab}
\newcommand{\lscac}{\Delta\scac}
\newcommand{\lscad}{\Delta\scad}
\newcommand{\lscae}{\Delta\scae}
\newcommand{\lscaf}{\Delta\scaf}
\newcommand{\lscag}{\Delta\scag}
\newcommand{\lscah}{\Delta\scah}
\newcommand{\lscai}{\Delta\scai}
\newcommand{\lscaj}{\Delta\scaj}
\newcommand{\lscak}{\Delta\scak}
\newcommand{\lscal}{\Delta\scal}
\newcommand{\lscam}{\Delta\scam}
\newcommand{\lscan}{\Delta\scan}
\newcommand{\lscao}{\Delta\scao}
\newcommand{\lscap}{\Delta\scap}
\newcommand{\lscaq}{\Delta\scaq}
\newcommand{\lscar}{\Delta\scar}
\newcommand{\lscas}{\Delta\scas}
\newcommand{\lscat}{\Delta\scat}

\newcommand{\lscaA}{\Delta\scaA}
\newcommand{\lscaD}{\Delta\scaD}
\newcommand{\lscaM}{\Delta\scaM}
\newcommand{\lscaN}{\Delta\scaN}

% Bar + Greek quantities
\newcommand{\balpha     }{\bar{\alpha}}
\newcommand{\bbeta      }{\bar{\beta}}
\newcommand{\bgamma     }{\bar{\gamma}}
\newcommand{\bdelta     }{\bar{\delta}}
\newcommand{\bepsilon   }{\bar{\epsilon}}
\newcommand{\bvareps    }{\bar{\varepsilon}}
\newcommand{\blambda    }{\bar{\lambda}}
\newcommand{\bxi        }{\bar{\xi}}
\newcommand{\bsigma     }{\bar{\sigma}}
\newcommand{\bvarsigma  }{\bar{\varsigma}}
\newcommand{\btau       }{\bar{\tau}}

% Tilde + Greek quantities
\newcommand{\tileps     }{\widetilde{\epsilon}}
\newcommand{\tillambda  }{\widetilde{\lambda}}
\newcommand{\tilsigma   }{\widetilde{\sigma}}

% Greek Vector
\newcommand{\vecalpha     }{\ensuremath{ \gvec{\alpha} }}
\newcommand{\vecbeta      }{\ensuremath{ \gvec{\beta} }}
\newcommand{\vecgamma     }{\ensuremath{ \gvec{\gamma} }}
\newcommand{\vecdelta     }{\ensuremath{ \gvec{\delta} }}
\newcommand{\vecepsilon   }{\ensuremath{ \gvec{\epsilon} }}
\newcommand{\vecvarepsilon}{\ensuremath{ \gvec{\varepsilon} }}
\newcommand{\veczeta      }{\ensuremath{ \gvec{\zeta} }}
\newcommand{\veceta       }{\ensuremath{ \gvec{\eta} }}
\newcommand{\vectheta     }{\ensuremath{ \gvec{\theta} }}
\newcommand{\vecvartheta  }{\ensuremath{ \gvec{\vartheta} }}
\newcommand{\veciota      }{\ensuremath{ \gvec{\iota} }}
\newcommand{\veckappa     }{\ensuremath{ \gvec{\kappa} }}
\newcommand{\veclam       }{\ensuremath{ \gvec{\lambda} }}
\newcommand{\vecmu        }{\ensuremath{ \gvec{\mu} }}
\newcommand{\vecnu        }{\ensuremath{ \gvec{\nu} }}
\newcommand{\vecxi        }{\ensuremath{ \gvec{\xi} }}
\newcommand{\vecpi        }{\ensuremath{ \gvec{\pi} }}
\newcommand{\vecvarpi     }{\ensuremath{ \gvec{\varphi} }}
\newcommand{\vecrho       }{\ensuremath{ \gvec{\rho} }}
\newcommand{\vecvarrho    }{\ensuremath{ \gvec{\varrho} }}
\newcommand{\vecsigma     }{\ensuremath{ \gvec{\sigma} }}
\newcommand{\vecvarsigma  }{\ensuremath{ \gvec{\varsigma} }}
\newcommand{\vectau       }{\ensuremath{ \gvec{\tau} }}
\newcommand{\vecupsilon   }{\ensuremath{ \gvec{\upsilon} }}
\newcommand{\vecphi       }{\ensuremath{ \gvec{\phi} }}
\newcommand{\vecvarphi    }{\ensuremath{ \gvec{\varphi} }}
\newcommand{\vecchi       }{\ensuremath{ \gvec{\chi} }}
\newcommand{\vecpsi       }{\ensuremath{ \gvec{\psi} }}
\newcommand{\vecomega     }{\ensuremath{ \gvec{\omega} }}
\newcommand{\vecUpsilon   }{\ensuremath{ \gvec{\Upsilon} }}

% Bar + Greek Vector quantities
\newcommand{\bveceps      }{\ensuremath{ \bar{\gvec{\epsilon}} }}
\newcommand{\bveceta      }{\ensuremath{ \bar{\gvec{\eta}} }}
\newcommand{\bveclam      }{\ensuremath{ \bar{\gvec{\lambda}} }}
\newcommand{\bvecsig      }{\ensuremath{ \bar{\gvec{\sigma}} }}
\newcommand{\bvecvarsigma }{\ensuremath{ \bar{\gvec{\varsigma}} }}
\newcommand{\bvectau      }{\ensuremath{ \bar{\gvec{\tau}} }}
\newcommand{\bvecupsilon  }{\ensuremath{ \bar{\gvec{\upsilon} }}}

% Tilde + Greek Vector quantities
\newcommand{\tilveceps    }{\widetilde{\vecepsilon}}
\newcommand{\tilvecsig    }{\widetilde{\vecsigma}}

% Dot + Greek Vector quantities
\newcommand{\tveclam}{\ensuremath{ \dot{\veclam }}}

% Linearization + Greek Vector quantities
\newcommand{\lveceps}{\Delta\vecepsilon}
\newcommand{\lveclam}{\Delta\veclam}
\newcommand{\lvecsig}{\Delta\vecsigma}
\newcommand{\lvectau}{\Delta\vectau}
\newcommand{\lvecxi }{\Delta\vecxi}

% Variation + Greek Vector quantities
\newcommand{\dveceps}{\delta\vecepsilon}
\newcommand{\dveclam}{\delta\veclam}
\newcommand{\dvecsig}{\delta\vecsigma}
\newcommand{\dvectau}{\delta\vectau}
\newcommand{\dvecxi }{\delta\vecxi}

% Linearization + Bar + Greek Vector quantities

\newcommand{\lbveclam}{\Delta\bar{\veclam}}

% Greek Tensor
\newcommand{\tenalpha     }{\ensuremath{ \gten{\alpha} }}
\newcommand{\tenbeta      }{\ensuremath{ \gten{\beta} }}
\newcommand{\tengamma     }{\ensuremath{ \gten{\gamma} }}
\newcommand{\tendelta     }{\ensuremath{ \gten{\delta} }}
\newcommand{\tenepsilon   }{\ensuremath{ \gten{\epsilon} }}
\newcommand{\teneps       }{\ensuremath{ \gten{\varepsilon} }}
\newcommand{\tenzeta      }{\ensuremath{ \gten{\zeta} }}
\newcommand{\teneta       }{\ensuremath{ \gten{\eta} }}
\newcommand{\tentheta     }{\ensuremath{ \gten{\theta} }}
\newcommand{\tenvartheta  }{\ensuremath{ \gten{\vartheta} }}
\newcommand{\teniota      }{\ensuremath{ \gten{\iota} }}
\newcommand{\tenkappa     }{\ensuremath{ \gten{\kappa} }}
\newcommand{\tenlambda    }{\ensuremath{ \gten{\lambda} }}
\newcommand{\tenmu        }{\ensuremath{ \gten{\mu} }}
\newcommand{\tennu        }{\ensuremath{ \gten{\nu} }}
\newcommand{\tenxi        }{\ensuremath{ \gten{\xi} }}
\newcommand{\tenpi        }{\ensuremath{ \gten{\pi} }}
\newcommand{\tenvarpi     }{\ensuremath{ \gten{\varphi} }}
\newcommand{\tenrho       }{\ensuremath{ \gten{\rho} }}
\newcommand{\tenvarrho    }{\ensuremath{ \gten{\varrho} }}
\newcommand{\tensig       }{\ensuremath{ \gten{\sigma} }}
\newcommand{\tenvarsigma  }{\ensuremath{ \gten{\varsigma} }}
\newcommand{\tentau       }{\ensuremath{ \gten{\tau} }}
\newcommand{\tenupsilon   }{\ensuremath{ \gten{\upsilon} }}
\newcommand{\tenphi       }{\ensuremath{ \gten{\phi} }}
\newcommand{\tenvarphi    }{\ensuremath{ \gten{\varphi} }}
\newcommand{\tenchi       }{\ensuremath{ \gten{\chi} }}
\newcommand{\tenpsi       }{\ensuremath{ \gten{\psi} }}
\newcommand{\tenomega     }{\ensuremath{ \gten{\omega} }}

\newcommand{\tenOmega     }{\ensuremath{ \gten{\Omega} }}

% Tilde + Greek Tensor
\newcommand{\tilteneps    }{\widetilde{\teneps}}
\newcommand{\tiltensig    }{\widetilde{\tensig}}

% Bar + Greek Tensor

\newcommand{\bteneps}{\ensuremath{ \bar{\teneps }}}
\newcommand{\btensig}{\ensuremath{ \bar{\tensig }}}

% Dot + Greek Tensor

\newcommand{\tteneps}{\ensuremath{ \dot{\teneps }}}
\newcommand{\ttensig}{\ensuremath{ \dot{\tensig }}}

% Linearization + Greek Tensor

\newcommand{\ltenalpha}{\Delta\tenalpha}
\newcommand{\ltenbeta }{\Delta\tenbeta}
\newcommand{\lteneps  }{\Delta\teneps}
\newcommand{\ltensig  }{\Delta\tensig}

% Dot + Greek quantities
\newcommand{\tgamma}{\ensuremath{ \dot{\gamma} }}
\newcommand{\txi}{\ensuremath{ \dot{\xi} }}
\newcommand{\tlam}{\ensuremath{ \dot{\lambda} }}
\newcommand{\tomega}{\ensuremath{ \dot{\omega} }}

% Linearization + Greek quantities
\newcommand{\lgamma}{\Delta\gamma}
\newcommand{\llambda}{\Delta\lambda}
\newcommand{\lxi}{\Delta\xi}
\newcommand{\lsigma}{\stackrel{\triangle}{\sigma}}
\newcommand{\ltau}{\stackrel{\triangle}{\tau}}

% Greek Hat
\newcommand{\hatalpha     }{\ensuremath{ \widehat{\alpha} }}
\newcommand{\hatbeta      }{\ensuremath{ \widehat{\beta} }}
\newcommand{\hatgamma     }{\ensuremath{ \widehat{\gamma} }}
\newcommand{\hatdelta     }{\ensuremath{ \widehat{\delta} }}
\newcommand{\hatepsilon   }{\ensuremath{ \widehat{\epsilon} }}
\newcommand{\hatvarepsilon}{\ensuremath{ \widehat{\varepsilon} }}
\newcommand{\hatzeta      }{\ensuremath{ \widehat{\zeta} }}
\newcommand{\hateta       }{\ensuremath{ \widehat{\eta} }}
\newcommand{\hattheta     }{\ensuremath{ \widehat{\theta} }}
\newcommand{\hatvartheta  }{\ensuremath{ \widehat{\vartheta} }}
\newcommand{\hatiota      }{\ensuremath{ \widehat{\iota} }}
\newcommand{\hatkappa     }{\ensuremath{ \widehat{\kappa} }}
\newcommand{\hatlambda    }{\ensuremath{ \widehat{\lambda} }}
\newcommand{\hatmu        }{\ensuremath{ \widehat{\mu} }}
\newcommand{\hatnu        }{\ensuremath{ \widehat{\nu} }}
\newcommand{\hatxi        }{\ensuremath{ \widehat{\xi} }}
\newcommand{\hatpi        }{\ensuremath{ \widehat{\pi} }}
\newcommand{\hatvarpi     }{\ensuremath{ \widehat{\varphi} }}
\newcommand{\hatrho       }{\ensuremath{ \widehat{\rho} }}
\newcommand{\hatvarrho    }{\ensuremath{ \widehat{\varrho} }}
\newcommand{\hatsigma     }{\ensuremath{ \widehat{\sigma} }}
\newcommand{\hatvarsigma  }{\ensuremath{ \widehat{\varsigma} }}
\newcommand{\hattau       }{\ensuremath{ \widehat{\tau} }}
\newcommand{\hatupsilon   }{\ensuremath{ \widehat{\upsilon} }}
\newcommand{\hatphi       }{\ensuremath{ \widehat{\phi} }}
\newcommand{\hatvarphi    }{\ensuremath{ \widehat{\varphi} }}
\newcommand{\hatchi       }{\ensuremath{ \widehat{\chi} }}
\newcommand{\hatpsi       }{\ensuremath{ \widehat{\psi} }}
\newcommand{\hatomega     }{\ensuremath{ \widehat{\omega} }}

% Greek Special
\newcommand{\hatteneps}{\ensuremath{ \widehat{\teneps} }}
\newcommand{\hattensig}{\ensuremath{ \widehat{\tensig} }}

% Fracture specials
\newcommand{\ionesi}{\scaI\utensig}
\newcommand{\itwosi}{\scaI\scaI\utensig}
\newcommand{\ithrsi}{\scaI\scaI\scaI\utensig}
\newcommand{\itwos}{\scaI\scaI\utens}
\newcommand{\ithrs}{\scaI\scaI\scaI\utens}

% Fracture specials
\newcommand{\onetwo}{\frac{1}{2}}
\newcommand{\thrtwo}{\frac{3}{2}}
\newcommand{\onethr}{\frac{1}{3}}
\newcommand{\twothr}{\frac{2}{3}}
\newcommand{\forthr}{\frac{4}{3}}
\newcommand{\onefor}{\frac{1}{4}}
\newcommand{\onesix}{\frac{1}{6}}
\newcommand{\oneeig}{\frac{1}{8}}
\newcommand{\onenin}{\frac{1}{9}}
\newcommand{\onetwe}{\frac{1}{12}}

% Differential geometry definitions
\newcommand{\tengf}{\teng^{\flat}}
\newcommand{\tengs}{\teng^{\sharp}}

% Linearization
\newcommand{\Lin}{^{Lin}}

% Subscript and superscript tiny letters
\newcommand{\uscan}{_{\mbox{\tiny{N}}}}

\newcommand{\ena}{\ensuremath{^{n+1}}}
\newcommand{\sena}{\ensuremath{^{1\,n+1}}}
\newcommand{\mena}{\ensuremath{^{2\,n+1}}}

% Subscript and superscript greek letters
\newcommand{\ea}{^{\alpha}}
\newcommand{\eb}{^{\beta}}
\newcommand{\ec}{^{\gamma}}
\newcommand{\ed}{^{\delta}}
\newcommand{\ex}{^{\xi}}

\newcommand{\eat}{^{\alpha T}}
\newcommand{\ebt}{^{\beta T}}
\newcommand{\ect}{^{\gamma T}}
\newcommand{\edt}{^{\delta T}}
\newcommand{\eet}{^{\epsilon T}}

\newcommand{\eaa}{^{\alpha\alpha}}
\newcommand{\eab}{^{\alpha\beta}}
\newcommand{\eac}{^{\alpha\gamma}}
\newcommand{\ead}{^{\alpha\delta}}
\newcommand{\eba}{^{\beta\alpha}}
\newcommand{\ebc}{^{\beta\gamma}}
\newcommand{\ebd}{^{\beta\delta}}
\newcommand{\ecb}{^{\gamma\beta}}
\newcommand{\ecd}{^{\gamma\delta}}
\newcommand{\edb}{^{\delta\beta}}
\newcommand{\ede}{^{\delta\epsilon}}
\newcommand{\eec}{^{\epsilon\gamma}}
\newcommand{\exx}{^{\xi\xi}}

\newcommand{\ua}{_{\alpha}}
\newcommand{\ub}{_{\beta}}
\newcommand{\uc}{_{\gamma}}
\newcommand{\ud}{_{\delta}}
\newcommand{\ue}{_{\epsilon}}
\newcommand{\ux}{_{\xi}}

\newcommand{\uaa}{_{\alpha\alpha}}
\newcommand{\uab}{_{\alpha\beta}}
\newcommand{\uac}{_{\alpha\gamma}}
\newcommand{\uba}{_{\beta\alpha}}
\newcommand{\ubb}{_{\beta\beta}}
\newcommand{\ubc}{_{\beta\gamma}}
\newcommand{\ubd}{_{\beta\delta}}
\newcommand{\ucd}{_{\gamma\delta}}
\newcommand{\ucb}{_{\gamma\beta}}
\newcommand{\ueb}{_{\epsilon\beta}}
\newcommand{\ued}{_{\epsilon\delta}}
\newcommand{\udb}{_{\delta\beta}}
\newcommand{\uta}{_{{\mbox{\tiny{T}}}\alpha}}
\newcommand{\utb}{_{{\mbox{\tiny{T}}}\beta}}
\newcommand{\utc}{_{{\mbox{\tiny{T}}}\gamma}}
\newcommand{\uxx}{_{\xi\xi}}

\newcommand{\uka}{_{,\alpha}}
\newcommand{\ukb}{_{,\beta}}
\newcommand{\ukc}{_{,\gamma}}
\newcommand{\ukx}{_{,\xi}}

\newcommand{\uakb}{_{\alpha,\beta}}
\newcommand{\uakc}{_{\alpha,\gamma}}
\newcommand{\ubkc}{_{\beta,\gamma}}
\newcommand{\ubkd}{_{\beta,\delta}}
\newcommand{\ubke}{_{\beta,\epsilon}}
\newcommand{\uckd}{_{\gamma,\delta}}
\newcommand{\udke}{_{\delta,\epsilon}}

\newcommand{\ukaa}{_{,\alpha\alpha}}
\newcommand{\ukab}{_{,\alpha\beta}}
\newcommand{\ukba}{_{,\beta\alpha}}
\newcommand{\ukbb}{_{,\beta\beta}}
\newcommand{\ukbc}{_{,\beta\gamma}}
\newcommand{\ukxx}{_{,\xi\xi}}
\newcommand{\uxkx}{_{\xi,\xi}}
\newcommand{\uxkxx}{_{\xi,\xi\xi}}

\newcommand{\ubkcd}{_{\beta,\gamma\delta}}

\newcommand{\uga}{_{g\alpha}}
\newcommand{\ugb}{_{g\beta}}
\newcommand{\ugc}{_{g\gamma}}
\newcommand{\ugd}{_{g\delta}}
\newcommand{\ugka}{_{g,\alpha}}
\newcommand{\ugkb}{_{g,\beta}}
\newcommand{\ugkc}{_{g,\gamma}}
\newcommand{\ugkx}{_{g,\xi}}
\newcommand{\ugakb}{_{g\alpha,\beta}}
\newcommand{\ugbkc}{_{g\beta,\gamma}}

\newcommand{\uana}{_{\alpha\,n+1}}
\newcommand{\ubna}{_{\beta\,n+1}}

\newcommand{\ukana}{_{,\alpha\,n+1}}
\newcommand{\ukano}{_{,\alpha\,n}}

\newcommand{\uano}{_{\alpha\,n}}
\newcommand{\ubno}{_{\beta\,n}}

%Subscript and superscript scalar letters 
\newcommand{\umN}{_{\mbox{\tiny{N}}}}
\newcommand{\umT}{_{\mbox{\tiny{T}}}}

%Subscript tensor quantities
\newcommand{\utenb}{_{\tenb}}
\newcommand{\utenp}{_{\tenp}}
\newcommand{\utens}{_{\tens}}
\newcommand{\utenC}{_{\tenC}}
\newcommand{\utenE}{_{\tenE}}

%Subscript tensor quantities actual time step
\newcommand{\utensna}{_{\tens\,n+1}}

%Subscript greek tensor quantities
\newcommand{\uteneps}{_{\teneps}}
\newcommand{\utenepse}{_{\teneps^e}}
\newcommand{\utenepsp}{_{\teneps^p}}
\newcommand{\utensig}{_{\tensig}}
\newcommand{\utensigsig}{_{\tensig\tensig}}

%Subscript greek tensor quantities actual time step
\newcommand{\utenepsna}{_{\teneps\,n+1}}
\newcommand{\utensigna}{_{\tensig\,n+1}}

%Mixture specials
\newcommand{\gvecx}{\grave{\vecx}}

% Subscript and superscript mixture letters

\newcommand{\mska}{_{s,\alpha}}
\newcommand{\mskb}{_{s,\beta}}
\newcommand{\mskc}{_{s,\gamma}}
\newcommand{\mfka}{_{f,\alpha}}
\newcommand{\mfkb}{_{f,\beta}}
\newcommand{\mfkc}{_{f,\gamma}}

% Subscript and superscript discretized mixture letters

\newcommand{\dmska}{_{Ag,\alpha}}
\newcommand{\dmskb}{_{Ag,\beta}}
\newcommand{\dmskc}{_{Ag,\gamma}}
\newcommand{\dmfka}{_{Ag,\alpha}}
\newcommand{\dmfkb}{_{Ag,\beta}}
\newcommand{\dmfkc}{_{Ag,\gamma}}

\newcommand{\gmska}{_{g,\alpha}}
\newcommand{\gmskb}{_{g,\beta}}
\newcommand{\gmskc}{_{g,\gamma}}
\newcommand{\gmfka}{_{g,\alpha}}
\newcommand{\gmfkb}{_{g,\beta}}
\newcommand{\gmfkc}{_{g,\gamma}}

%Summation over discretized values
\newcommand{\sumgp}{\sum_{g=1}^{n_{gp}}}
\newcommand{\sumni}{\sum_{I=1}^{n_{I}}}
\newcommand{\sumnj}{\sum_{J=1}^{n_{J}}}
\newcommand{\sumseg}{\sum_{seg}}
\newcommand{\sumel}{\sum_{e=1}^{n_{el}}}

%Summation over greek letters
\newcommand{\suma}{\sum_{\alpha=1}^{n_{\alpha}}}

%Contact GAUSS point quantities
\newcommand{\gng}{\scag_{N\,g}}
\newcommand{\gtg}{g_{T\,g}}
\newcommand{\lng}{\lambda_{N\,g}}
\newcommand{\ltg}{\lambda_{T\,g}}
\newcommand{\ttg}{t_{T\,g}}

% Differential Quantities
\renewcommand{\d}[1]{\text{$\hspace{0.1cm}$d $\hspace{-0.11cm}#1$}}
\newcommand{\del}{\ensuremath{\partial}}
\newcommand{\divx}[1]{\text{$\hspace{0.1cm}$div$\left(#1\right)$}}
\newcommand{\divX}[1]{\text{$\hspace{0.1cm}$Div$\left(#1\right)$}}
\newcommand{\grad}[1]{\ensuremath{ \boldsymbol{\nabla}{#1}}}
\newcommand{\gradx}[1]{\ensuremath{ \boldsymbol{\nabla}_{\vecx}{#1}}}
\newcommand{\gradX}[1]{\ensuremath{ \boldsymbol{\nabla}_{\vecX}{#1}}}
\newcommand{\parder}[2]{\ensuremath{ \frac{\del #1}{\del #2} }}
\newcommand{\tder}[1]{\ensuremath{ \frac{\d{#1}}{\d{} \hspace{0.05cm}{t}} }}
\newcommand{\dx}{\ensuremath{ \d{\vecx} }}
\newcommand{\dX}{\ensuremath{ \d{\vecX} }}
\newcommand{\da}{\ensuremath{ \d{a} }}
\newcommand{\dA}{\ensuremath{ \d{A} }}
\newcommand{\dv}{\ensuremath{ \d{v} }}
\newcommand{\dV}{\ensuremath{ \d{V} }}
\newcommand{\dxis}{\ensuremath{ \d{\xi} }}
\newcommand{\lda}{\stackrel{\triangle}{\da}}

\newcommand{\dr}{\ensuremath{ \d{r} }}
\newcommand{\dphi}{\ensuremath{ \d{\phi} }}
\newcommand{\dz}{\ensuremath{\d{z}}}
\newcommand{\du}{\ensuremath{\d{u}}}
\newcommand{\dy}{\ensuremath{\d{y}}}
\newcommand{\dxscal}{\ensuremath{\d{x}}}

% Functions
\renewcommand{\cos}[1]{ \text{cos}\hspace{0.0cm}\left( {#1} \right) }
\renewcommand{\sin}[1]{ \text{sin}\hspace{0.0cm}\left( {#1} \right) }
\renewcommand{\ln}[1]{\text{$\hspace{0.1cm}$ln$\left(#1\right)$}}
\renewcommand{\exp}[1]{\ensuremath{ \,\text{exp}{\left( #1 \right)} }}

% Symbols
\newcommand{\define}{\ensuremath{\stackrel{\mathrm{def}}{=}}}
\newcommand{\p}{\ensuremath{ ^{\prime} }}
\newcommand{\pp}{\ensuremath{ ^{\prime \prime} }}
\newcommand{\first}{$\ensuremath{ 1^{\text{st}} }\,$}
\newcommand{\second}{$\ensuremath{ 2^{\text{nd}} }\,$}
\newcommand{\lin}[1]{\ensuremath{ \calL[#1] }}
%\newcommand{\lin}[1]{\ensuremath{ \calL[#1 ; \calR_o] }}

% Tensor definitions
\newcommand{\tenfour}[1]{ \ensuremath {\boldsymbol{\mathcal{#1}} } }
\newcommand{\tr}[1]{\text{$\hspace{0.1cm}$tr$\left(#1\right)$}}
\newcommand{\dev}{\ensuremath{ ^{\prime} }}
\renewcommand{\det}[1]{\text{$\hspace{0.1cm}$det$\left(#1\right)$}}
\newcommand{\inv}{\ensuremath{ ^{-1} }}

\renewcommand{\it}{\ensuremath{ ^{-T} }}
\newcommand{\sym}{\ensuremath{ ^{\text{sym}} }}
\newcommand{\skw}{\ensuremath{ ^{\text{skw}} }}
\newcommand{\adj}{\ensuremath{ ^{\sharp}  }}
\newcommand{\mg}[1]{\ensuremath{ \left\| #1 \right\| }}
\newcommand{\mgv}[1]{\ensuremath{ \left| #1 \right| }}
\newcommand{\s}{\ensuremath{ ^{2}  }}
\newcommand{\ione}[1]{\ensuremath{ I_{#1} }}
\newcommand{\itwo}[1]{\ensuremath{ I\hspace{-0.1cm}I_{#1} }}
\newcommand{\ithree}[1]{\ensuremath{ I\hspace{-0.1cm}I\hspace{-0.1cm}I_{#1} }}
\newcommand{\ionep}[1]{\ensuremath{ I_{#1}\p }}
\newcommand{\itwop}[1]{\ensuremath{ I\hspace{-0.1cm}I_{#1}\p }}
\newcommand{\ithreep}[1]{\ensuremath{ I\hspace{-0.1cm}I\hspace{-0.1cm}I_{#1}\p }}
\newcommand{\ionepp}[1]{\ensuremath{ I_{#1}\pp }}
\newcommand{\itwopp}[1]{\ensuremath{ I\hspace{-0.1cm}I_{#1}\pp }}
\newcommand{\ithreepp}[1]{\ensuremath{ I\hspace{-0.1cm}I\hspace{-0.1cm}I_{#1}\pp }}

\newcommand{\gus}{\frac{\partial\scag}{\partial\tensig}}
\newcommand{\guss}{\frac{\partial^2\scag}{\partial^2\tensig}}

% Special Characters
\newcommand{\percent}{\ensuremath{ \%  }}
\newcommand{\IE}{\ensuremath{I\hspace{-0.12cm}E  }}
\newcommand{\II}{\ensuremath{1\hspace{-0.12cm}1  }}
\newcommand{\tenIE}{\ensuremath{ \ten{\IE}  }}
\newcommand{\tenII}{\ensuremath{ \ten{\II}  }}
\newcommand{\hattenIE}{\ensuremath{ \widehat{\tenIE}  }}
\newcommand{\IC}{\ensuremath{C\hspace{-0.22cm}C  }}
\newcommand{\cc}{\ensuremath{c\hspace{-0.22cm}c  }}
\newcommand{\tenIC}{\ensuremath{ \ten{\IC}  }}
\newcommand{\tencc}{\ensuremath{ \ten{\cc}  }}
\newcommand{\lsup}[1]{\ensuremath{ {}^{[#1]} \hspace{-0.05cm} }}
\newcommand{\lsub}[1]{\ensuremath{ {}_{[#1]} \hspace{-0.05cm} }}
\newcommand{\lsupb}[2]{\ensuremath{ {}^{#1}_{#2} \hspace{-0.05cm} }}
\newcommand{\Phihat}{\ensuremath{ \widehat{\Phi} }}
\newcommand{\Psihat}{\ensuremath{ \widehat{\Psi} }}
\renewcommand{\s}{\ensuremath{ ^*} }
\renewcommand{\k}{\ensuremath{ ^K} }
\newcommand{\kp}{\ensuremath{ ^{K+1}} }
\newcommand{\km}{\ensuremath{ ^{K-1}} }
\newcommand{\tol}[1]{\ensuremath{ \text{TOL}_{#1} } }
\newcommand{\pen}{\ensuremath{ \calK }}
\newcommand{\tenone}{\ensuremath{ \ten{1} }}
\newcommand{\err}{\ensuremath{ \pounds }}
\newcommand{\li}{\ensuremath{\left[\hspace{-0.15cm}\left[ }}
\newcommand{\ri}{\ensuremath{\right]\hspace{-0.15cm}\right] }}
\renewcommand{\t}{\ensuremath{ ^{\scaT} }}
\newcommand{\tria}{\ensuremath{ ^{tr} }}

% Isoparametric Mapping
\newcommand{\iso}[1]{\ensuremath{ \widetilde{#1} }}
\newcommand{\tenFisox}{\ensuremath{ \tenF_{\vecx} } }
\newcommand{\tenFisoX}{\ensuremath{ \tenF_{\vecX} } }
\newcommand{\tenCisox}{\ensuremath{ \tenC_{\vecx} } }
\newcommand{\tenCisoX}{\ensuremath{ \tenC_{\vecX} } }
\newcommand{\Jisox}{\ensuremath{ J_{\vecx} } }
\newcommand{\JisoX}{\ensuremath{ J_{\vecX} } }
\newcommand{\calRiso}{ \ensuremath{ \iso{\calR}}}
\newcommand{\dzeta}{\ensuremath{ \d{\veczeta} } }
\newcommand{\daiso}{\ensuremath{ \d{\iso{a}} }}
\newcommand{\dviso}{\ensuremath{ \d{\iso{v}} }}
\newcommand{\vecniso}{\ensuremath{ \iso{\vecn} } }

%Special contact quantities and varation and linearization
\newcommand{\dgn}{\delta \bar{g}_{\mbox{\tiny{N}}}}
\newcommand{\lgn}{\Delta \bar{g}_{\mbox{\tiny{N}}}}
\newcommand{\ldgn}{\Delta\delta \bar{g}_{\mbox{\tiny{N}}}}

\newcommand{\dbxi}{\delta \bar{\xi}}
\newcommand{\lbxi}{\Delta \bar{\xi}}
\newcommand{\ldbxi}{\Delta\delta \bar{\xi}}

%integration current configuration
\newcommand{\intbco}{\int\limits_{\varphi\left(B^1\right)}}
%{\int\limits_{\varphi\left(B^1\right)}}
\newcommand{\intbct}{\int\limits_{\varphi\left(B^2\right)}}
\newcommand{\intbcc}{\int\limits_{\varphi\left(B_c\right)}}
\newcommand{\intpbco}{\int\limits_{\varphi\left(\partial B^1\right)}}
\newcommand{\intpbct}{\int\limits_{\varphi\left(\partial B^2\right)}}
\newcommand{\intpbcon}{\int\limits_{\varphi\left(\partial B^1_n\right)}}
\newcommand{\intpbctn}{\int\limits_{\varphi\left(\partial B^2_n\right)}}
\newcommand{\intpbcc}{\int\limits_{\Gamma_c}}
%integration reference configuration
\newcommand{\intbro}{\int\limits_{B^1}}
\newcommand{\intbrt}{\int\limits_{B^2}}
\newcommand{\intbrc}{\int\limits_{B_c}}

\newcommand{\intprco}{\int\limits_{\partial B^1}}
\newcommand{\intprct}{\int\limits_{\partial B^2}}
\newcommand{\intprcon}{\int\limits_{\partial B^1_n}}
\newcommand{\intprctn}{\int\limits_{\partial B^2_n}}
\newcommand{\intprcc}{\int\limits_{\partial B_c}}

%Mortar mass matrix
\newcommand{\mnab}{n_{AB}}
\newcommand{\mnac}{n_{AC}}
\newcommand{\mnad}{n_{AD}}
\newcommand{\mnae}{n_{AE}}

%Mortar shell quantities
\newcommand{\hthr}{\frac{h}{3}}
\newcommand{\hsix}{\frac{h}{6}}

%Mortar local and intrinsic basis

\newcommand{\pna}{_{p\,n+1}}
\newcommand{\pno}{_{p\,n}}
\newcommand{\Ina}{_{I\,n+1}}
\newcommand{\Ino}{_{I\,n}}
\newcommand{\Jna}{_{J\,n+1}}
\newcommand{\Jno}{_{J\,n}}
\newcommand{\sal}{^{1\,\alpha}}
\newcommand{\sbl}{^{1\,\beta}}
\newcommand{\mal}{^{2\,\alpha}}
\newcommand{\xisg}{\left(\xi^1_{g\,n+1}\right)}
\newcommand{\ximg}{\left(\xi^2_{g\,n+1}\right)}
\newcommand{\xisp}{\left(\xi^1_{p\,n+1}\right)}
\newcommand{\ximp}{\left(\xi^2_{p\,n+1}\right)}
\newcommand{\xisgo}{\left(\xi^1_{g\,n}\right)}
\newcommand{\xiso}{\left(\xi^1_{p\,n}\right)}
\newcommand{\ximo}{\left(\xi^2_{p\,n}\right)}
\newcommand{\xise}{\left(\xi^1_g\left(\eta\right)\right)}
\newcommand{\xime}{\left(\xi^2_g\left(\eta\right)\right)}
\newcommand{\xipe}{\left(\xi^1_p\left(\eta\right)\right)}
\newcommand{\xisa}{\left(\xi^1_A\right)}
\newcommand{\xima}{\left(\xi^2_A\right)}
\newcommand{\xisq}{\left(\xi^1_Q\right)}
\newcommand{\ximq}{\left(\xi^2_Q\right)}

\newcommand{\etag}{\left(\eta\right)}
\newcommand{\xig}{\left(\xi_g\right)}

\newcommand{\xisf}{\left(\xi^1_1\right)}
\newcommand{\ximf}{\left(\xi^2_1\right)}
\newcommand{\xiss}{\left(\xi^1_2\right)}
\newcommand{\xims}{\left(\xi^2_2\right)}

%Time integration indexes
\newcommand{\ana}{\ensuremath{ _{\alpha\,n+1} }}
\newcommand{\bna}{\ensuremath{ _{\beta\,n+1} }}

\newcommand{\sna}{\ensuremath{ _{s\,n+1} }}
\newcommand{\nna}{\ensuremath{ _{\mbox{\tiny{N}}\,n+1} }}
\newcommand{\tna}{\ensuremath{ _{\mbox{\tiny{T}}\,n+1} }}
\newcommand{\tnaa}{\ensuremath{_{\mbox{\tiny{T}}\alpha\,n+1\,}}}
\newcommand{\tnab}{\ensuremath{_{\mbox{\tiny{T}}\,n+1\,\beta}}}
\newcommand{\tnax}{\ensuremath{_{\mbox{\tiny{T}}\,n+1\,\xi}}}

\newcommand{\nA}{\ensuremath{ _{\mbox{\tiny{N}} A} }}
\newcommand{\Ng}{\ensuremath{ _{\mbox{\tiny{N}} g} }}
\newcommand{\tA}{\ensuremath{ _{\mbox{\tiny{T}} A} }}
\newcommand{\ta}{\ensuremath{ _{\mbox{\tiny{T}} \alpha} }}
\newcommand{\Tag}{\ensuremath{ _{\mbox{\tiny{T}} \alpha g} }}
\newcommand{\taA}{\ensuremath{ _{\mbox{\tiny{T}}\alpha A} }}
\newcommand{\tbA}{\ensuremath{ _{\mbox{\tiny{T}}\beta A} }}

\newcommand{\tana}{\ensuremath{ _{\mbox{\tiny{T}} A\,n+1} }}

\newcommand{\taana}{\ensuremath{ _{\mbox{\tiny{T}}\alpha A\,n+1} }}
\newcommand{\tbana}{\ensuremath{ _{\mbox{\tiny{T}}\beta A\,n+1} }}

\newcommand{\taano}{\ensuremath{ _{\mbox{\tiny{T}}\alpha A\,n} }}

\newcommand{\iana}{\ensuremath{ _{i A\,n+1} }}
\newcommand{\nana}{\ensuremath{ _{\mbox{\tiny{N}} A\,n+1} }}
\newcommand{\nano}{\ensuremath{ _{\mbox{\tiny{N}} A\,n} }}
\newcommand{\dana}{\ensuremath{ _{\mbox{\tiny{D}} A\,n+1} }}
\newcommand{\tno}{\ensuremath{ _{\mbox{\tiny{T}}\,n} }}
\newcommand{\tano}{\ensuremath{ _{\mbox{\tiny{T}}A\,n} }}
\newcommand{\na}{\ensuremath{ _{n+1} }}
\newcommand{\ina}{\ensuremath{ _{i\,n+1} }}
\newcommand{\jna}{\ensuremath{ _{j\,n+1} }}

\newcommand{\gna}{\ensuremath{ _{g\,n+1} }}
\newcommand{\qna}{\ensuremath{ _{q\,n+1} }}
\newcommand{\gno}{\ensuremath{ _{g\,n} }}

\newcommand{\Ana}{\ensuremath{ _{A\,n+1} }}
\newcommand{\Bna}{\ensuremath{ _{B\,n+1} }}
\newcommand{\Cna}{\ensuremath{ _{C\,n+1} }}
\newcommand{\Ano}{\ensuremath{ _{A\,n} }}
\newcommand{\Bno}{\ensuremath{ _{B\,n} }}
\newcommand{\Cno}{\ensuremath{ _{C\,n} }}

\newcommand{\nBna}{\ensuremath{ _{\mbox{\tiny{N}}B\,n+1} }}
\newcommand{\tBna}{\ensuremath{ _{\mbox{\tiny{T}}B\,n+1} }}

%Shell mortar quatities
\newcommand{\xina}{\left(\vecxi_{n+1}\right)}
\newcommand{\xino}{\left(\vecxi_{n}\right)}
\newcommand{\xigna}{\left(\vecxi_{g\,n+1}\right)}
\newcommand{\xigno}{\left(\vecxi_{g\,n}\right)}

\newcommand{\agna}{\ensuremath{ _{\alpha\,g\,n+1} }}
\newcommand{\kagna}{\ensuremath{ _{,\alpha\,g\,n+1} }}
\newcommand{\agno}{\ensuremath{ _{\alpha\,g\,n} }}
\newcommand{\kagno}{\ensuremath{ _{,\alpha\,g\,n} }}

\newcommand{\ogna}{\ensuremath{ _{1\,g\,n+1} }}
\newcommand{\kogna}{\ensuremath{ _{,1\,g\,n+1} }}
\newcommand{\ogno}{\ensuremath{ _{1\,g\,n} }}
\newcommand{\kogno}{\ensuremath{ _{,1\,g\,n} }}

\newcommand{\tgna}{\ensuremath{ _{2\,g\,n+1} }}
\newcommand{\ktgna}{\ensuremath{ _{,2\,g\,n+1} }}
\newcommand{\tgno}{\ensuremath{ _{2\,g\,n} }}
\newcommand{\ktgno}{\ensuremath{ _{,2\,g\,n} }}

\newcommand{\aAna}{\ensuremath{ _{\alpha\,A\,n+1} }}
\newcommand{\bAna}{\ensuremath{ _{\beta\,A\,n+1} }}
\newcommand{\aBna}{\ensuremath{ _{\alpha\,B\,n+1} }}

\newcommand{\nnAna}{\ensuremath{ _{33\,A\,n+1} }}
\newcommand{\aaAna}{\ensuremath{ _{\alpha\alpha\,A\,n+1} }}
\newcommand{\abAna}{\ensuremath{ _{12\,A\,n+1} }}
\newcommand{\anAna}{\ensuremath{ _{\alpha3\,A\,n+1} }}

\newcommand{\iAna}{\ensuremath{ _{i\,A\,n+1} }}
\newcommand{\jAna}{\ensuremath{ _{j\,A\,n+1} }}
\newcommand{\kAna}{\ensuremath{ _{k\,A\,n+1} }}
\newcommand{\kBna}{\ensuremath{ _{k\,B\,n+1} }}

\newcommand{\dthetc}{\delta\bar{\vartheta}^1}
\newcommand{\dthets}{\delta\vartheta^2}
\newcommand{\dphic}{\delta\bar{\varphi}^1}
\newcommand{\dphis}{\delta\varphi^2}

\newcommand{\Dthetc}{\Delta\bar{\vartheta}^1}
\newcommand{\Dthets}{\Delta\vartheta^2}
\newcommand{\Dphic}{\Delta\bar{\varphi}^1}
\newcommand{\Dphis}{\Delta\varphi^2}

\newcommand{\dxi}{\delta\bar{\xi}}
\newcommand{\dxia}{\delta\bar{\xi}^{\alpha}}
\newcommand{\dxib}{\delta\bar{\xi}^{\beta}}
\newcommand{\dxic}{\delta\bar{\xi}^{\gamma}}
\newcommand{\dxid}{\delta\bar{\xi}^{\delta}}
\newcommand{\dxie}{\delta\bar{\xi}^{\epsilon}}
\newcommand{\dxif}{\delta\bar{\xi}^{\eta}}

\newcommand{\Dxi}{\Delta\bar{\xi}}
\newcommand{\Dxia}{\Delta\bar{\xi}^{\alpha}}
\newcommand{\Dxib}{\Delta\bar{\xi}^{\beta}}
\newcommand{\Dxic}{\Delta\bar{\xi}^{\gamma}}
\newcommand{\Dxid}{\Delta\bar{\xi}^{\delta}}
\newcommand{\Dxie}{\Delta\bar{\xi}^{\epsilon}}
\newcommand{\Dxif}{\Delta\bar{\xi}^{\eta}}

\newcommand{\Ddxi}{\Delta\delta\bar{\xi}}
\newcommand{\Ddxia}{\Delta\delta\bar{\xi}^{\alpha}}
\newcommand{\Ddxib}{\Delta\delta\bar{\xi}^{\beta}}
\newcommand{\Ddxic}{\Delta\delta\bar{\xi}^{\gamma}}
\newcommand{\Ddxid}{\Delta\delta\bar{\xi}^{\delta}}
\newcommand{\Ddxie}{\Delta\delta\bar{\xi}^{\epsilon}}
\newcommand{\Ddxif}{\Delta\delta\bar{\xi}^{\eta}}

\newcommand{\dvecaa}{\delta\bar{\veca}_{\alpha}^1}
\newcommand{\dvecab}{\delta\bar{\veca}_{\beta}^1}

\newcommand{\dvecaat}{\delta\bar{\veca}^{1\alpha}}
\newcommand{\dvecabt}{\delta\bar{\veca}^{1\beta}}

\newcommand{\Dvecaa}{\Delta\bar{\veca}_{\alpha}^1}
\newcommand{\Dvecab}{\Delta\bar{\veca}_{\beta}^1}
\newcommand{\Dvecac}{\Delta\bar{\veca}_{\gamma}^1}

\newcommand{\Dvecaat}{\Delta\bar{\veca}^{1\alpha}}
\newcommand{\Dvecabt}{\Delta\bar{\veca}^{1\beta}}

\newcommand{\dvecn}{\delta\bar{\vecn}^1}
\newcommand{\Dvecn}{\Delta\bar{\vecn}^1}
\newcommand{\Ddvecn}{\Delta\delta\bar{\vecn}^1}

\newcommand{\dvecua}{\delta\bar{\vecu}_{,\alpha}^2}
\newcommand{\dvecub}{\delta\bar{\vecu}_{,\beta}^2}
\newcommand{\dvecuc}{\delta\bar{\vecu}_{,\gamma}^2}
\newcommand{\dvecud}{\delta\bar{\vecu}_{,\vartheta}^2}
\newcommand{\dvecue}{\delta\bar{\vecu}_{,\theta}^2}

\newcommand{\Dvecua}{\Delta\bar{\vecu}_{,\alpha}^2}
\newcommand{\Dvecub}{\Delta\bar{\vecu}_{,\beta}^2}
\newcommand{\Dvecuc}{\Delta\bar{\vecu}_{,\gamma}^2}
\newcommand{\Dvecud}{\Delta\bar{\vecu}_{,\vartheta}^2}
\newcommand{\Dvecue}{\Delta\bar{\vecu}_{,\theta}^2}

\newcommand{\dvecuab}{\delta\bar{\vecu}_{,\alpha\beta}^2}
\newcommand{\dvecuac}{\delta\bar{\vecu}_{,\alpha\gamma}^2}
\newcommand{\dvecubc}{\delta\bar{\vecu}_{,\beta\gamma}^2}
\newcommand{\dvecuad}{\delta\bar{\vecu}_{,\alpha\vartheta}^2}
\newcommand{\dvecuae}{\delta\bar{\vecu}_{,\alpha\theta}^2}

\newcommand{\Dvecuab}{\Delta\bar{\vecu}_{,\alpha\beta}^2}
\newcommand{\Dvecuac}{\Delta\bar{\vecu}_{,\alpha\gamma}^2}
\newcommand{\Dvecubc}{\Delta\bar{\vecu}_{,\beta\gamma}^2}
\newcommand{\Dvecuad}{\Delta\bar{\vecu}_{,\alpha\vartheta}^2}
\newcommand{\Dvecuae}{\Delta\bar{\vecu}_{,\alpha\theta}^2}

\newcommand{\dvecus}{\delta\vecu^1}
\newcommand{\Dvecus}{\Delta\vecu^1}
\newcommand{\vecxa}{\bar{\vecx}_{,\alpha}^2}
\newcommand{\vecxb}{\bar{\vecx}_{,\beta}^2}
\newcommand{\vecxab}{\bar{\vecx}_{,\alpha\beta}^2}

\newcommand{\gthet}{g_{\vartheta}}
\newcommand{\dgthet}{\delta g_{\vartheta}}
\newcommand{\Dgthet}{\Delta g_{\vartheta}}

\newcommand{\thetg}{\theta_G}
\newcommand{\dthetg}{\delta\theta_G}
\newcommand{\Dthetg}{\Delta\theta_G}

\newcommand{\gphi}{g_{\varphi}}
\newcommand{\dgphi}{\delta g_{\varphi}}
\newcommand{\Dgphi}{\Delta g_{\varphi}}

\newcommand{\tta}{t_{T\alpha}}
\newcommand{\ttb}{t_{T\beta}}
\newcommand{\ttc}{t_{T\gamma}}

\newcommand{\ttx}{t_{T\xi}}

\newcommand{\ttat}{t_{T}^{\alpha}}

\newcommand{\Dtta}{\Delta t_{T\alpha}}
\newcommand{\Dttb}{\Delta t_{T\beta}}
\newcommand{\Dttc}{\Delta t_{T\gamma}}

\newcommand{\ttra}{t_{t\alpha}^{tr}}
\newcommand{\ttrb}{t_{t\beta}^{tr}}
\newcommand{\ttrc}{t_{t\gamma}^{tr}}

\newcommand{\ttrat}{t_{T}^{trial\alpha}}
\newcommand{\ttrbt}{t_{T}^{trial\beta}}
\newcommand{\ttrct}{t_{T}^{trial\gamma}}
\newcommand{\ttrdt}{t_{T}^{trial\vartheta}}

\newcommand{\Dttra}{\Delta t_{T\alpha}^{trial}}
\newcommand{\Dttrb}{\Delta t_{T\beta}^{trial}}
\newcommand{\Dttrc}{\Delta t_{T\gamma}^{trial}}

\title{Model-based reinforcement corrosion prediction: Continuous calibration with Bayesian optimization and corrosion wire sensor data}

\author{A. Potnis$^1$, M. Macier$^2$, T. Leusmann$^2$, D. Anton$^1$, H. Wessels$^1$, D. Lowke$^3$}

\institute{(1)  Division Data-driven modeling of mechanical systems, Institute for Applied Mechanics, Technische Universtität Braunschweig, Pockelsstraße 3, 38106 Braunschweig,
\email{h.wessels@tu-braunschweig.de}\\
(2) Institute of Building Materials, Concrete Construction and Fire Safety, Technische Universtität Braunschweig, Beethovenstraße 52, 38106 Braunschweig, \email{t.leusmann@ibmb.tu-bs.de}\\
(3) TUM School of Engineering and Design, Technische Universität München, \email{lowke@tum.de}}

\maketitle

\thispagestyle{empty}

\abstract{Chloride-induced corrosion significantly contributes to the degradation of reinforced concrete structures, making accurate predictions of chloride migration and its effects on the materials durability critical. This paper explores two modeling approaches to estimate the effective diffusion coefficient for chloride transport. The first approach follows Gehlen’s interpretable diffusion model \cite{Gehlen} which is based on established physical principles and incorporates time and temperature dependencies in predicting chloride migration. The second approach is a neural network-based approach, where the neural network approximates the effective diffusion coefficient. In a subsequent step, the calibrated models are used to predict the penetration depth of the critical chloride content. The uncertainty in the critical chloride content is also taken into account here. The models are calibrated utilizing experimental data measured by a wire sensor, which is installed in a concrete test bridge. The calibration results are compared to effective diffusion coefficients derived from drilling dust samples. A comparison of both approaches reveals the advantages of the physics-based model in terms of transparency and interpretability, while the neural network model demonstrates flexibility and adaptability in data-driven predictions. This study emphasizes the importance of combining traditional and machine learning-based methods for improving the accuracy of chloride migration predictions in reinforced concrete.}


\section{Introduction}
\begin{comment}
\begin{itemize}
    \item \HW{@iBMB (Maurice, Thorsten)}: short motivation, sketch the introduction
    \item \HW{@iBMB (Maurice, Thorsten)} damage mechanisms overview, chloride migration in detail
\end{itemize}
\end{comment}

The deterioration of infrastructure in Germany and many European countries poses a growing risk to public safety and economic growth. Recent incidents such as the collapse of the Polcevera Viaduct in Genoa, the demolition of the Rahmede Valley Bridge near Lüdenscheid, and the failure of the Carola Bridge in Dresden underscore the consequences of ageing infrastructure and deferred maintenance. The primary cause of infrastructure deterioration is chloride-induced corrosion. In winter, de-icing agents are used to maintain road safety and prevent weather-related traffic disruptions. However, these agents release chlorides, which penetrate the concrete through its porous system or cracks. Once the chlorides reach a critical concentration on the reinforcement, they cause localized pitting corrosion, resulting in significant weakening of the reinforcement.
It is often only when such damage becomes visible that sampling for chloride content is performed, typically using dust samples from boreholes. By this time, however, costly repair measures such as removal and replacement of chloride-contaminated concrete are usually required to restore the integrity of the structure. The use of corrosion sensors, such as wire sensors, provides a more proactive approach by detecting the critical corrosion-causing chloride penetration front before it reaches the reinforcement, allowing for timely and less invasive maintenance interventions.
The wire sensor developed at the  Institute for Building Materials, Concrete Structures and Fire Safety (iBMB) of the TU Braunschweig indicates whether thin iron wires placed between concrete surface and reinforcement are corroded through by measuring the electrical resistance. The sensor consists of four parallel steel wires with a thickness of 0.065 mm and enables simple monitoring of the corrosion risk in the area of the concrete cover of the reinforcement. When the depassivation front (chloride or carbonation) reaches the steel wire, it usually corrodes very quickly. Corrosion-induced wire breakage can be detected by a significant, sudden increase in the sensor resistance, either manually with a standard ohmmeter or automatically with a resistance bridge circuit. The wire sensor is, therefore, a qualitative, non-reversible electrical corrosion sensor with digital measurement information indicating the presence of a wire break.
In this study, data from wire sensors installed in the 'Concerto' test bridge is used to calibrate two modeling approaches: the interpretative Gehlen model and a neural network-based model. These models are then applied to predict the time-to-corrosion, with results showing improved accuracy as more data became available. A comparative analysis of both approaches highlighted a trade-off between interpretability and flexibility. Additionally, uncertainty in estimating the critical chloride content for corrosion is considered by accounting for bounds, enhancing model reliability in predicting corrosion progression.

\subsection{'Concerto' Test Bridge}
\begin{figure}[h!]
    \centering
    \includegraphics[width=0.8\textwidth]{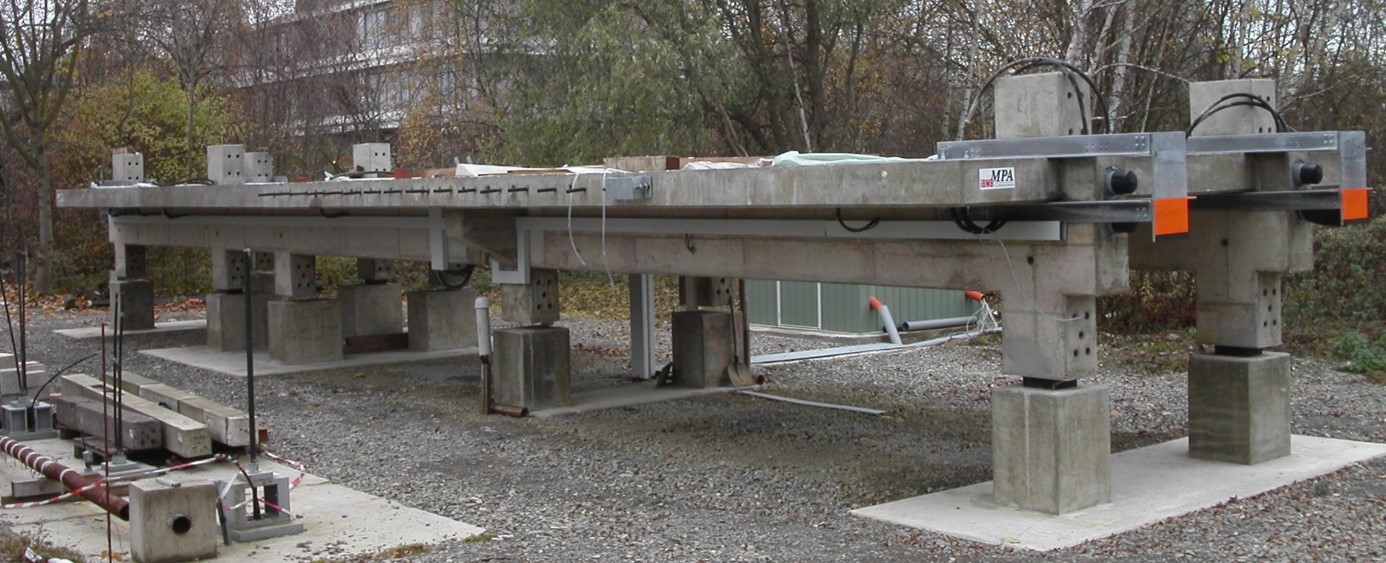}
    \caption{The 'Concerto' test bridge in 2005}
    \label{fig:bridge}
\end{figure}

The 'Concerto' test bridge is a true-to-scale, near-realistic section of a prestressed concrete bridge. It was constructed in August 2005 on the premises of the Institute for Building Materials, Concrete Structures, and Fire Safety (iBMB) at the TU Braunschweig for long-term testing of sensors and measurement concepts under realistic conditions. The structure (see Figure~\ref{fig:bridge}) is designed as a two-web slab-girder bridge with a total length of 18.5 meters, a slab width of 4.0 meters, and a web height of 0.8 meters. The bridge's substructure was built using concrete of strength class C30/37, while the superstructure consists of C20/25 concrete with 270 kg/m³ of CEM I 32.5 R, 60 kg/m³ of limestone powder, and a water-cement ratio of 0.58. The lower quality of concrete in the superstructure is one of several deliberately integrated typical weak points and defects found in real structures. With built-in sensors and the capability to configure various support positions (single-span girder with cantilever, continuous girder, or single-span girder) and load case combinations, different structural loading scenarios can be simulated, allowing the investigation of their effects on the structure's durability (see for example \cite{Holst.2007}). Thus, 'Concerto' serves as a central research platform for validating and further developing sensor technology and monitoring methods for use on real structures.\cite{HaraldBudelmann.2006}

\subsection{Investigations at 'Concerto' and laboratory experiments}
In addition to the investigation of monitoring techniques for locating prestress steel fractures and the subsequent installation of RFID wire sensors \cite{Dressler.2015} a major focus  of the investigations carried out at ‘Concerto’ was on monitoring corrosion activity using the wire sensor developed at iBMB \cite{Holst.2010}. Localised corrosive damage in the area of the wire sensor embedded in the bridge slab was induced by intermittent exposure to a 3 wt.\% NaCl solution in the period from May 2006 to June 2010. No exposure took place between October 2007 and May 2008. The sensor data collected during this period forms the basis for parameter identification and sensitivity analyses.
Furthermore, chloride concentration profiles were created in 2007 and 2024 with the help of concrete dust analyses from dry drilling. In addition, a rapid chloride migration test on the basis of \cite{BundesanstaltfurWasserbauBAW.2019} was carried out in 2024 on a drill core from the ‘Concerto’ bridge concrete that was not contaminated with chlorides. These standard tests are used to determine comparative values for the chloride diffusion coefficient.

\section{Diffusion model}\label{sec:physicalPhenomena}

The constant presence of \textit{NaCl} (sodium chloride) solution over a reinforced concrete block can cause corrosion through the production of chloride ions, as described by the dissociation reaction:
\begin{equation}
\textit{NaCl} \rightarrow\textit{Na}^+ + \textit{Cl}^-.
\end{equation}
The corrosion of rebars placed inside the concrete block is then driven by the concentration of chloride ions. This process can be modeled according to Fick’s second law of diffusion
\begin{equation}\label{eq:diffusion}
    \frac{\partial C}{\partial t} = D \frac{\partial^2 C}{\partial x^2},
\end{equation}
where $C$ is the concentration of chloride ions, $t$ is time, $x$ is the spatial coordinate, and $D$ denotes the diffusion coefficient.

The analytical solution applied in this research to model chloride diffusion in concrete structures is based on the work by Gehlen (2000) \cite{Gehlen} and builds on Fick’s second law of diffusion with a time-dependent diffusion coefficient. Starting from the standard one-dimensional diffusion equation, Gehlen modifies the classical form to incorporate environmental effects through a time-varying diffusion coefficient $D(t)$. This leads to the generalized solution:
\begin{equation}
    C(x, t) = C_{S, \Delta x} \cdot \left[ 1 - \text{erf} \left( \frac{x - \Delta x}{2 \cdot \sqrt{D_{\text{Eff,C}}(t) \cdot t}} \right) \right].
\label{eq:analytical_a}
\end{equation}
Here, $D_{\text{Eff,C}}(t)$ is the effective diffusion coefficient, which includes environmental conditions, and $erf$ denotes the error function. For an explanation of the variables $\Delta x$ and $C_{S, \Delta x}$, we refer to table \ref{tab:parameters_a}. The model is adapted from the frameworks developed by Bakker (1994) \cite{Bakker1994} and CEB Task Group V (1997)\cite{CEB1997}, which account for the variable weathering effects on the diffusion process.

The prerequisite for corrosion is that a critical chloride content $C_{crit}$ is exceeded. Equating the critical concentration $C_{crit}$ with the concentration $C (x,t)$ in \eqref{eq:analytical_a} allows to identify the effective diffusion coefficient $D_{\text{Eff,C}}(t)$ for given values of $C_{S,\Delta x}$, $\Delta x$ and $t$. In the following, we consider \eqref{eq:analytical_a} with two different approximations of the effective diffusion coefficient:\\

\noindent \textbf{Approach A - Gehlen:} In Gehlen’s model\cite{Gehlen}, the effective diffusion coefficient \( D_{\text{Eff},C}(t) \) is calculated as:
\begin{equation}\label{eq:d_eff}
\begin{aligned}
    D_{\text{Eff},C}(t) &= k_e \cdot k_t \cdot D_{RCM,0} \cdot A(t), \\
    D_{\text{Eff}, C}(t;b_e, D_{\text{t}}, a) &= \underbrace{e^{\left(b_e \cdot \left( \frac{1}{T_{\text{ref}}} - \frac{1}{T_{IST}(t)}\right)\right)}}_{k_e} \underbrace{k_t \cdot D_{\text{RCM}, 0}}_{D_{\text{t}}} \left( \frac{t_0}{t} \right)^a,
\end{aligned}
\end{equation}
where the corresponding terms are defined in the table \ref{tab:parameters_a}.

\begin{table}[h!]
    \centering
    \begin{tabular}{|C{0.15\linewidth} | p{0.8\linewidth}|}
        \hline
        \textbf{symbol} & \textbf{description} \\ 
        \hline 
        \multicolumn{2}{|c|}{\textbf{observable quantities}} \\
        \hline
        $t$ [s] & concrete age \\
        $x$ [m] & depth, with a corresponding chloride Concentration \( C(x,t) \)\\
        \hline 
        \multicolumn{2}{|c|}{\textbf{parameters from Gehlen's approach}} \\
        \hline
        $D_{\text{RCM},0}$ [m\(^2\)/s] & chloride migration coefficient of water-saturated concrete, determined at the reference time \( t_0 \) on specifically manufactured and preconditioned test specimens\\
        $k_t$ [-] & transfer parameter to account for deviations between chloride migration coefficients determined under accelerated conditions (Rapid Chloride Migration – \( D_{\text{RCM},0} \)) and diffusion coefficients determined under natural conditions, e.g., in the laboratory (Chloride Profiling Method – \( D_{\text{CPM},0} \))\\
        $A(t)$ [-] & aging function to account for time-dependent reduction of diffusion \\
        \hspace{0.2cm} $a$ [-] & exponent for considering the time dependence of \( D_{\text{Eff},C}(t) \), aging exponent \\
        \hspace{0.2cm} $t_0$ [s] & reference time  \\
        $k_e$ [-] & parameter to account for the temperature dependence of \( D_{\text{Eff},C}(t) \)\\
        \hspace{0.2cm} $T_{IST}$ [K] & internal temperature of the concrete at time $t$ \\
        \hspace{0.2cm} $T_{ref}$ [K] & reference temperature  \\
        \hspace{0.2cm} $b_e$ [K] & regression parameter \\
        \hline 
        \multicolumn{2}{|c|}{\textbf{other parameters}} \\
        \hline
        $C_{Crit}$ [kg/m\(^3\)] & critical chloride concentration value, above which corrosion initiates \\
        $C_{S, \Delta x}$ [kg/m\(^3\)] & chloride concentration at depth $\Delta x$ as a function of the chloride exposure at time $t$ \\
        $\Delta x$ [m] & depth, where intermittent chloride exposure causes chloride concentrations to deviate from typical Fick's behavior. \\
        \hline
    \end{tabular}
    \caption{List of parameters and quantities for chloride diffusion modeling using Gehlen's approach}
    \label{tab:parameters_a}
\end{table}

This formulation considers time-dependent aging effects and temperature dependence to provide a realistic model of chloride diffusion in concrete. Here, the semicolon denotes parameterization in $a$, $b_e$ and $D_{\text{t}}$. These parameters can be identified utilizing experimental data, as described in \ref{sec:method}\\

\noindent \textbf{Approach B - neural network:} Gehlen \cite{Gehlen} has introduced a time and temperature dependent ansatz for the diffusion coefficient based on domain-specific considerations. Alternatively, a purely data-driven ansatz based on a feed-forward neural network (FFNN), parameterized in weights and biases $\vectheta$ is considered:
\begin{equation}
    D^{NN}\left(T_{\text{IST}}(t), t; \, \vectheta\right) = \text{FFNN}\left(T_{\text{IST}}(t), t; \, \vectheta\right)
\end{equation}

\section{Parameter identification and sensitivity analysis} \label{sec:method}
\begin{figure}[h!]
    \centering
    \includegraphics[width=0.7\linewidth]{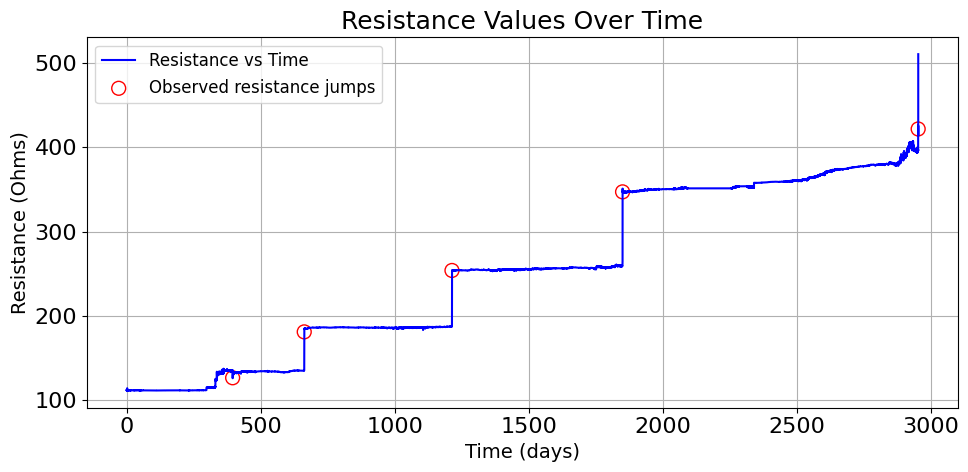}
    \caption{Resistance jumps over time period}
    \label{fig:R-jumps}
\end{figure}

We consider measurement data from the Concerto research bridge which is collected using built-in wire sensors. The sensor wires measure the effective resistance at specific timestamps which, in turn, depends on whether the wires are still intact or corroded. From this raw data, jumps in resistances can be identified, indicating corrosion. From the given data, there are five jumps in resistance as seen in Figure \ref{fig:R-jumps}, see also Table~\ref{tab:depth_time}. While five distinct resistance jumps were observed, the first identified point was excluded from further analysis due to inconsistencies in the data. This point was found to exhibit a high degree of noise, making it difficult to reliably determine the exact time of corrosion initiation. Including this point would introduce high uncertainties and potentially skew the optimization process, leading to less reliable results. Therefore, the focus of the optimization was restricted to the remaining four data points, ensuring a more accurate and robust result.

\begin{figure}[ht]
    \centering
    \includegraphics[width=0.99\linewidth]{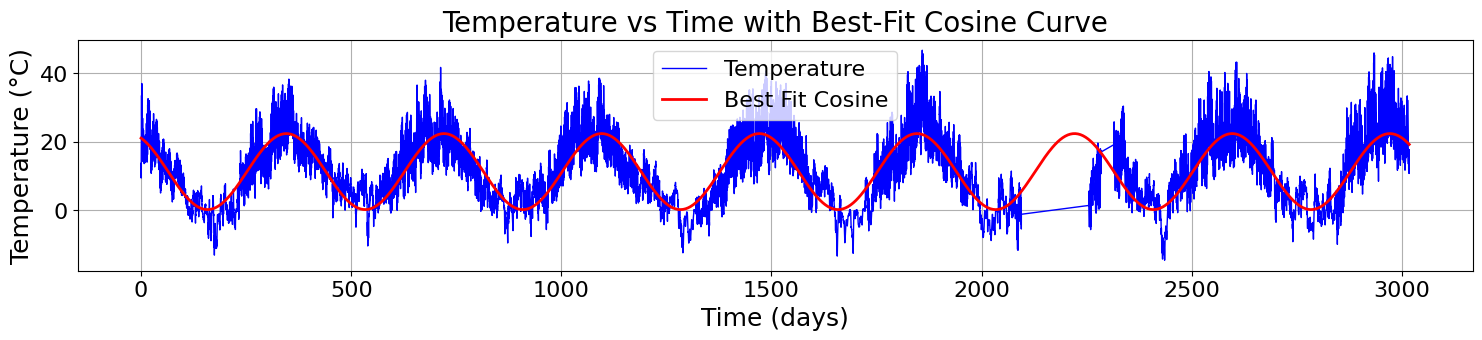}
    \caption{Temperature vs. time with best-fit cosine curve}
    \label{fig:temperature_vs_time}
\end{figure}  

The internal temperature of the concrete $T_{\text{IST}}$ is modeled as a function of time $t$ based on the data collected from the temperature sensor. Based on the data, we choose a cosine ansatz to describe the temperature profile over time and determine the unknown parameters by regressing the approach on the data. The regression results in the following analytical equation:
\begin{equation}\label{eq:T-IST}
    T_{\text{IST}}(t) = 11.10 \cdot \cos{2\pi \cdot \frac{t + 2,542,453.44}{32,407,303.30}} + 284.39
\end{equation}
This equation, shown in Figure \ref{fig:temperature_vs_time}, represents the variation of internal temperature over time and is used as an input for the subsequent optimization process. 

Further we make the following assumption: 
\begin{enumerate}
    \item The jumps in resistance shown in Figure \ref{fig:R-jumps} indicate that the iron wire has completely corroded. 
    \item Twenty-four days prior to the failure of the wire, the wire had just started corroding.\cite{Holst.2010} Considering the cross section area of the wires to be $0.065~mm$, the concentration of chloride ions was 0.6 M\% related to the cement mass at this time \cite{Gehlen}. As the Concerto bridge has a total of 270~kg/m³ of cement, a concentration of chloride ions of 0.6~M\% related to the cement mass is equivalent to 1.62~kg/m³. 
\end{enumerate}
\begin{table}[ht]
    \centering
    \begin{tabular}{|c|c|c|c|c|c|}
        \hline
        $x$~[m] & $t$~[s] (24 days prior failure) & Days & $T_{\text{IST}}$~[$^{\circ}$C] & date & time \\
        \hline
        0.01 & 32,110,680 & 372 & 23.0151 & 2006-08-09 & 13:32:35 \\
        \hline
        0.015 & 55,194,877 & 639 & 23.0151 & 2007-05-03 & 17:49:12  \\
        \hline
        0.02 & 102,827,181 & 1190 & 8.68677 & 2008-11-05 & 01:00:56 \\
        \hline
        0.025 & 157,826,993 & 1827 & 32.9442 & 2010-08-03 & 15:14:28 \\
        \hline
        0.03 & 253,131,439 & 2930 & 27.6982 & 2013-08-10 & 16:41:53  \\
        \hline
    \end{tabular}
    \caption{Depth of wires and corresponding time values with ambient temperature. The first data point was not included in the parameter estimation since it was found to exhibit a high degree of uncertainty.}
    \label{tab:depth_time}
\end{table}

In the following, we set the hyperparameters introduced in Table~\ref{tab:parameters_a} to the values listed in Table~\ref{tab:set_hyperparameters}. Given these hyperparameters and the observed data defined in Table~\ref{tab:depth_time}, the trainable parameters for the two concurrent approaches can be identified as outlined in the following. \\
\begin{table}[h!]
    \centering
    \begin{tabular}{|l|l|}
    \hline
    \textbf{hyperparameter} & \textbf{value} \\ \hline
     $\Delta x$ &  0~m \\ \hline
     $C_{S, \Delta x}$ &  18.19~kg/m\(^3\) (see Appendix~\ref{sec: Csdx calculations}) \\ \hline
     $t_0$ & 2.419 × $10^6$~s (28 days) \cite{Gehlen} \\ \hline
     $C_{Crit}$ & 0.6 \text{M\% }(\text{range:} 0.2 \text{to} 2 \text{M\%})
 \\ \hline
     $T_{\text{ref}}$ & 293.15~K \\ \hline
    \end{tabular}
    \caption{Hyperparameter values}
    \label{tab:set_hyperparameters}
\end{table}

\subsection{Approach A - Gehlen}
\label{Approach-A}
Here, the task is to identify the unknown parameters $a, b_e, D_{\text{t}}$ in Equation \eqref{eq:d_eff}. We can further simplify the one-dimensional analytical solution \eqref{eq:analytical_a} by isolating the error function
\begin{equation}
\begin{aligned}
    C(x, t) &= C_{S, \Delta x} \cdot \left[ 1 - \text{erf} \left( \frac{x - \Delta x}{2 \cdot \sqrt{D_{\text{Eff,C}}(t) \cdot t}} \right) \right] \\
   \Leftrightarrow \text{erf} \left( \frac{x - \Delta x}{2 \cdot \sqrt{D_{\text{Eff,C}}(t) \cdot t}} \right) &= 1 - \frac{C(x, t)}{C_{S, \Delta x}}. \\
\end{aligned}
\end{equation}
We then use the inverse error function ($\operatorname{erf}^{-1}$) to isolate the argument
\begin{equation}
\frac{x - \Delta x}{2 \cdot \sqrt{D_{\text{Eff,C}}(t) \cdot t}} = \text{erf}^{-1}\left(1 - \frac{C(x, t)}{C_{S, \Delta x}}\right).
\end{equation}
When substituting  $D_{\text{Eff,C}}(t)$ by the expression in \eqref{eq:d_eff} and rearranging, we obtain
\begin{equation}
\begin{aligned}
    x &= \Delta x + \text{erf}^{-1}\left(1 - \frac{C(x, t)}{C_{S, \Delta x}}\right) \cdot 2 \cdot \sqrt{D_{\text{Eff,C}}(t) \cdot t} \\
    &= \Delta x + \text{erf}^{-1}\left(1 - \frac{C(x, t)}{C_{S, \Delta x}}\right) \cdot 2 \cdot \sqrt{e^{\left(b_e \cdot \left( \frac{1}{T_{\text{ref}}} - \frac{1}{T(t)}\right)\right)} \cdot D_{\text{t}} \cdot \left(\frac{t_0}{t}\right)^a \cdot t}.
\end{aligned}
\label{eqn:simplified_x}
\end{equation}
Expression \ref{eqn:simplified_x} can be further used to identify parameters $a, b_e, {D_{\text{t}}}$ as an optimization problem of the form
\begin{equation}\label{eq:optimization}
\begin{aligned}
\left\{a, b_e,{D_{\text{t}}}\right\}^* &= \operatorname{arg\,min}_{a,b_e,{D_{\text{t}}}} \frac{1}{n}\sum_{i=0}^{n} \left[ x_i - \Delta x - \text{erf}^{-1}\left(1 - \frac{C_{Crit}}{C_{S, \Delta x}}\right) \cdot ...  \right. \\
&\quad \left. ... 2 \cdot \sqrt{e^{\left(b_e \cdot \left( \frac{1}{T_{\text{ref}}} - \frac{1}{T_i}\right)\right)} \cdot D_{\text{t}} \cdot \left(\frac{t_0}{t_i}\right)^a \cdot t_i} \right]^2,
\end{aligned}
\end{equation}
The aim of this optimization problem is to find the optimal parameters $\left\{a, b_e,{D_{\text{t}}}\right\}^*$ such that the the mean squared error (MSE) between the depths $x$ according to \ref{eqn:simplified_x} and the depths $\hat{x}_i$ of the measurement points listed in Table~\ref{tab:depth_time} is minimized. The parameters to be sought then enter the optimization problem through the depth $x$ calculated according to \ref{eqn:simplified_x}. We further use the hyperparameters summarized in Table~\ref{tab:set_hyperparameters}.

The problem is solved using the \href{https://github.com/bayesian-optimization/BayesianOptimization}{\texttt{bayes\_opt}} package \cite{Bayesian_Optimisation_Package} to perform Bayesian optimization on the parameters \(a\), \(D_{\text{t}}\), and \(b_e\) as described in more detail in Algorithm \ref{alg:1}. Before solving the optimization problem using real data, clean data was used to perform a sanity check as briefly discussed in Appendix \ref{sec:sanity_check}

\begin{algorithm}[h!]
\caption{Bayesian Optimization for chloride migration parameters}

\begin{algorithmic}[1]

\State \textbf{Input:} Data; bounds for parameters $a$, $D_{t}$, and $b_e$

\State \textbf{Objective function:} Negative MSE between predicted depth $\hat{x}$ and expected depth $x$:
\[
f(x) = -\frac{1}{N} \sum_{i=1}^{N} (\hat{x}_i - x_i)^2
\]

\State \textbf{Output:} Optimized parameters $a^{\text{opt}}$, $D_{t}^{\text{opt}}$, and $b_{e}^{\text{opt}}$

\State \textbf{Initialization:}
\State Choose $n$ initial random points $\alpha_1, \alpha_2, \dots, \alpha_n$ with $\alpha = \{a, D_{t}, b_e\}$ within respective bounds for $a$, $D_{t}$, and $b_e$.
\State Evaluate the objective function $f(x)$ at these points and store initial loss values.

\Repeat
\State Train a Gaussian Process (GP) as a surrogate model for the loss function based on stored parameter evaluations.
\State Compute the Expected Improvement (EI) acquisition function:
\[
    \text{EI}(x) = \mathbb{E}[\max(0, f(x) - f(x^+))]
\]
\State Select the next set of parameters such that the EI function is maximized.
where \( x^+ \) represents the location of the current best evaluation of the objective function, guiding the search to improve upon the best-known solution.

\State Evaluate the objective function $f(x)$ at the new parameters and update the surrogate model.
\State Update the best observed values of $a$, $D_{t}$, and $b_e$ if the loss improves.
\Until{Convergence or stopping criterion}

\State \textbf{Stopping Criterion:} Stop the optimization if the difference in the objective function values over the last four iterations is less than a set tolerance. Additionally, the acquisition function's behavior is monitored in the parameter space, as seen in Figures \ref{fig:termination_criteria_a} and \ref{fig:termination_criteria_b}

\end{algorithmic}
\label{alg:1}
\end{algorithm}

\subsubsection*{Dimensionality reduction proposal}
From the results of the sensitivity analysis briefly discussed in Appendix \ref{sec:Uncertainty_propagation}, it becomes evident that the parameter $b_e$ contributes minimally to the overall uncertainty in the model output. As shown by the first-order Sobol indices (Figure \ref{fig:sobol_1}), the sensitivity of $b_e$ is only 0.5\%, compared to the much larger contributions of $a$ and $D_{\text{t}}$, which account for 64.3\% and 35.1\% of the uncertainty, respectively. This trend is further confirmed by the total Sobol indices (Figure \ref{fig:sobol_2}), where the total contribution of $b_e$ is negligible.

Given this low sensitivity, a potential strategy to enhance the efficiency of the model is to fix the value of $b_e$ during future optimization processes. By doing so, we can reduce the dimensionality of the parameter space, which could lead to faster convergence of the optimization algorithm and reduce computational overhead. Additionally, fixing $b_e$ may help streamline the analysis by focusing on the more impactful parameters, $a$ and $D_{\text{t}}$.

It is important to note that this proposal is currently a theoretical suggestion and has not been implemented in the current model. Future work could explore the impact of fixing $b_e$ to assess the trade-offs between model accuracy and computational efficiency. Given the limited sensitivity of $b_e$, it is likely that fixing this parameter would not significantly affect the model’s performance, while potentially improving the overall efficiency of the optimization process.

\subsection{Approach B - Neural network}
\label{Approach-B}
In this data driven approach, a neural network is employed to predict the effective diffusion coefficient, which is crucial for modeling chloride migration in cement-based materials. The loss function is similar to the one used in Bayesian optimization to minimize the error in predicting the depth of chloride migration at different times. In this approach, however, the effective diffusion coefficient is not replaced by Gehlen's model but approximated by a neural  network. The network is designed to take time $t$ and temperature $T(t)$ as inputs. The network architecture consists of fully connected layers with non-linear activation functions to capture the underlying patterns in the data. The optimization process is performed using the Adam optimizer, with the training loop terminating once the model achieves a convergence criterion of \( 10^{-7} \). The algorithm is detailed in Algorithm \ref{alg:2}. For an in-depth introduction to artificial neural networks and deep learning, the reader is referred to \cite{goodfellow_deepLearning_2016}.

\begin{algorithm}[h!]
\caption{Neural network for predicting effective diffusion coefficient}
\begin{algorithmic}[1]

\State \textbf{Input:} Time $t$; internal temperature of concrete $T_{\text{IST}}(t)$

\State \textbf{Output:} Effective diffusion coefficient $D_{\text{Eff,C}}$ as a function of time $t$ and temperature $T_{\text{IST}}(t)$

\State \textbf{Step 1: Normalization}

\State Normalize the input features (time $t$ and temperature $T_{\text{IST}}(t)$(approximated temperature using the regression model in \ref{eq:T-IST})) using joint normalization to ensure stability during training, as $T_{\text{IST}}(t)$ is a function of $t$. 

Joint normalization scales both $t$ and $T_{\text{IST}}(t)$ together, taking into account their dependency. This process computes a common mean and standard deviation across both features and applies these values to normalize them. By aligning their scales, joint normalization reduces variance in the input space and helps the model learn more effectively, as it treats correlated inputs consistently.

\State \textbf{Step 2: Define Neural Network}
\State Approximate the effective diffusion coefficient $D_{\text{Eff,C}}$ by a feedforward neural network $NN$ with two hidden layers of 10 neurons each and ReLU activation functions in the hidden layers:
\[
D_{\text{Eff,C}}(t, T_{\text{IST}}(t); \mathbf{\theta}) \approx NN(t, T_{\text{IST}}(t); \mathbf{\theta}),
\]
% \[
% \mathbf{z}_1 = \text{ReLU}(\mathbf{W}_1 [t, T_{\text{IST}}(t)]^T + \mathbf{b}_1),
% \]
% \[
% \mathbf{z}_2 = \text{ReLU}(\mathbf{W}_2 \mathbf{z}_1 + \mathbf{b}_2),
% \]
% \[
% \hat{y} = \mathbf{W}_3 \mathbf{z}_2 + \mathbf{b}_3,
% \]
where $\mathbf{\theta}$ are the trainable parameters of the neural network. In order to improving numerical stability, in the output layer, a logarithmic scaling for $D_{\text{Eff,C}}$ is used as the values are very small (around $10^{-13}$).

\State \textbf{Step 3: Loss Function}
\State Define the loss function:
\[
f(\mathbf{x}, \mathbf{\theta}) = \frac{1}{N} \sum_{i=1}^{N} (\hat{x}_i(\mathbf{\theta}) - x_i)^2,
\]
where $\hat{x}(\mathbf{\theta})$ is the depth predicted using the rearranged form of the chloride migration equation \ref{eq:analytical_a}:
\[
\hat{x}(\mathbf{\theta}) = \Delta x + \left( 2 \sqrt{NN(t, T_{\text{IST}}(t); \mathbf{\theta}) \cdot t} \cdot \text{erf}^{-1} \left(1 - \frac{(C(x, t) = C_{\text{crit}})}{C_{S, \Delta x}} \right) \right)
\]
Note that $D_{\text{Eff,C}}$ is replaced by the neural network according to step 2. The parameters $C_{\text{crit}}$, $C_{S, \Delta x}$, and $\Delta x$ are constants used only in the loss function.

\State \textbf{Step 4: Optimization}
\State The network's parameters $\mathbf{\theta}$ are optimized using the Adam optimizer:
\[
    \vecW^*,\vecb^* = \arg\min_{\vecW, \vecb} \frac{1}{N} \sum_{i=1}^{N} \left( x_{i} - \left( \Delta x + 2 \sqrt{D_{\text{Eff,C}}^{NN}(T(t_i), t_i; \vecW, \vecb) \cdot t_i} \cdot \text{erf}^{-1} \left( 1 - \frac{C_{\text{crit}}}{C_{S, \Delta x}} \right) \right) \right)^2
\]

\State \textbf{Step 5: Stopping Criterion}
\State Terminate training when the difference in loss values between successive iterations is below a tolerance of $10^{-7}$, indicating convergence.

\end{algorithmic}
\label{alg:2}
\end{algorithm}

\section{Diffusion coefficient from standard tests}

In October 2007, the iBMB collected two drilling dust samples from the area around the wire sensor and determined the chloride content. Sampling was carried out in 5 mm increments to a depth of 60 mm. The mean value from the two drilling dust samples per layer was determined and related to the cement content of 270 kg/m³ (ratio of concrete to cement of $k = 8.89$).

In September 2024, Lowke Schiessl Ingenieure GmbH carried out another sampling of drilling dust from the Concerto Bridge. Sampling was carried out at three points in the area around the wire sensor. A drill with a diameter of 18 mm was used for the drill holes. The drilling dust was collected in layers of 10 mm using a suction device. The drilling depth was 60 mm. Two neighbouring boreholes were drilled per sampling point. The drilling dust from each layer from both boreholes was mixed and further processed as one sample. To analyse the drilling dust samples, the chloride content of the samples was also related to the cement content.

\begin{figure}[h!]
    \centering
    \includegraphics[width=0.8\textwidth]{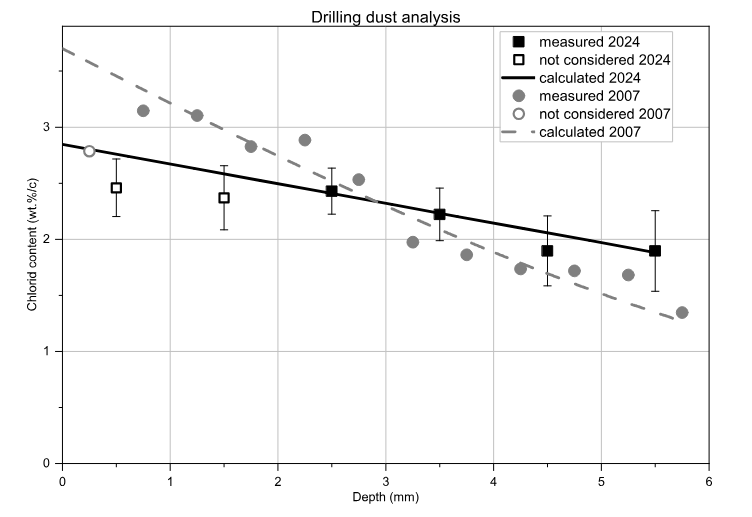}
    \caption{Measured chloride content from the drilling dust analysis or from the regression analysis as a function of depth}
    \label{fig:chlorid_content_1}
\end{figure}

Figure~\ref{fig:chlorid_content_1} shows the chloride profiles of the drilling dust samples from 2007 and 2024. These are the mean values of the sampling points. The standard deviation for the chloride profile from 2024 is shown as an error bar.
The effective diffusion coefficient $D_{\text{Eff,C}}$ at the time of the respective drilling dust extraction is calculated according to Equation~\ref{eq:analytical_a}, by adjusting the calculated chloride concentration $C_S,\Delta x$ to the assumed chloride profile from the drilling dust analysis. A non-linear regression analysis using the least squares method was applied for this purpose \cite{DINEN12390-11}. The inherent chloride content of the concrete was neglected.

The chloride concentration on the concrete surface was assumed to be 3.70 wt.\% in 2007 and 2.84 wt.\% in 2024. By adjusting $C_S$ in each case, $D_{\text{Eff,C}}$ could be determined in such a way that the best possible agreement between the measured and calculated chloride profiles was achieved. In the regression analyses, the uppermost 5 mm (2007) and the uppermost 20 mm (2024) of the chloride profiles were not taken into account. This area deviates from Fick's 2nd law due to carbonation and chloride discharge as a result of diffusion and convection. The period prior to May 2006, when exposure began, was also not taken into account.

The effective diffusion coefficient $D_{\text{Eff,C}}$ at the time of $t = 509$ days (2007) is $0.48 \times 10^{-12}$ m²/s. The coefficient of determination of the regression was $R^2 = 0.944$. For the observation time in 2024 ($t = 6692$ days), the effective diffusion coefficient is $D_{\text{Eff,C}} = 0.12 \times 10^{-12}$ m²/s, with a coefficient of determination of $R^2 = 0.903$. According to \cite{DINEN12390-11}, an $R^2 \geq 0.950$ is required under standard test conditions. For tests on existing structures, this value may be deviated from due to transport processes that are not exclusively diffusion-dependent.

In 2005, concrete cubes with an edge length of 200 mm were produced as accompanying specimens during concreting and have since been stored next to the bridge. A Rapid Chloride Migration (RCM) test was carried out on two of these cubes to determine the chloride migration coefficient $D_{\text{RCM}}$. The carbonation depth of the cubes was 10 mm. This layer was removed by sawing off 15 mm from the top and bottom sides of the cubes. A drill core with a diameter of 100 mm was taken from each of the two cubes, which was then divided into three samples with a height of 50 mm each by means of two saw cuts. A total of six samples were stored underwater for one week before the test. The RCM test was carried out and analysed in accordance with the specifications of the BAW data sheet \cite{BundesanstaltfurWasserbauBAW.2019}. The result showed an average chloride migration coefficient of $17.6 \times 10^{-12}$ m²/s with a standard deviation of $2.0 \times 10^{-12}$ m²/s.

\section{Results and discussion}
This section presents the findings obtained from the two primary models considered in this study: The physics-based model by Gehlen (Approach A) \ref{Approach-A} and the neural network-based model (Approach B) \ref{Approach-B}. The results from both approaches are analyzed with a particular focus on the calibration process, parameter optimization, and overall performance in predicting chloride migration.

For Approach A, Bayesian optimization is utilized to identify key parameters in the diffusion model. The model is calibrated progressively by increasing the number of data points to refine the predictions. The results showcase the importance of accurate parameter estimation in improving model reliability.

In Approach B, a neural network is employed to predict the effective diffusion coefficient. This data-driven method highlights the potential of machine learning in capturing complex patterns in corrosion behavior. The performance of the neural network model is assessed and compared to the physics-based approach.

\subsection{Approach A - Gehlen}

In Approach A, we utilize the physics-based model by Gehlen to estimate chloride migration in concrete. This method is rooted in Fick’s second law of diffusion. Bayesian optimization is employed to calibrate key parameters of the model, specifically the parameters $a$, $D_{\text{t}}$, and $b_e$, as discussed earlier in subsection \ref{Approach-A} . The optimization operates within expert-defined bounds \cite{Gehlen, BundesanstaltfurWasserbauBAW.2019}, as shown in Equation \ref{eq:expert_param}, which inform the calibration process.
\begin{equation}\label{eq:expert_param}
    \begin{aligned}
        a &\in [0.1,\, 0.9], \\
        D_{\text{t}} &\in [1e^{-12},\, 30e^{-12}], \\
        b_e &\in [1000, \,5200].
    \end{aligned}
\end{equation}
The model calibration is performed progressively. Initially, only one data point is used, and subsequent data points are incorporated incrementally to refine the model’s accuracy. Table \ref{tab:calibration_data} presents the progression of calibrations, where the inclusion of more data points improves the precision of parameter estimation. This iterative process allows for a continuous refinement, leading to more reliable predictions for chloride diffusion.

\begin{figure}[H]
    \newcommand{\subfigureWidth}{0.49\linewidth}
    \centering

    \begin{subfigure}{\subfigureWidth}
        \centering
        \includegraphics[width=1\linewidth]{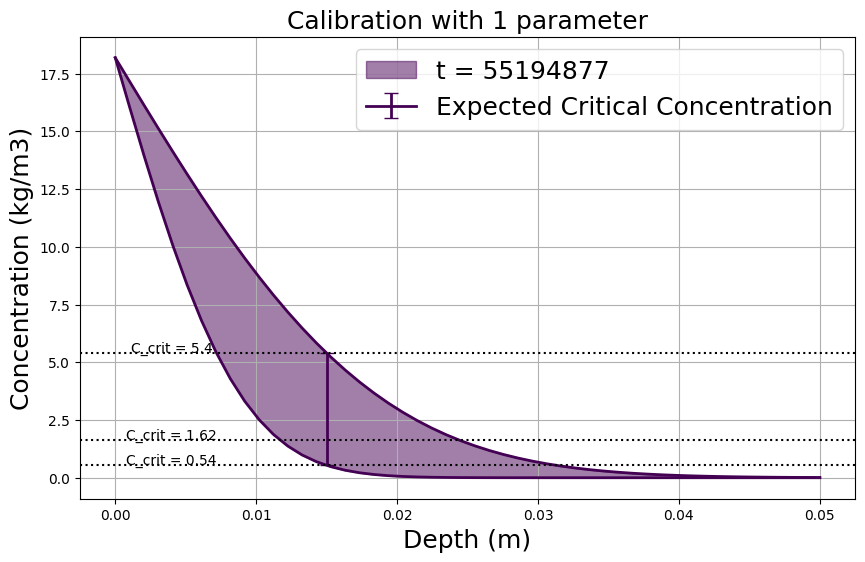}
        \caption{Calibration with one data pair}
        \label{fig:calibration_one_data_pair}
    \end{subfigure}
    \hfill
    \begin{subfigure}{\subfigureWidth}
        \centering
        \includegraphics[width=1\linewidth]{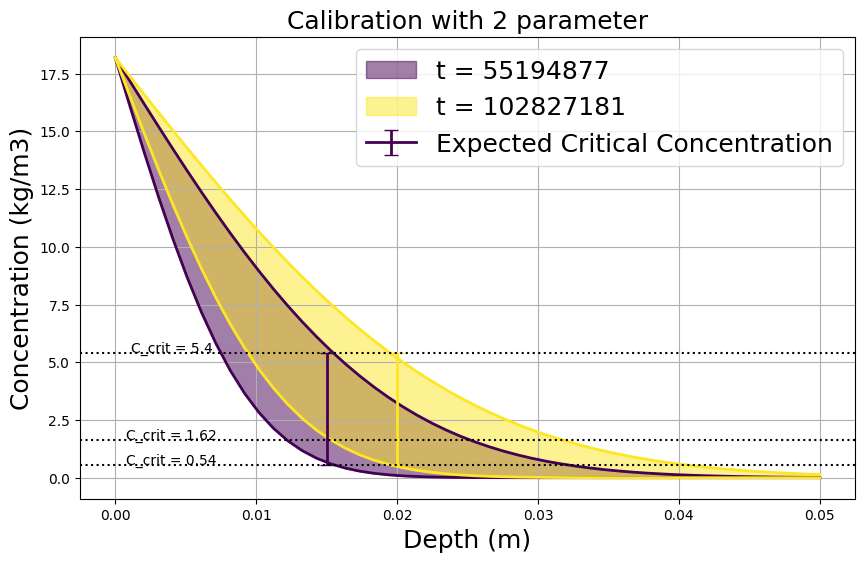}
        \caption{Calibration with two data pairs}
        \label{fig:calibration_two_data_pairs}
    \end{subfigure}

    \vskip\baselineskip

    \begin{subfigure}{\subfigureWidth}
        \centering
        \includegraphics[width=1\linewidth]{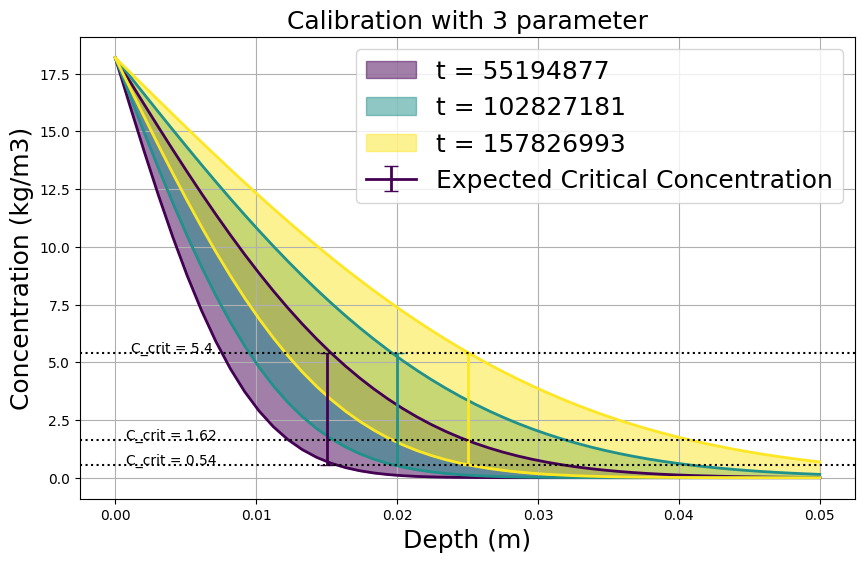}
        \caption{Calibration with three data pairs}
        \label{fig:calibration_three_data_pairs}
    \end{subfigure}
    \hfill
    \begin{subfigure}{\subfigureWidth}
        \centering
        \includegraphics[width=1\linewidth]{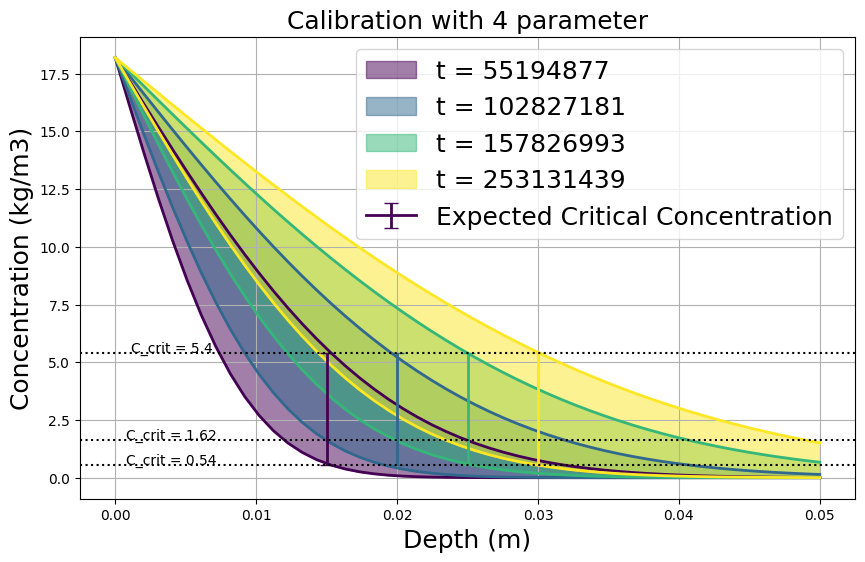}
        \caption{Calibration with four data pairs}
        \label{fig:calibration_four_data_pairs}
    \end{subfigure}

    \caption{Calibration plots: $0.54 < C_{\text{Crit}} < 5.4$ (approach A)}
    \label{fig:calibration_collage}
\end{figure}

To account for uncertainties in the critical chloride concentration, a range of values from $0.54$ to $5.4\, \text{kg/m}^3$ \cite{Gehlen} is considered during the optimization. As seen in Figure \ref{fig:calibration_collage}, the calibration results show that the predicted depth aligns increasingly well with the observed data as more calibration points are added. Figures \ref{fig:calibration_one_data_pair} to \ref{fig:calibration_four_data_pairs} demonstrate the improvement in predictions with one, two, three, and four calibration data points, respectively.

\begin{figure}[h!]
    \centering
    \begin{subfigure}[b]{0.49\textwidth}
        \centering
        \includegraphics[width=\linewidth]{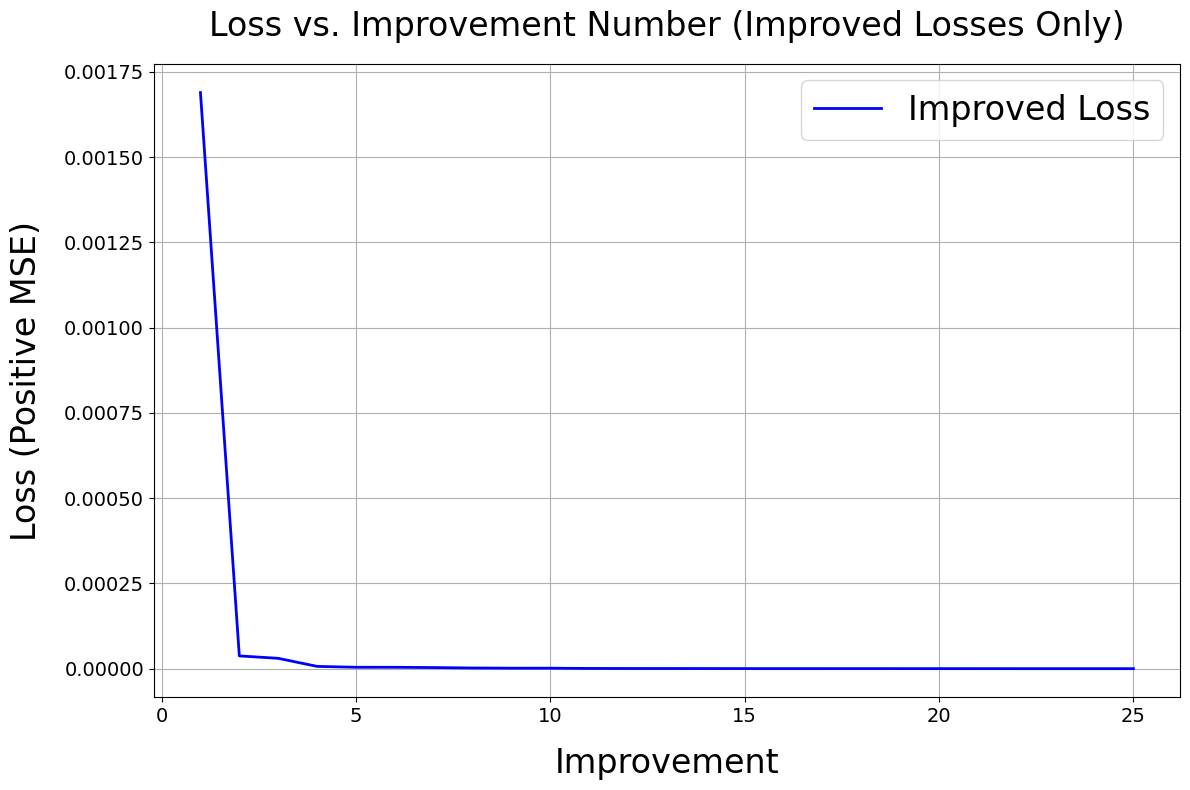}
        \caption{Convergence of Loss with Improvement Steps}
        \label{fig:termination_criteria_a}
    \end{subfigure}
    \hfill
    \begin{subfigure}[b]{0.45\textwidth}
        \centering
        \includegraphics[width=\linewidth]{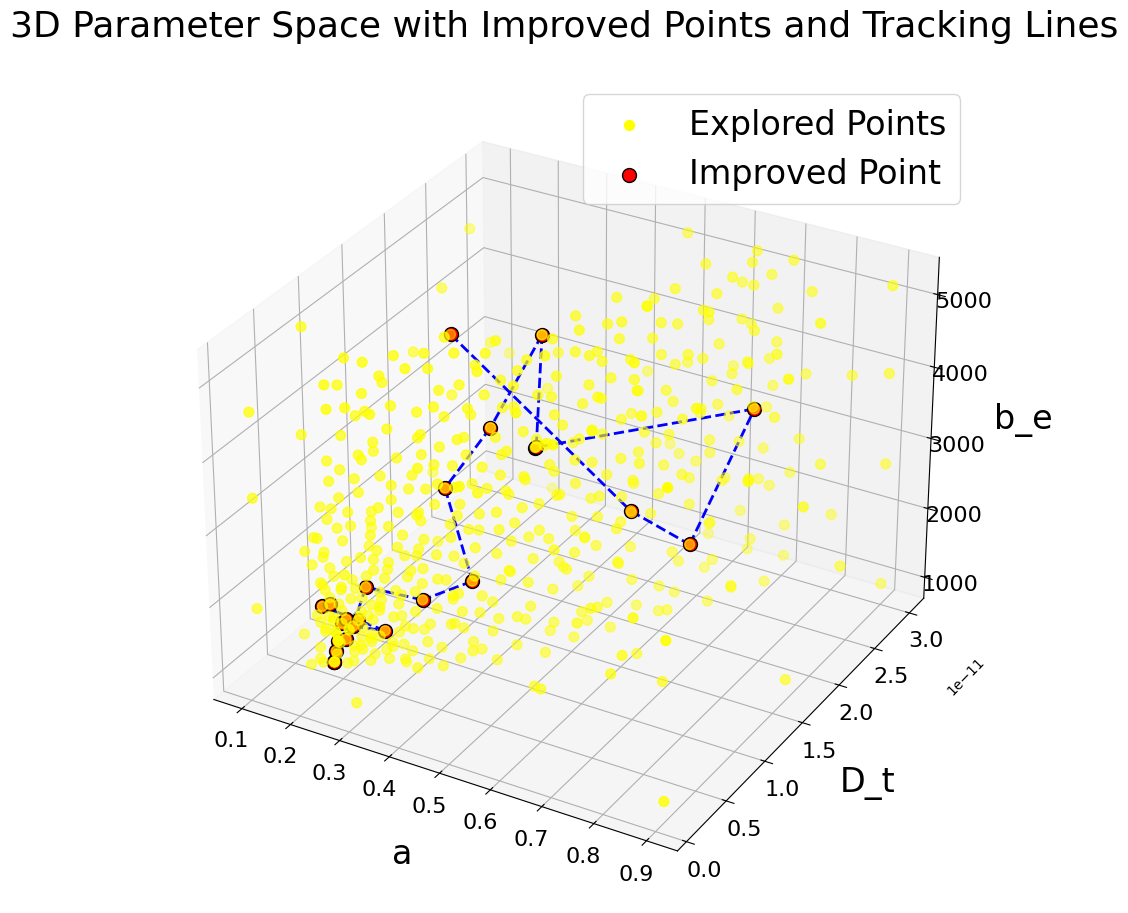}
        \caption{Behavior of Acquisition Function }
        \label{fig:termination_criteria_b}
    \end{subfigure}
    \caption{Termination Criteria}
    \label{fig:termination_criteria_combined}
\end{figure}

Additionally, Figure \ref{fig:termination_criteria_combined} shows the termination criteria achieved during the Bayesian optimization. Subfigure \ref{fig:termination_criteria_a} displays the convergence of the loss function, indicating that the optimization process successfully reduces the error with each iteration. Subfigure \ref{fig:termination_criteria_b} illustrates the behavior of the acquisition function, highlighting the convergence in parameter selection.

\begin{figure}[h!]
    \centering
    \begin{subfigure}[b]{0.49\textwidth}
        \centering
        \includegraphics[width=\linewidth]{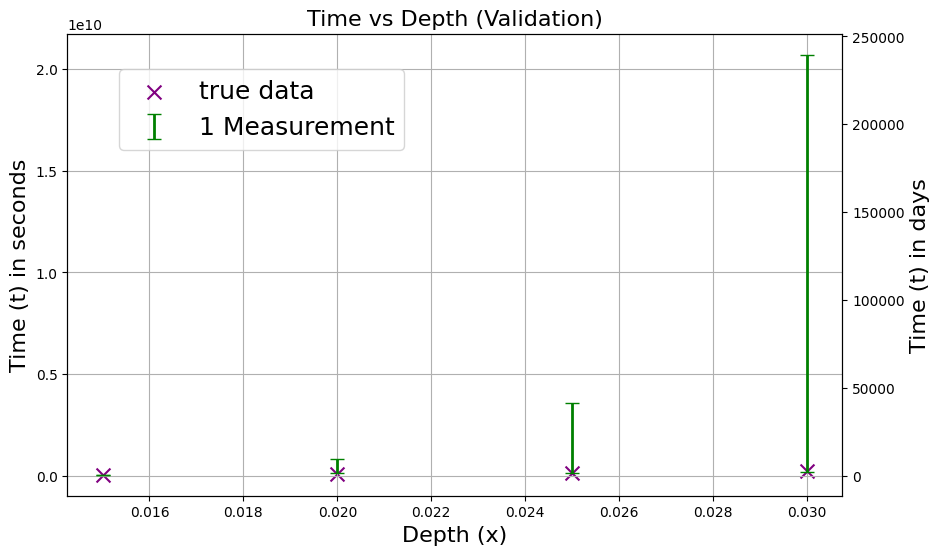}
    \end{subfigure}
    \hfill
    \begin{subfigure}[b]{0.49\textwidth}
        \centering
        \includegraphics[width=\linewidth]{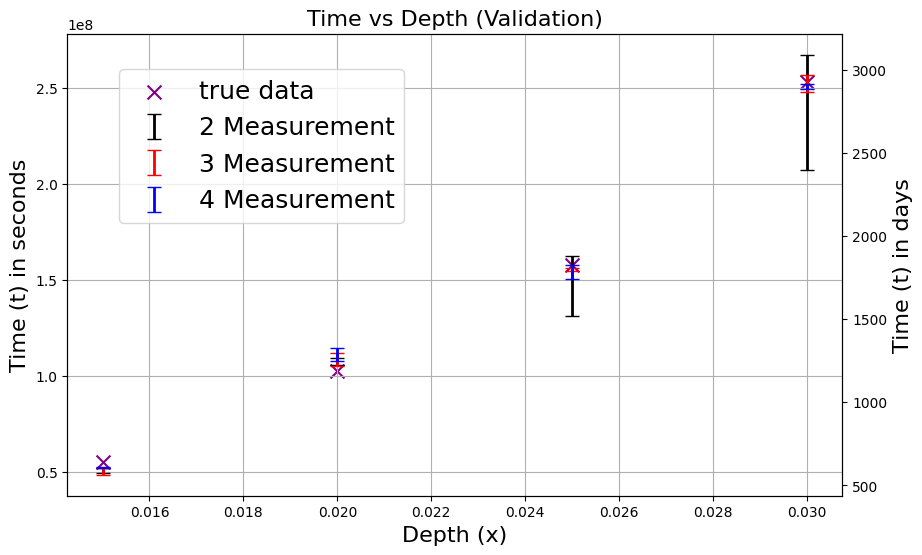}
    \end{subfigure}
    \caption{Calibration plot: Time vs. Depth validation (approach A)}
    \label{fig:calibration II}
\end{figure}

\begin{figure}[H]
    \centering
    \includegraphics[width=0.7\linewidth]{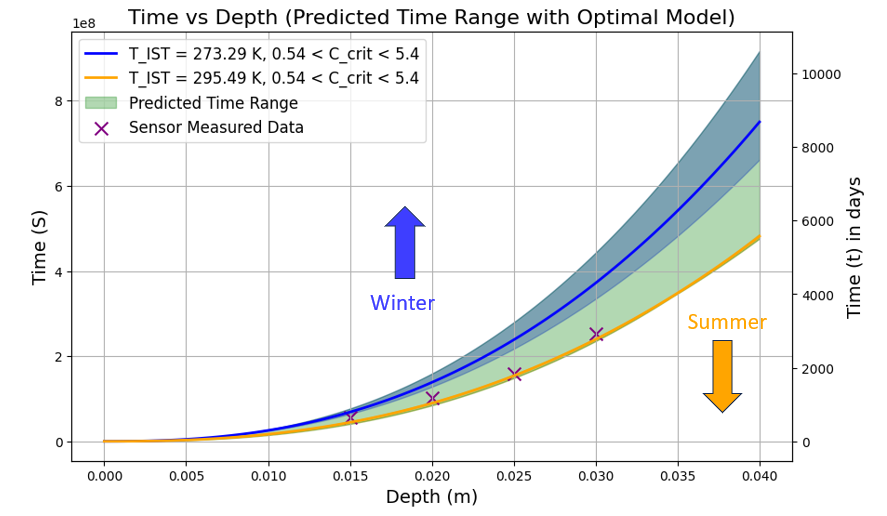}
    \caption{Calibration plot: Time vs. Depth prediction (approach A)}
    \label{fig:calibration II_prediction}
\end{figure}

Figure \ref{fig:calibration II} provides a key observation: as more data points are included, the accuracy of the predicted time-depth relationship improves significantly, showing that the model is capable of effectively simulating chloride ingress when appropriately calibrated. In Figure \ref{fig:calibration II_prediction}, the calibrated model is then used for prediction, taking into account uncertainties in both $C_{\text{Crit}}$ and temperature. From the temperature approximation in Equation \ref{eq:T-IST}, the maximum and minimum bounds are 295.49 K and 273.29 K, respectively.

Thus, Approach A shows that the progressive calibration of the model, paired with Bayesian optimization, leads to increasingly accurate predictions.

\begin{table}[ht]
\centering
\begin{tabular}{|c|c|c|c|c|c|}
\hline
$t~[\text{sec}]$ & $x_{\text{exp}}~[\text{m}]$ & $x_{\text{calc}}~[\text{m}] $ & $a~[-]$ & $D_{\text{t}}~[\text{m}^2/\text{s}]$ & $b_{e}[\text{K}]$ \\ \hline
\multicolumn{6}{|c|}{Calibration using one Data pair} \\ \hline
55194877 & 0.015 & 0.015 & \multirow{1}{*}{$\approx 0.78$} & \multirow{1}{*}{$\approx 8.69 \times 10^{-12}$} & \multirow{1}{*}{$\approx 1000.0$} \\ \hline
\multicolumn{6}{|c|}{Calibration using two data pairs} \\ \hline
55194877 & 0.015 & 0.015 & \multirow{2}{*}{$\approx 0.1$} & \multirow{2}{*}{$\approx 1.18 \times 10^{-12}$} & \multirow{2}{*}{$\approx 2006.0$} \\ \cline{1-3}
102827181 & 0.02 & 0.02 &  &  &  \\ \hline
\multicolumn{6}{|c|}{Calibration using three data pairs} \\ \hline
55194877 & 0.015 & 0.015 & \multirow{3}{*}{$\approx 0.21$} & \multirow{3}{*}{$\approx 1.63 \times 10^{-12}$} & \multirow{3}{*}{$\approx 1500.33$} \\ \cline{1-3}
102827181 & 0.02 & 0.02 &  &  &  \\ \cline{1-3}
157826993 & 0.025 & 0.025 &  &  &  \\ \hline
\multicolumn{6}{|c|}{Calibration using four data pairs} \\ \hline
55194877 & 0.015 & 0.015 & \multirow{4}{*}{$\approx 0.16$} & \multirow{4}{*}{$\approx 1.29 \times 10^{-12}$} & \multirow{4}{*}{$\approx 1010.70 $} \\ \cline{1-3}
102827181 & 0.02 & 0.02 &  &  &  \\ \cline{1-3}
157826993 & 0.025 & 0.025 &  &  &  \\ \cline{1-3}
253131439 & 0.03 & 0.03 &  &  &  \\ \hline
\end{tabular}
\caption{Data obtained from calibrations for $C_{\text{Crit}}$ = 1.62}
\label{tab:calibration_data}
\end{table}

\subsection{Approach B - Neural network}
In Approach B, a data-driven methodology is applied using a feed-forward neural network (FFNN) to predict the effective diffusion coefficient $D_{\text{Eff,C}}$as a function of time and temperature. Unlike Approach A, which relies on domain-specific models and physics-based principles, this neural network approach seeks to capture the underlying patterns in the data through machine learning, allowing for greater flexibility in dealing with complex scenarios that might not be well-explained by traditional models.

The neural network model is trained on available data, where the internal temperature of the concrete, $T_{\text{IST}}(t)$, is modeled analytically using a cosine function, as shown in Equation \ref{eq:T-IST}. This function serves as one of the primary inputs for the neural network, alongside time $t$.

The network is trained to minimize a custom loss function that measures the error in predicting the depth of chloride migration, with the rearranged form of the chloride migration equation being used for this calculation. Constants such as $C_{\text{Crit}}$, $C_{\text{S,}\Delta x}$, and $\Delta x$ are only used in the loss function and are not part of the trainable parameters. 

\begin{figure}[h]
    \newcommand{\subfigureWidth}{0.45\linewidth}
    \centering

    \begin{subfigure}{\subfigureWidth}
        \centering
        \includegraphics[width=1\linewidth]{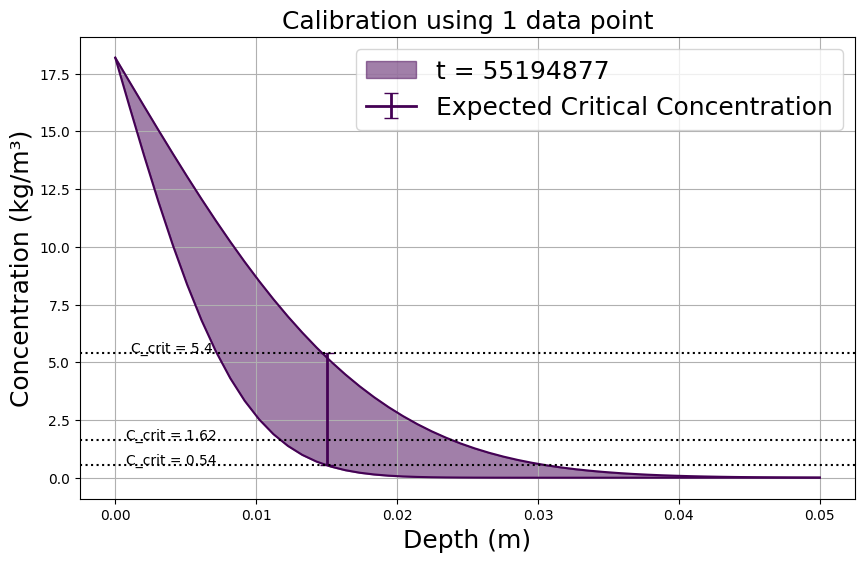}
        \caption{Calibration with one data pair}
        \label{fig:calibration_one_data_pair_NN}
    \end{subfigure}
    \hfill
    \begin{subfigure}{\subfigureWidth}
        \centering
        \includegraphics[width=1\linewidth]{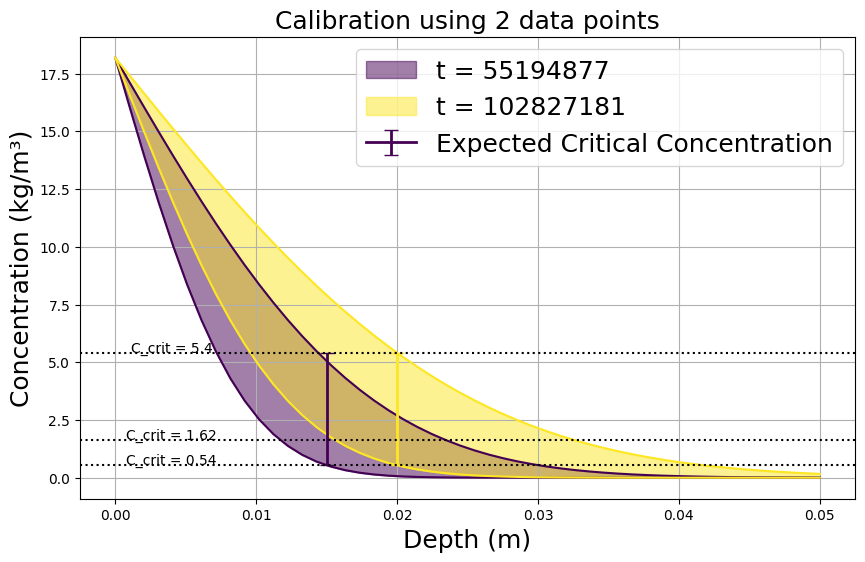}
        \caption{Calibration with two data pairs}
        \label{fig:calibration_two_data_pairs_NN}
    \end{subfigure}

    \vskip\baselineskip

    \begin{subfigure}{\subfigureWidth}
        \centering
        \includegraphics[width=1\linewidth]{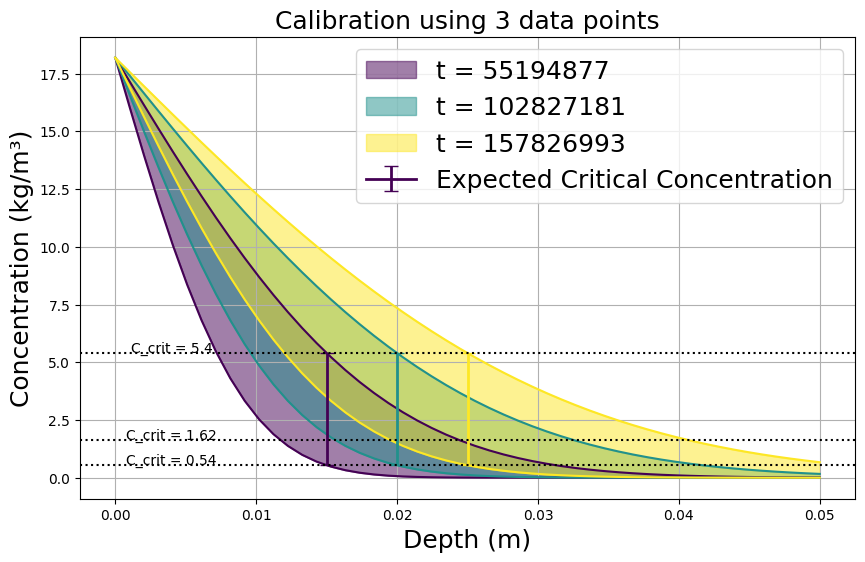}
        \caption{Calibration with three data pairs}
        \label{fig:calibration_three_data_pairs_NN}
    \end{subfigure}
    \hfill
    \begin{subfigure}{\subfigureWidth}
        \centering
        \includegraphics[width=1\linewidth]{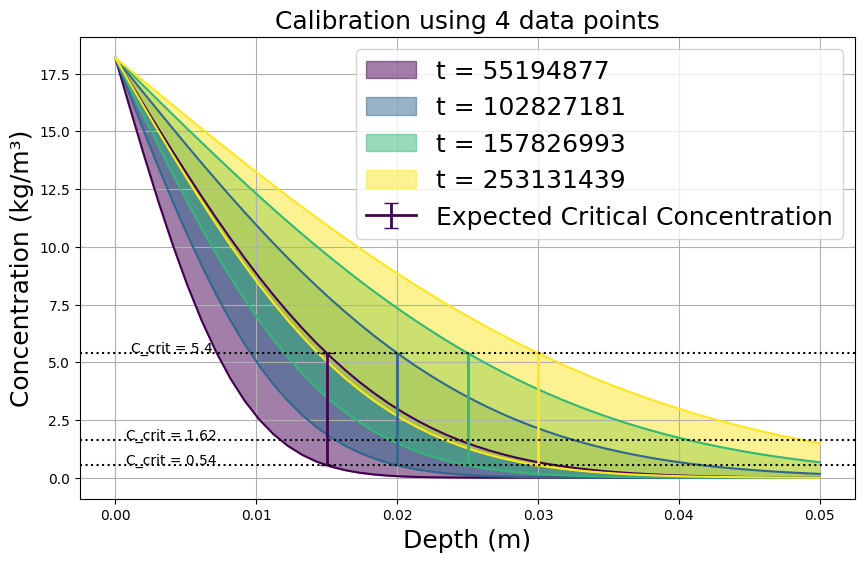}
        \caption{Calibration with four data pairs}
        \label{fig:calibration_four_data_pairs_NN}
    \end{subfigure}

    \caption{Calibration plots: $0.54 < C_{\text{Crit}} < 5.4$ (approach B)}
    \label{fig:calibration_collage_NN}
\end{figure}

Algorithm \ref{alg:2} states the complete process. The results of the neural network predictions for the effective diffusion coefficient and further calculations of chloride concentration versus depth are shown in Figure \ref{fig:calibration_collage_NN}. 
As in approach A, we take the uncertainty in $C_{\text{Crit}}$ into account by considering the values $0.54~kg/m^3$ and $5.4~kg/m^3$ in addition to the mean value of $1.62~kg/m^3$.

The neural network approach provides a valuable comparison to the physics-based method, demonstrating the potential of machine learning in handling more complex or unstructured data. The comparison between the two approaches is explored further in the next subsection. 

Furthermore, the authors would like to point out that they are aware that the neural network is very likely overfitting and not generalizing well, as it was only trained with four data points. Nevertheless, we have used the example to demonstrate the feasibility and showcase the basic process. It can be assumed that the generalization of the neural network increases as the amount of data increases.

\begin{figure}[h]
    \centering
    \includegraphics[width=0.6\linewidth]{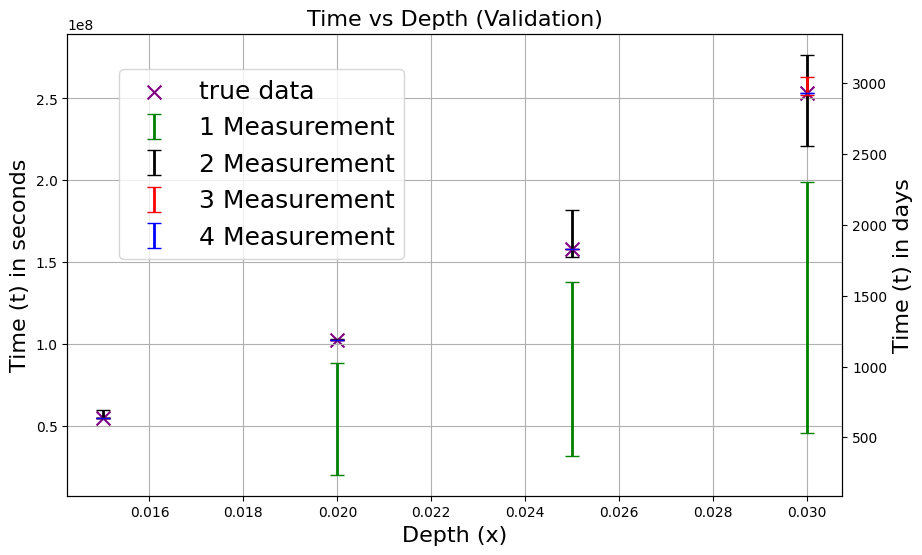}
    \caption{Calibration plot: Time vs. Depth (approach B)}
    \label{fig:calibration_II_NN}
\end{figure}

\begin{figure}[h]
    \centering
    \includegraphics[width=0.7\linewidth]{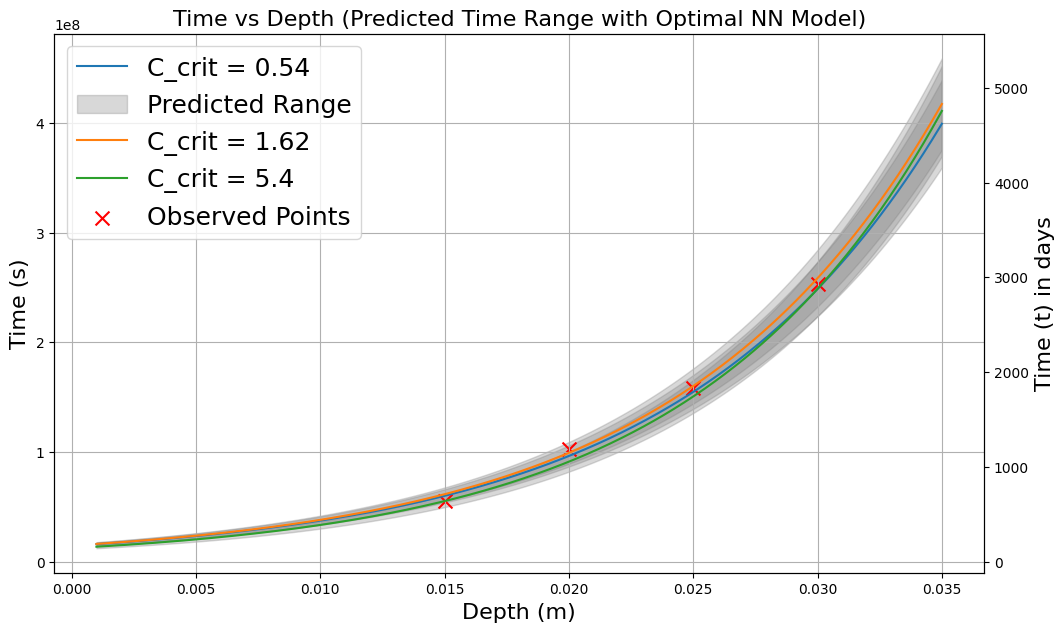}
    \caption{Calibration plot: Time vs. Depth prediction (approach B)}
    \label{fig:calibration II_prediction_NN}
\end{figure}

\subsection{Comparison} 

\begin{figure}[h]
    \centering
    \includegraphics[width=0.7\linewidth]{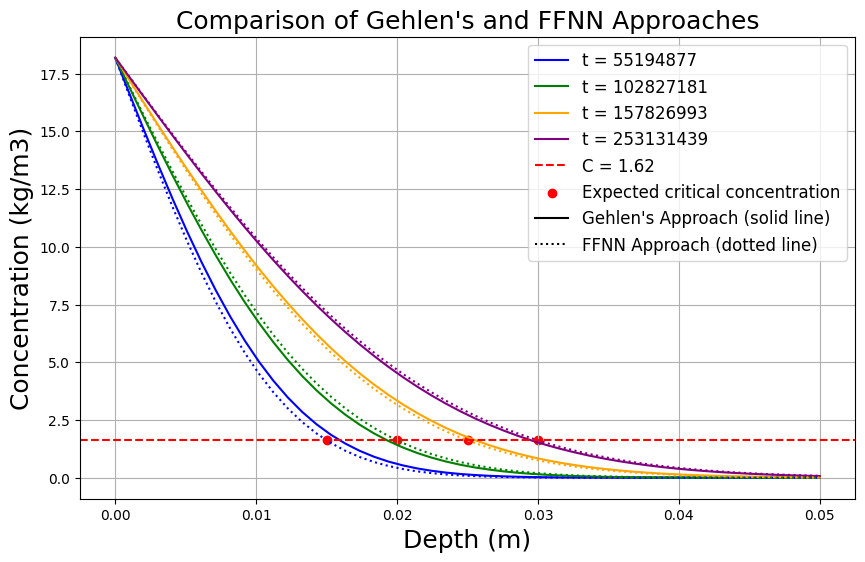}
    \caption{Comparison of chloride concentration predictions using physics-based and data-driven approaches for $C_{\text{Crit}}= 1.62$}
    \label{fig:comparison}
\end{figure}

In this subsection, two concurrent approaches—Approach A, based on Gehlen's physics-driven model, and Approach B, using a neural network—for predicting chloride migration related parameters are examined. Both approaches exhibit different characteristics and capabilities in handling the data and producing reliable predictions.

\begin{enumerate}
    \item \textbf{Accuracy:} Approach A relies on a domain-specific physical model, progressively calibrated to improve prediction accuracy. As more data points are introduced, the model's predictions for the chloride migration depth align closely with experimental results. Approach B, using a neural network, captures patterns in the data, allowing it to model more complex relationships that may not be explicitly encoded in traditional models. However, since it is purely data-driven, its prediction accuracy and generalization performance depends on the quality and quantity of the available training data.

    \item \textbf{Model Flexibility:} Approach A is constructed using predefined physical laws, which provide a structured framework for prediction. In contrast, Approach B is inherently flexible and can model complex, nonlinear behaviors within the data. This flexibility allows it to handle a wider variety of input conditions.

    \item \textbf{Ease of Implementation:} The process of setting up Approach A involves tuning parameters such as $a$, $D_{\text{t}}$, and $b_e$ using Bayesian optimization. This process requires detailed knowledge of the system and careful calibration with experimental data. Approach B, on the other hand, uses a predefined neural network architecture and relies more heavily on data availability. Once trained, the neural network can generalize to new data quickly without the need for additional recalibration.

    \item \textbf{Interpretability:} Approach A offers interpretability since it is derived from well-known physical principles, allowing the results to be validated against known behaviors of chloride diffusion. The parameters in Approach A directly correspond to physical quantities, making the results easy to understand and explain. In contrast, Approach B functions more as a data-driven "grey box", where the internal workings of the model may be less transparent, though it can still provide accurate predictions. The neural network may not be explicitly interpretable in terms of traditional physical parameters, in comparison to approach A.

    \item \textbf{Convergence and Performance:} Approach A uses Bayesian optimization for parameter calibration, which leads to steady convergence as demonstrated in the termination criteria (Figure \ref{fig:termination_criteria_combined}). Approach B is trained using gradient-based optimization techniques such as the Adam optimizer, which aim to minimize the error during training. The performance of both approaches depends on the calibration or training process, but each approach converges based on its respective methodology.

\begin{table}[h]
\centering
\begin{tabular}{|c|c|c|c|}
\hline
\textbf{($x_i$, $t_i$)} & \textbf{True $C_{\text{Crit}}$} & \textbf{Approach A Prediction} & \textbf{Approach B Prediction} \\ \hline
(0.015, 55194877)  & 1.62 & 1.9881 & 1.61981744 \\ \hline
(0.02, 102827181)  & 1.62 & 1.4169 & 1.62008777 \\ \hline
(0.025, 157826993) & 1.62 & 1.7524 & 1.61999942 \\ \hline
(0.03, 253131439)  & 1.62 & 1.5257 & 1.62000299 \\ \hline
\end{tabular}
\caption{Comparison of predictions from Approach A and Approach B for different values of $x_i$ and $t_i$.}
\label{tab:comparison}
\end{table}

\end{enumerate}
Both Approach A and Approach B provide valuable insights into chloride migration prediction, utilizing different strategies. Approach A, with its basis in physics, emphasizes structure and interpretability, while Approach B effectively uses the flexibility of neural networks to model complex data patterns. The selection of an approach depends on the specific requirements of the task, such as the availability of data, the need for interpretability, and the complexity of the system being modeled.

\subsection{Comparison with standard tests}

\begin{figure}[H]
    \centering
    \includegraphics[width=0.8\textwidth]{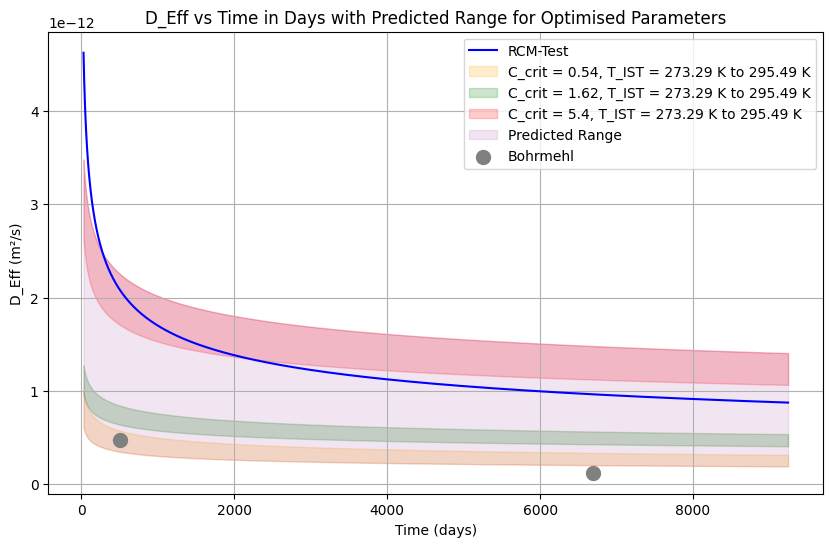}
    \caption{Comparison of $D_{\text{Eff,C}}$ from the RCM test, the wire sensor, and drilling dust samples over time}
    \label{fig:chlorid_content}
\end{figure}

The rapid chloride migration test in accordance with \cite{BundesanstaltfurWasserbauBAW.2019} at the drill core without chloride exposure provides an average chloride diffusion coefficient $D_{\text{RCM}}$ of $17.6 \times 10^{-12}~m^2/s$. This can be used in Equation \eqref{eq:d_eff} with the parameters $a = 0.3$, $k_{t} = 1.0$ and $b_{e} = 4800~K$ from \cite{Gehlen} to determine the effective diffusion coefficient $D_{\text{Eff,C}}(t)$, see blue curve in Figure \ref{fig:chlorid_content}.\\
The chloride content profiles created in 2007 and 2024 are used directly in Equation 2 to determine the parameter $D_{\text{Eff,C}}$ at time $t = 509$ days to $0.48 \times 10^{-12}~m^2/s$ (2007) and time $t = 6692$ days to $0.12 \times 10^{-12}~m^2/s$ (2024). These measured values are represented by grey dots in Figure \ref{fig:chlorid_content}.\\
The standard method according to \cite{BundesanstaltfurWasserbauBAW.2019} provides rather conservative values. The values for the diffusion coefficient determined on the basis of the chloride profiles indicate that the critical corrosion-inducing chloride content must be at the lower limit with  $C_{\text{crit}} = 0.54~kg/m^3$. This shows good agreement with the values from the parameter adjustment. Here it becomes clear that additional structural investigations in the form of concrete dust analyses can be used to help classify the model parameters identified on the basis of sensor data.

\section{Conclusion}

This study addresses the need for accurate predictions of chloride migration in reinforced concrete structures to mitigate the risks of chloride-induced corrosion. Two distinct modeling approaches for the diffusion coefficient were considered: Gehlen’s empirical and a neural network-based ansatz.

The two concurrent approaches mainly differ in the number of their parameters: Gehlen's model offers interpretability and aligns closely with established theory, making it well-suited for scenarios where transparency and validation against known behaviors are essential. Progressive calibration through Bayesian optimization further enhances its reliability. In contrast, the neural network model demonstrates flexibility and adaptability, effectively capturing complex patterns in chloride migration. However, its increased number of parameters makes it sensitive to overfitting, particularly with limited datasets, as demonstrated in this study.
Both approaches underscore the importance of accounting for uncertainties, such as variations in critical chloride content, to improve the robustness of predictions. %The findings have practical implications for proactive infrastructure monitoring, enabling timely interventions and reducing repair costs.

In addition to the current isotropic effective diffusion coefficient used here, future work will focus on developing an appropriate ansatz for an anisotropic effective diffusion coefficient. This extension will enhance predictive capabilities for corrosion by accounting for varying diffusion rates along spatial dimensions. Specifically, the diffusion coefficient \( D(t, x, T_{\text{IST}}(t)) \) can be modeled as a function of depth \( x \), along with time \( t \) and temperature \( T_{\text{IST}}(t) \).
\begin{comment}
The governing equation for this scenario will be derived from Fick’s second law in 1D with a variable diffusion coefficient, as shown below:
\begin{equation}
\frac{\partial C}{\partial t} = \frac{\partial}{\partial x} \left( D(t, x, T_{\text{IST}}(t)) \frac{\partial C}{\partial x} \right)
\end{equation}
where:
\begin{itemize}
    \item \( C = C(x,t) \) represents the concentration as a function of position \( x \) and time \( t \)
    \item \( D(t, x, T_{\text{IST}}(t)) \) is the diffusion coefficient, which varies over time \( t \), depth \( x \), and temperature \( T_{\text{IST}}(t) \), reflecting anisotropic behavior in the material.
\end{itemize}
\end{comment}
To optimize the parameters governing the anisotropic diffusion coefficient, the diffusion equation must be solved numerically. Finite Element Method or  Finite Difference Methods, as well as surrogate models based on Gaussian Processes or (physics-informed) neural networks can be applied to obtain numerical solutions for the governing equation.
%along with Bayesian Optimization, could be used to efficiently calibrate the anisotropic model by approximating the behavior of the diffusion process without requiring full-scale simulations for every iteration. 

\section*{Data availability}

All data and code for this project will be published at \url{zenodo.org} upon acceptance of this manuscript.

\bibliographystyle{ieeetr}
\bibliography{literature}

\appendix

\section{Calculations for \texorpdfstring{$C_{S,\Delta X}$}{C_{S,DeltaX}}}\label{sec: Csdx calculations}

The concentration of NaCl in the solution was determined based on a 3\% NaCl solution by mass, which means that there are 3 grams of NaCl in every 100 grams of solution. The molar mass of NaCl is approximately 58.44 g/mol.

Given the 3\% solution, we can calculate the molarity of NaCl. Assuming that the density of the solution is approximately 1 g/mL (similar to water for dilute solutions), 10 liters of solution corresponds to 10,000 grams. The mass of NaCl in this solution is calculated as follows:

\begin{equation}
\text{Mass of NaCl} = 0.03 \times 10,000 \, \text{g} = 300 \, \text{g}
\end{equation}

Next, the moles of NaCl in 300 grams are determined by dividing the mass of NaCl by its molar mass:

\begin{equation}
\text{Moles of NaCl} = \frac{300 \, \text{g}}{58.44 \, \text{g/mol}} \approx 5.13 \, \text{mol}
\end{equation}

The molarity of the NaCl solution, which is the number of moles per liter of solution, is then given by:

\begin{equation}
\text{Molarity of NaCl solution} = \frac{5.13 \, \text{mol}}{10 \, \text{L}} = 0.513 \, \text{M}
\end{equation}

Since each NaCl molecule dissociates into one Na\(^+\) ion and one Cl\(^-\) ion, the concentration of chloride ions in the solution is equal to the molarity of NaCl, which is 0.513 M.

To convert this chloride ion concentration into kilograms per cubic meter (kg/m\(^3\)), we first recognize that the molar mass of Cl\(^-\) is approximately 35.45 g/mol. The mass of chloride ions per liter of solution is calculated as:

\begin{equation}
\text{Mass of Cl}^- = 0.513 \, \text{mol/L} \times 35.45 \, \text{g/mol} \approx 18.19 \, \text{g/L}
\end{equation}

Finally, to convert grams per liter into kilograms per cubic meter, we observe that 1 g/L is equivalent to 1 kg/m\(^3\). Thus, the concentration of chloride ions is:

\begin{equation}
18.19 \, \text{g/L} = 18.19 \, \text{kg/m}^3
\end{equation}

This result gives the concentration of chloride ions in the solution in terms of mass per volume, which can be used in subsequent calculations for modeling chloride diffusion.

\section{Sensitivity Analysis}
\label{sec:Uncertainty_propagation}
Prior to running the Sobol sensitivity analysis on Gehlen's model \ref{eq:d_eff}, a sanity check was performed using a simple dummy model (as discussed in Appendix \ref{Sensitivity_sanity} ), defined as $f(X) = 2X_1^2 + X_2 + 3X_3^3$, to ensure the correctness of the methodology and implementation.  

The following description outlines the step-by-step procedure used to calculate the Sobol indices for Gehlen's model \ref{eq:d_eff}:\\
\subsection{Input}
\label{sobol_Input}
\begin{itemize}
    \item \textbf{Model function} \( f(\mathbf{X}) \): The function representing the system to analyze, defined as:
    \begin{equation}
    f(\mathbf{X}) = C_{S_{\Delta x}} \cdot \left( 1 - \text{erf} \left( \frac{x}{2 \sqrt{\exp{ b_e \left( \frac{1}{T_{\text{ref}}} - \frac{1}{T_{\text{IST}}} \right)} \cdot D_{\text{t}} \cdot \left( \frac{t_0}{t} \right)^a \cdot t}} \right) \right)
    \end{equation}
    where the hyperparameters $C_{S_{\Delta x}}$, $T_{\text{ref}}$ and $t_0$ are defined in Table~\ref{tab:set_hyperparameters}.

    \item \textbf{Input parameters} \( \mathbf{X} = \{b_e, D_{\text{t}}, a\} \):  
    The parameters are assumed to follow a uniform distribution within their respective ranges specified in equation~\ref{eq:expert_param}.
    
    \item \textbf{Data values for each run}:
    We run the sensitivity analysis for the four data points with the depths $x = [0.015, 0.02, 0.025, 0.03]$ and the corresponding time and ambient temperature listed in Table~\ref{tab:depth_time}.
    
    \item \textbf{Number of samples} \( N \): The number of Sobol samples (preferably a power of 2), here 8192 ($13^2$) samples are considered.
\end{itemize}

\subsection{Output}
\begin{itemize}
    \item \textbf{First-order Sobol indices} \( S_1 \): The individual contribution of each parameter \( X_i \) (i.e., \( b_e \), \( D_{\text{t}} \), and \( a \)) to the variance of the model output. This index measures the effect of each parameter independently, assuming no interaction with other parameters and is defined as
    \begin{equation}
    S_i = \frac{V_{X_i}(\mathbb{E}[Y | X_i])}{V(Y)}.
    \end{equation}
    Here \( V_{X_i}(\mathbb{E}[Y | X_i]) \) is the variance of the conditional expectation of \( Y \) given \( X_i \), representing the impact of \( X_i \) alone on \( Y \).
    
    \item \textbf{Total-order Sobol indices} \( S_T \): The total contribution of each parameter \( X_i \) to the output variance, including all possible interactions (permutations) of \( X_i \) with other parameters. This index captures both the independent effect of \( X_i \) and its interactions with other parameters (e.g., pairwise, triple-wise combinations). The total-order Sobol index \( S_{T_i} \) is calculated as:
    \begin{equation}
    S_{T_i} = 1 - \frac{V_{\sim i}(\mathbb{E}[Y | X_{\sim i}])}{V(Y)},
    \end{equation}
    where \( V_{\sim i}(\mathbb{E}[Y | X_{\sim i}]) \) is the variance of \( Y \) when all parameters except \( X_i \) are fixed. \( S_{T_i} \) thus represents the total impact of \( X_i \) on \( Y \), including its interaction with every subset of the other parameters.
\end{itemize}

\subsection{Algorithm}
\begin{enumerate}
    \item \textbf{Define the Problem}: \\
    Specify the input parameters (i.e., \( b_e \), \( D_{\text{t}} \), and \( a \)) and their respective distribution (refer Subsection \ref{sobol_Input}).

    \item \textbf{Generate Sobol Samples}:
    \begin{itemize}
        \item Sobol samples are quasi-random, low-discrepancy sequences that ensure good coverage of the input space with fewer samples than purely random sampling.\cite{stanford_montecarlo}
        \item For each input parameter, generate \( N \) samples and collect them in $\mathbf{X}$. The Sobol sampling algorithm ensures that the points are distributed uniformly, covering the range for each input parameter effectively.
    \end{itemize}

    \item \textbf{Evaluate the Model for Each Combination of \( x \), \( t \), and \( T_{\text{IST}} \)}: \\
    For each combination of \( x \), \( t \), and \( T_{\text{IST}} \), evaluate the model \( f(\mathbf{X}) \) for all Sobol samples to obtain the output \( Y \):
    \begin{equation}
        Y = f(\mathbf{X}).
    \end{equation}

\item \textbf{Perform Sobol Analysis}: \\
    For each combination \( j \), where \( j \) corresponds to a specific combination of depth \( x \), time \( t \), and temperature \( T_{\text{IST}} \) as listed in Table~\ref{tab:depth_time}, calculate:
        \begin{itemize}
            \item First-order Sobol indices \( S_1^{(j)} \) for \( b_e \), \( D_{\text{t}} \), and \( a \), representing the independent effect of each parameter for the \( j \)-th combination of data points.
            \item Total-order Sobol indices \( S_T^{(j)} \) for \( b_e \), \( D_{\text{t}} \), and \( a \), representing the total contribution of each parameter, including all interactions, for the \( j \)-th combination of data points.
        \end{itemize}

    \item \textbf{Validate Variance Decomposition}: \\
    For each combination \( j \), check if the sum of first-order Sobol indices \( S_1 \) approximates the total variance \( V(Y) \) by verifying:
    \begin{equation}
    \text{Variance Sum}^{(j)} = \sum_{i=1}^{k} S_1^{(j)}(X_i) \times V(Y^{(j)}) \approx V(Y^{(j)}),
    \end{equation}
    where:
    \begin{itemize}
        \item \( j \) refers to a specific combination of \( x \), \( t \), and \( T_{\text{IST}} \) as defined in Table~\ref{tab:depth_time}.
        \item \( X_i \) represents the \( i \)-th input parameter (\( b_e \), \( D_{\text{t}} \), or \( a \)).
        \item \( V(Y^{(j)}) \) is the variance of the model output \( Y \) for the \( j \)-th combination.
        \item \( S_1^{(j)}(X_i) \) is the first-order Sobol index for the \( i \)-th parameter and the \( j \)-th combination.
    \end{itemize}
    If the variance decomposition does not sum to the total variance \( V(Y^{(j)}) \) for any combination \( j \), print a warning indicating that the decomposition might be incorrect for that specific run.
    
    \item \textbf{Average Sobol Indices Across All Runs}: \\
    Compute the average first-order \( S_1 \) and total-order \( S_T \) indices across all combinations of \( x \), \( t \), and \( T_{\text{IST}} \) (indexed by \( j \)) for each input parameter (\( b_e \), \( D_{\text{t}} \), and \( a \))(indexed by \( i \)):
    \begin{equation}
    \text{Average } S_1(X_i) = \frac{1}{m} \sum_{j=1}^m S_1^{(j)}(X_i), \quad \text{Average } S_T(X_i) = \frac{1}{m} \sum_{j=1}^m S_T^{(j)}(X_i),      
    \end{equation}
    where:
    \begin{itemize}
        \item \( X_i \) represents the \( i \)-th input parameter (\( b_e \), \( D_{\text{t}} \), or \( a \)).
        \item \( m \) is the total number of combinations of \( x \), \( t \), and \( T_{\text{IST}} \) as defined in Table~\ref{tab:depth_time}.
        \item \( S_1^{(j)}(X_i) \) and \( S_T^{(j)}(X_i) \) are the first-order and total-order Sobol indices for the \( i \)-th parameter and the \( j \)-th combination of data points.
    \end{itemize}
    Finally, report the average indices for each parameter (\( b_e \), \( D_{\text{t}} \), and \( a \)) across all runs.

    \item \textbf{Plot Results}: \\
    Plot the averaged first-order Sobol indices \( S_1 \) and total-order Sobol indices \( S_T \) as bar charts.

    \item \textbf{Return Results}: \\
    Return a dictionary containing \( S_1 \) and \( S_T \) averaged over all runs.
\end{enumerate}

\begin{figure}[H]
    \centering
    \begin{subfigure}[b]{0.45\textwidth}
        \centering
        \includegraphics[width=\linewidth]{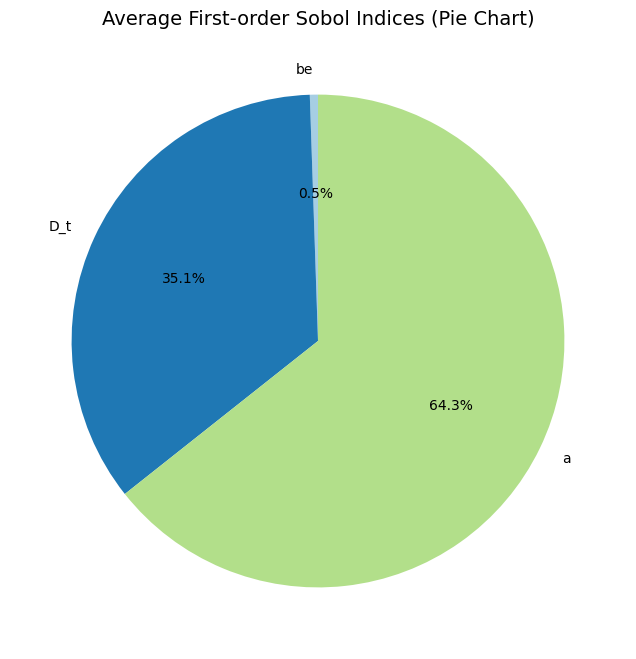}
        \caption{First-order Sobol indices for the input parameters.}
        \label{fig:sobol_1}
    \end{subfigure}
    \hfill
    \begin{subfigure}[b]{0.45\textwidth}
        \centering
        \includegraphics[width=\linewidth]{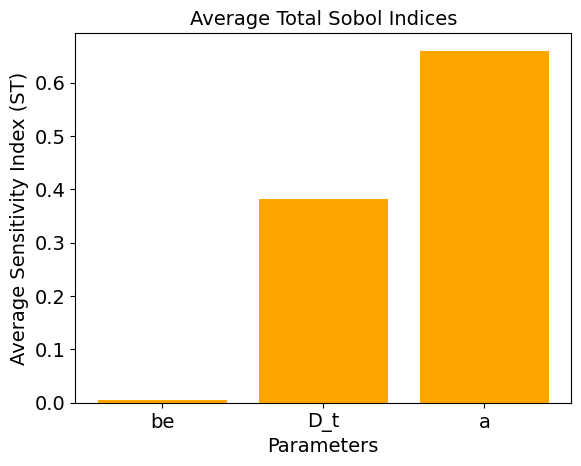}
        \caption{Total Sobol indices, showing overall contribution to variance. }
        \label{fig:sobol_2}
    \end{subfigure}
    \caption{Sobol indices}
    \label{fig:sobol_both}
\end{figure}

\noindent The Sobol sensitivity analysis revealed that the first-order as well as the total-order Sobol index for $a$ was the highest, followed by $D_{\text{t}}$, with the least contribution by $b_{\text{e}}$, confirming its dominant role in both main effects and interactions. Figures \ref{fig:sobol_1} and \ref{fig:sobol_2} illustrate the contribution of each input parameter to the output uncertainty.

\section{Sanity check with clean data}
\subsection{Bayesian Optimization}
\label{sec:sanity_check}
\begin{figure}[H]
    \centering
    \includegraphics[width=0.5\linewidth]{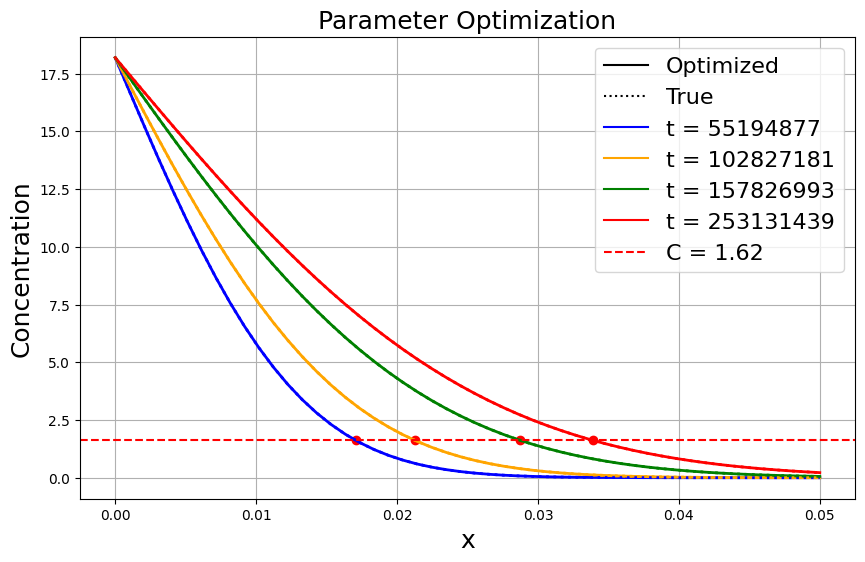}
    \caption{Sanity Check:True vs. Predicted results}
    \label{fig:sanity_check_plot}
\end{figure}
\begin{figure}[H]
    \centering
    \begin{subfigure}[b]{0.49\textwidth}
        \centering
        \includegraphics[width=\linewidth]{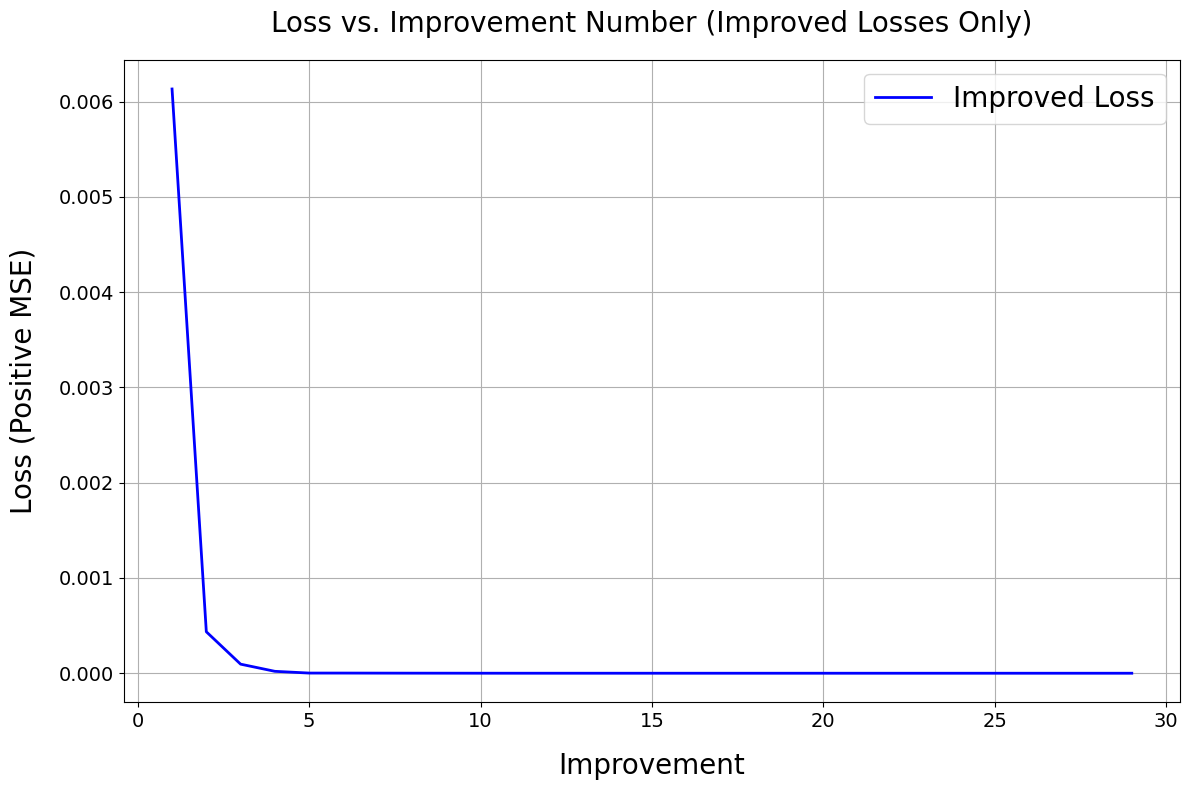}
        \caption{Convergence of loss with improvement steps}
    \end{subfigure}
    \hfill
    \begin{subfigure}[b]{0.49\textwidth}
        \centering
        \includegraphics[width=\linewidth]{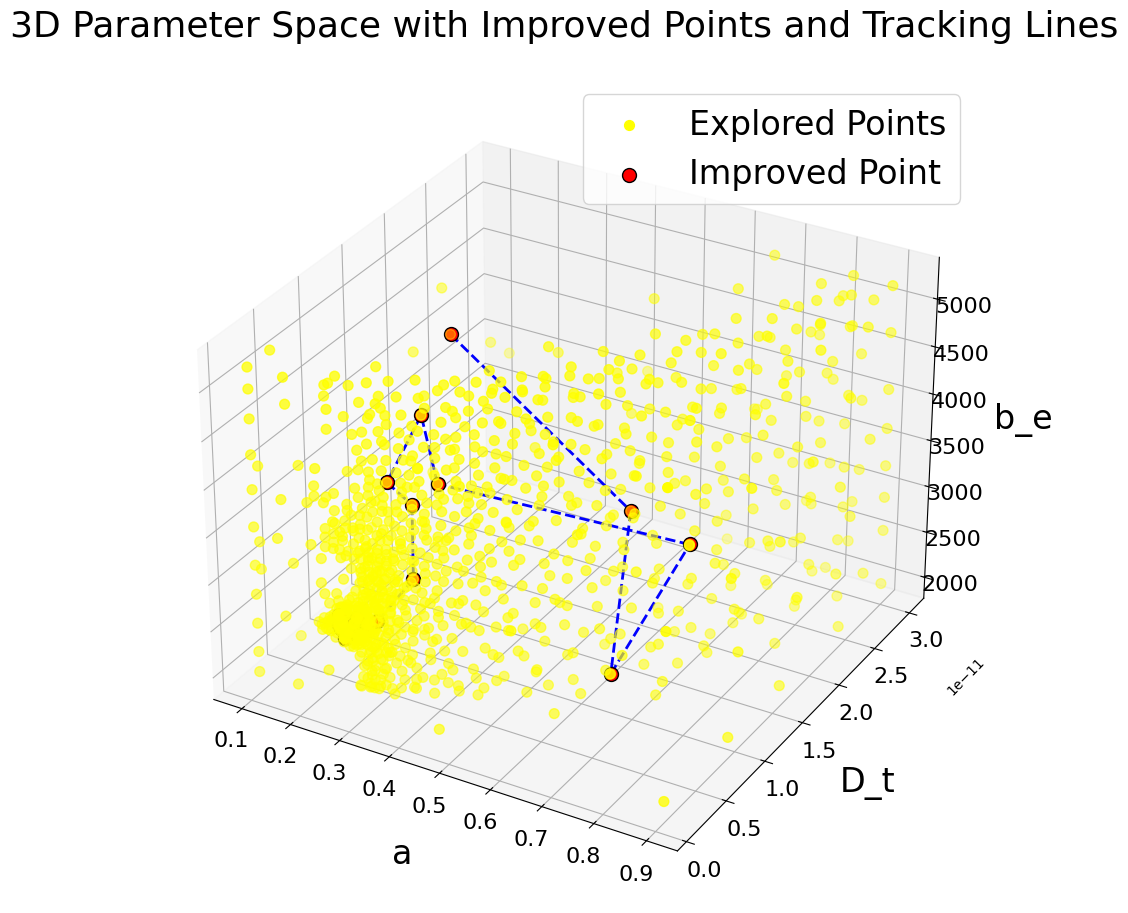}
        \caption{Behavior of acquisition function }
    \end{subfigure}
    \caption{Termination Criteria During Sanity Check}
    \label{fig:sanity_termination}
\end{figure}

\begin{table}[H]
    \centering
    \begin{tabular}{|c|c|c|}
        \hline
        \textbf{Parameter} & \textbf{True Value} & \textbf{Optimized Value} \\ \hline
        \(a\) & 0.2 & 0.20010373848673135 \\ \hline
        \(D_{\text{t}}\) & \(2 \times 10^{-12}\) & \(2.0015664373556944 \times 10^{-12}\) \\ \hline
        \(b_e\) & 2050 & 2054.4906785267476 \\ \hline
    \end{tabular}
    \caption{Comparison of True and Optimized Parameters}
    \label{tab:sanity comparison}
\end{table}
During the sanity check, the parameters \(a\), \(D_{\text{t}}\), and \(b_e\) are assumed to be 0.2, 2e-11, 2050 respectively. These parameters are used to calculate the depth where the critical chloride content is reached for the four last four point in time listed in Table \ref{tab:depth_time} by using equation \ref{eq:analytical_a}. The forward evaluation of the model yields the following depth
\begin{equation}
\textbf{x}_{calculated} = [0.01705657, 0.02125531, 0.02872792, 0.03388954].
\end{equation}
Bayesian optimization is then employed to validate if it can re-identify the original assumed parameters using in advance calculated depth and the corresponding concentration data as defined in Table \ref{tab:depth_time}. The optimized parameters are compared with the assumed ones to verify whether the optimization process can accurately reproduce the initial assumptions as shown in Figure \ref{fig:sanity_check_plot} and Table \ref{tab:sanity comparison}, and the behavior of termination criteria is also confirmed as shown in Figure \ref{fig:sanity_termination} . This comparison ensures that the model behaves as expected, thus validating the optimization approach.

\subsection{Sensitivity Analysis} 
\label{Sensitivity_sanity}
Before conducting the Sobol sensitivity analysis on Gehlen's model, a sanity check was performed using the dummy model to validate the methodology and ensure proper implementation.
\begin{equation}
f(\textbf{X}) = 2X_1^2 + X_2 + 3X_3^3,
\end{equation}
This dummy model was chosen deliberately because its structure allows us to predict the expected contributions of each parameter to the output variance. Specifically:
\begin{itemize}
    \item The term \( 3X_3^3 \) dominates the model because it is cubic, meaning it grows much faster than the quadratic term \( 2X_1^2 \) or the linear term \( X_2 \). As a result, \( X_3 \) is expected to have the highest first-order and total Sobol indices.
    \item The term \( 2X_1^2 \) is quadratic, so its influence on the output is expected to be moderate, placing it second in terms of contribution.
    \item The term \( X_2 \) is linear and has no higher-order effects (e.g., interactions or nonlinearity), so it contributes the least to the output variance.
\end{itemize}
The Sobol analysis results matched these expectations:
\begin{itemize}
    \item Parameter \( X_3 \) had the highest first-order and total Sobol indices, indicating its dominant influence on the output, as predicted by its cubic contribution.
    \item Parameter \( X_1 \) ranked second, consistent with its quadratic nature.
    \item Parameter \( X_2 \) had the lowest indices, aligning with its linear and non-dominant contribution.
\end{itemize}
These results confirmed that the Sobol sensitivity analysis correctly captured the relative contributions of the parameters, validating the implementation. The analysis provided confidence in proceeding with the main study using this robust methodology.
\begin{figure}[H]
    \centering
    \begin{subfigure}[b]{0.45\textwidth}
        \centering
        \includegraphics[width=\linewidth]{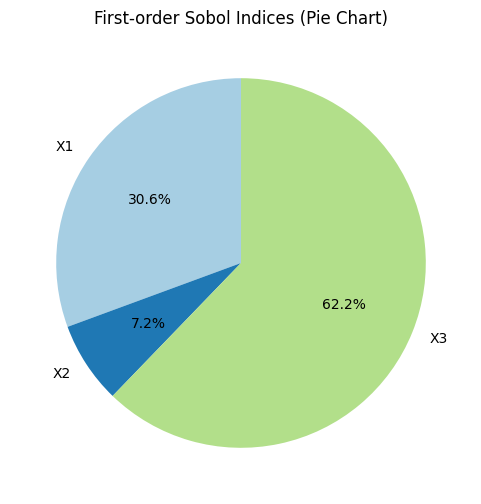}
        \caption{First-order Sobol indices for the input parameters.}
    \end{subfigure}
    \hfill
    \begin{subfigure}[b]{0.45\textwidth}
        \centering
        \includegraphics[width=\linewidth]{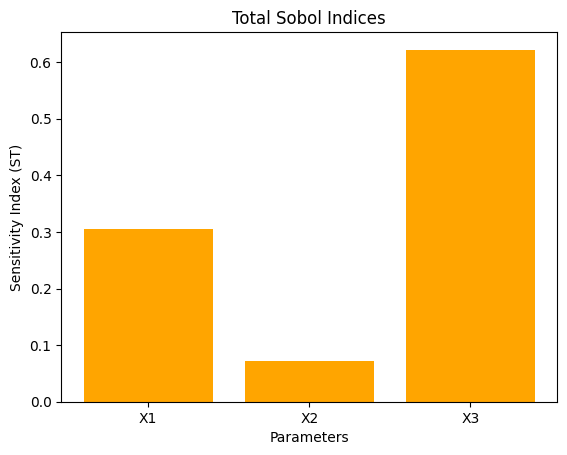}
        \caption{Total Sobol indices, showing overall contribution to variance. }
    \end{subfigure}
    \caption{Sobol indices sanity check}
    \label{fig:sobol_sanity}
\end{figure}

%End of document
\end{document}